\documentclass[%
 reprint,
unsortedaddress,
 amsmath,amssymb,
 aps,
prb,
floatfix,
]{revtex4-2}

\usepackage{graphicx}
\usepackage{dcolumn}
\usepackage{bm}

\begin{document}
\title{Many-Body Physics in Small Systems: Observing the Onset and Saturation of Correlation in Linear Atomic Chains}
\author{Emily Townsend}
\email[]{emily.townsend@nist.gov}
\affiliation{Nanoscale Device Characterization Division and Joint Quantum Institute, National Institute of Standards and Technology, Gaithersburg, Maryland 20899-8423, USA; and University of Maryland, College Park, Maryland 20742, USA}
\author{Tom\'a\v s Neuman}
\affiliation{Centro de F\'isica de Materiales CFM - MPC, Centro Mixto CSIC-UPV/EHU, 20018 San Sebasti\'an-Donostia, Basque Country, Spain}
\affiliation{Donostia International Physics Center (DIPC), 20018 San Sebasti\'an-Donostia, Basque Country, Spain}
\affiliation{Nanoscale Device Characterization Division and Joint Quantum Institute, National Institute of Standards and Technology, Gaithersburg, Maryland 20899-8423, USA; and University of Maryland, College Park, Maryland 20742, USA}
\affiliation{Institut de Physique et Chimie des Matériaux de Strasbourg, Université de Strasbourg, 23, rue du Loess 67034, Strasbourg France}
\author{Alex Debrecht}
\affiliation{Nanoscale Device Characterization Division and Joint Quantum Institute, National Institute of Standards and Technology, Gaithersburg, Maryland 20899-8423, USA; and University of Maryland, College Park, Maryland 20742, USA}
\affiliation {Department of Physics and Astronomy,  University of Rochester, Rochester, New York 14627, USA}
\author{Javier Aizpurua}
\affiliation{Centro de F\'isica de Materiales CFM - MPC, Centro Mixto CSIC-UPV/EHU, 20018 San Sebasti\'an-Donostia, Basque Country, Spain}
\affiliation{Donostia International Physics Center (DIPC), 20018 San Sebasti\'an-Donostia, Basque Country, Spain}
\author{Garnett Bryant}
\affiliation{Nanoscale Device Characterization Division and Joint Quantum Institute, National Institute of Standards and Technology, Gaithersburg, Maryland 20899-8423, USA; and University of Maryland, College Park, Maryland 20742, USA}

\date{\today}

\begin{abstract}
The exact study of small systems can guide us toward relevant measures for extracting information about many-body physics as we move to larger and more complex systems capable of quantum information processing or quantum analog simulation.  We use exact diagonalization to study many electrons in short 1-D atom chains represented by long-range extended Hubbard-like models.  We introduce a novel measure, the Single-Particle Excitation Content (SPEC) of an eigenstate and show that the dependence of SPEC on eigenstate number reveals the nature of the ground state (ordered phases), and the onset and saturation of correlation between the electrons as Coulomb interaction strength increases.  We use this SPEC behavior to identify five regimes as interaction is increased: a non-interacting single-particle regime, a regime of perturbative Coulomb interaction in which the SPEC is a nearly universal function of eigenstate number, the onset and saturation of correlation, a regime of fully correlated states in which hopping is a perturbation and SPEC is a different universal function of state number, and the regime of no hopping.  
In particular, the behavior of the SPEC shows that when electron-electron correlation plays a minor role, all of the lowest energy eigenstates are made up primarily of single-particle excitations of the ground state, and as the Coulomb interaction increases, the lowest energy eigenstates increasingly contain many-particle excitations. In addition, the SPEC highlights a fundamental, distinct difference between a non-interacting system and one with minute, very weak interactions.
While SPEC is a quantity that can be calculated for small exactly diagonalizable systems, it guides our intuition for larger systems, suggesting the nature of excitations and their distribution in the spectrum. Thus, this function, like correlation functions or order parameters, provides us with a window of intuition about the behavior of a physical system.
\end{abstract}
\maketitle

\section{\label{Introduction}Introduction}
Quantum simulation of small physically realizable systems (e.g. chains of precision-placed atoms on surfaces or dopant atoms in silicon) provides an opportunity to learn about many-body physics at larger scales.  While larger scale quantum simulators with fifty to hundreds of atoms are becoming possible \cite{Bohnet2016, Zhang2017, Bernien2017},
the majority are still much smaller  \cite{Tarruell2018, Hensgens2017, Salfi2016, Wyrick2019, Le2017}, particularly in solid-state realizations.  Studying these smaller systems theoretically has the advantage that we can exactly diagonalize their Hamiltonian, inspect the full spectrum of their eigenstates, and learn what both the ground and the excited states of that spectrum can reveal about the nature of the system. 
 Even for small systems we gain insight into many-body behavior, both from a theoretical perspective and with an eye toward experimental realization of quantum simulators for these systems.  Hubbard model realizations are becoming a common stepping stone on the road to building universal quantum computers \cite{Preskill2018a, McArdle2020, Wecker2015, Loss1998}, and are being developed in ultracold atom systems in optical lattices \cite{Tarruell2018} and solid state systems, such as gate-defined quantum dots \cite{Hensgens2017}, and donor dots in semiconductors \cite{Salfi2016, Wyrick2019, Le2017}.  
In this work we use exact diagonalization of a small spinless electron system to find all of the many-body eigenstates, which gives us access to a wide range of exact quanties.  We calculate the extent to which each of the eigenstates consists entirely of single particle excitations of the ground state of the system, which we refer to as the single particle excitation content (SPEC) of an eigenstate. Expanding our scope to all of the eigenstates provides a significant new perspective beyond what can be learned from examining only the ground state, in addition to confirming previously known ground-state behaviors in a new way.  The SPEC of the excited states makes visible a fundamental difference between an unperturbed Hamiltonian and one with a minute perturbation, no matter how small.  It also provides a division of the parameter-space of our Hamiltonian into regimes which we can identify as those with different ground state behaviors.

We work with a linear chain of atoms, half-filled with spinless electrons, which we describe using a long-range extended Hubbard model.  By extended we mean that unlike a typical Hubbard Hamiltonian which has only on-site interactions between the electrons and a hopping kinetic energy, we use a Coulombic ($\sim 1/r$ with $r$ the distance between charges) interaction between electrons and between electrons and the atomic cores. 
We vary the ratio of Coulomb interaction strength, $\lambda_{ee}$, to the hopping, $t$, to examine the different regimes of behavior that this model gives rise to.   As the Coulomb interaction is turned on it causes correlation between the electrons and then eventually strong Wigner crystallization that isolates electrons to individual sites of the lattice in an every-other-site pattern \cite{Wigner1934}.

The range of the electron-electron interaction, whether short or long range, plays an important role in defining the physics of  interacting systems. 
Long-range interactions allow the transfer of information and the spread of entanglement to exceed the Lieb-Robinson bound \cite{Eldredge2017}, which describes entanglement spread under only local interactions, and also implies our ability to efficiently simulate a one-dimensional (1-D) system classically, e.g. using density matrix renormalization group (DMRG) or matrix product states \cite{Eisert2010}.  In a 1-D system with only nearest-neighbor hopping and interaction, information transfer will be local, the system integrable, and the system will not thermalize following a quench, whereas next nearest-neighbor hopping and interactions break integrability, leading to quantum chaotic behavior and thermalization \cite{Santos2010}. 

The strength of the interaction relative to the hopping is a key parameter which defines the phases of these sytems. One-dimensional fermion systems have been studied extensively, e.g. \cite{Luttinger1963, Giamarchi2003, Schonhammer2013}, often with an emphasis on short-range interactions.  When the fermions experience long-range Coulomb repulsion, Schulz \cite{Schulz1993} showed using bosonization that the ground state is a Wigner crystal (WC)-like state for a continuous, infinite 1-D region.  The defining feature of this WC-like state is quasi-long-range order (quasi-LRO) in which the density-density correlation function shows an incipient charge density wave that decays slower than a power-law,  $\sim e^{-\sqrt{\alpha \ln x}}$ with $\alpha \sim t/\lambda_{ee}$, so with a stronger interaction the quasi-LRO decays more slowlly.    While Schulz showed that this is the ground state at any strength of the Coulomb interaction, once a finite lattice of atomic sites is introduced there will be several different ground-state phases as one tunes the strength of interaction \cite{Capponi2000, Valenzuela2003, Li2019, Ren2020}.  Different authors disagree on how to name these phases, but they broadly agree on many of their characteristics.  Here we briefly describe the previous work, but delay discussing these characteristics until the results section for ease of comparison. 

In 1978, Hubbard \cite{Hubbard1978} considered a Hubbard model with long-range but convex interactions and no hopping, which allows analytical solution by considering how to minimize the energy of placing $m_e$ classical electrons on $N_s$ sites.  He named this ground state a generalized Wigner lattice.  At half fiilling the doubly degenerate ground states have electrons only on either odd or even sites.  This corresponds to the $t=0$ limit that we will discuss with our model.

When hopping is included, numerical solutions are typically needed. In 2000 Capponi {\it et al.} \cite{Capponi2000}  considered spinless fermions with Coulomb repulsion on a lattice of varying lengths with periodic boundary conditions using exact diagonalization. By considering the thermodynamic limit of infinite chain length they investigated whether the system would be insulating or metallic, and looked at how their numerical results departed from  analytical predictions for a Luttinger liquid.  
More recently, Li {\it et al.} \cite{Li2019} extended the work of Capponi by studying larger lattices also with periodic boundary conditions, with long-range interactions of varying power laws, including a Coulombic $1/r$, using density matrix renormalization group, rather than exact diagonalization.
Finally, Ren {\it et al.} \cite{Ren2020} used DMRG to study the phase diagram of the XXZ model of an anisotropic spin chain (a static 1-D lattice of spin-1/2 particles with long-range interactions via Pauli spin matrices, also with periodic boundary conditions).  While their model should map to the interacting fermion model of Capponi \cite{Capponi2000} and Li \cite{Li2019}, they do observe differences in the phase diagram, including a region of intermediate strength Coulomb interaction with the ground state in a phase that corresponds to Luttinger liquid behavior in the fermion system, a phase absent from Li \cite{Li2019} and Ren's \cite{Ren2020} studies of fermions with long-range interactions.  

In 2003, Valenzuela {\it et al.} \cite{Valenzuela2003} used a variational ansatz wave function to describe a smooth crossover between Hubbard's generalized Wigner lattice behavior and a state with weak charge density modulation as well as delocalized charge.  (For fillings smaller than $1/2$ they also obtain a phase similar to to Schultz's with quasi-LRO.) 
 Their variational ansatz differs from the numerical approaches discussed so far, yielding only an approximation for the ground state, but not any  excited eigenstates.

This paper is organized as follows:  Subsection \ref{subsect:MM} on the model and methods describes the details of the Hamiltonian we study and defines both single-particle excitations and what we mean by the ``single particle excitation content'' of an eigenstate.  In section \ref{sect:SPEC} we then   describe the behavior of the SPEC and show how SPEC allows us to identify five regimes of behavior for different interaction strengths.  These regimes are the non-interacting case, the no hopping case, perturbative regimes around each of these cases, and the intermediate regime of onset and saturation of correlation.  We finish with a conclusion in section \ref{sect:Conc}.

\subsection{Model and Methods} \label{subsect:MM}
We consider linear chains consisting of $N_s$ atoms 
 at fixed sites (indexed by $i$ and $j$, at positions $x_i$, $x_j$ with unit spacing) with $m_e$ spinless electrons moving from site to site via a nearest-neighbor hopping, $t$. 
  An electron on site $i$  interacts Coulombically with the other electrons ($V_{\rm ee}$) and with the nuclei of each of the atoms ($V_{\rm nuc}$).  Results presented here are for charge neutral systems, in which each site has a nuclear charge of $Z = m_e / N_s$, so the attractive interaction between an electron at $x_i$ and each of the atoms (at positions $x_j$) is 

\begin{equation}
V_{\rm nuc}(x_i) = \sum_j \frac{-\lambda_{\rm nuc} Z }{ (|x_i-x_j| + \zeta_{\rm nuc})},
\end{equation}
while the repulsive interaction between electrons at $x_i$ and $x_j$ consists of a direct Coulomb interaction reduced by exchange which we assume affects only nearest neighbor electrons:
 \begin{equation}
V_{\rm ee}(x_i,x_j) = \frac{\lambda_{\rm ee}(1-f_{\rm ex}\delta_{|i-j|,1}) }{(|x_i-x_j| + \zeta_{\rm ee})}
\end{equation}
where $\lambda_{\rm ee} $ and $\lambda_{\rm nuc}$ are scale factors accounting for the strength of these interactions,  including any dielectric screening as well as the size of the lattice spacing.  Variables $\zeta_{\rm ee}$ and $\zeta_{\rm nuc}$ are cutoffs that account for the spread of the electron orbital on a site.  In all results presented here  $\zeta_{\rm ee} = \zeta_{\rm nuc} = 0.5$, or half a lattice spacing.
We assume each site has a single accessible orbital, so with spinless electrons each site can accommodate only one electron. 
The fraction by which the nearest-neighbor electron interaction is reduced due to exchange is $f_{\rm ex} = 0.2$ 
 in the results presented here (but see Supplemental Information \cite{SuppInfo} for results for other values of the exchange fraction, as well as modifications of the range of the Coulomb/nuclear interaction, filling factor and system size).

The full Hamiltonian is then: 
\begin{equation}
\hat{{\cal H}} = \sum_{i = 1}^{N_s} \left( - t(\hat{c}_i^\dagger \hat{c}_{i+1} + \hat{c}_{i+1}^\dagger \hat{c_i}) +V_{\rm nuc}\hat{n}_i + \sum_{j=1}^{i-1}V_{\rm ee}\hat{n}_i\hat{n}_j \right)
\end{equation}
We express the Hamiltonian in a many-electron site basis and solve for the many-electron eigenstates and energies by direct diagonalization (LAPACK dsyev).
  For comparison with theories of bulk materials, the value of the ratio $\lambda_{\rm ee}/t $ corresponds to the ratio of the Wigner-Seitz radius ($r_s = L/2m_e,$) to the Bohr radius ()$a_0 = \hbar^2/me^2$) (in which $L$ is the length of the system, $m_e$ is the number of electons, $\hbar$ is Planck's constant and $m$ and $e$ are the mass and charge of the electron).  Thus small  $\lambda_{\rm ee}/t $ corresponds to the limit of high electron density and small Wigner-Seitz radius, in which hopping is relatively more important than the Coulomb interactions.

\subsubsection{Single-particle excitations}

 For zero Coulomb interaction, the many-electron eigenstates $\Psi_N (x_1, ...,x_{m_e}) $ (where $N$ is the many-electron eigenstate index and $x_1, ...,x_{m_e}$ are the position coordinates for the $m_e$ electrons) are each a single determinant of $m_e$ single-electron eigenstates (non-interacting modes) that are the solutions of the same system with one electron, $\phi_{n}(x)$ (where $n$ labels the single-particle eigenstates that make up the $N$th many-electron eigenstate and $x$ is the position coordinate for a single electron):
\begin{equation}
\langle x_1, ...,x_n | \Psi_N^{\lambda_{\rm ee} = 0} \rangle = \frac{1}{{\sqrt{ {m_e}! } }} \left\vert 
{\begin{array}{ccc}
\phi_{n_1}(x_1) & \cdots & \phi_{n_{m_e}} (x_1) \\
  \vdots  & &  \vdots  \\
\phi_{n_1}(x_{m_e}) & \cdots & \phi_{n_{m_e}}(x_{m_e}). \\
\end{array}} \right\vert
\label{eqn:det}
\end{equation}

When the Coulomb interaction is turned on, the many-electron eigenstates are superpositions of many determinants with increasing departure from the single-determinant behavior as the interaction strength increases.

The single-particle excitation of the many-electron ground state that takes a single particle from site $j$ to site $i$ can be written as 
$$
c_i^\dagger c_j | \Psi_{GS} \rangle, i\neq j
$$
where $c_i^\dagger$ ($c_i$) is an operator that creates (destroys) a particle at site $i$. 
The set of all single-particle excitations is defined by the above states for all values of $i$ and $j$.  The full set can be equivalently defined by the states:
$$
a_m^\dagger a_n | \Psi_{GS} \rangle, m \neq n
$$
where  $a_m^\dagger$ ($a_m$) instead creates (destroys) an electron in the non-interacting single-electron state $\phi_m$.  Similarly, a two-particle excitation of the ground state consists of $ a_m^\dagger a_n^\dagger a_p a_q | \Psi_{GS} \rangle, m \neq n \neq p \neq q$ (or a similar construction with site creation and destruction operators).

A many-body excited state can be characterized by the number of one-, two-, three-, ... particle excitations that make it up. We show in this paper that the single-particle excitation content of the many-body excited states plays an important role, providing a new way to characterize the effects of Coulomb interactions on the many-body states. We focus on the extent to which different many-electron eigenstates consist of single-particle excitations of the interacting ground state.  For this purpose we define an orthonormal basis $|u_i \rangle$ that spans the full set of all single-particle excitations, and a projection operator, $\hat{P}_{SPE} = \sum_i |u_i \rangle \langle u_i |$, that projects onto that subspace.  Computationally we find the spanning orthonormal basis by Gram-Schmidt decomposition:  For each particular vector representing an excitation of the ground state, $c_i^\dagger c_j | \Psi_{GS} \rangle$, we create a basis vector $|u_{i'} \rangle$ by normalizing the vector that consists of the components of the excitation vector that are orthogonal to the ground state and to all previous basis vectors $|u_{i''}\rangle (i''<i')$. (In the non-interacting case single-particle excitations will be eigenstates of the Hamiltonian and will already be orthogonal, but this is not true when interactions are present.) Because some of those components may be small, the normalization of the orthogonalized basis state has the effect of magnifying the components that are kept to make a new orthonormal basis state.  We apply a cutoff, only including a new orthonormal basis state if the sum of the magnitudes of all of the components to be kept to make the new orthogonal basis state is, before normalization, greater than $\epsilon_{\text{G-S}}$.  As we will discuss later, the choice of cutoff can affect our estimate of single-particle excitation content when interactions are weak and the interacting ground state includes many small single-electron excitations of the non-interacting ground state.
Correct choice of cutoff is thus a regularization of the theory needed to get physically meaningful results:  The cutoff needs to be chosen to exclude numerical error in the non-interacting and no-hopping cases, however the most inclusive cutoff is the most accurate in other cases. If a more exclusive cutoff were used, the order that the single-particle excitations are included in the Gram-Schmidt process could affect (slightly) the SPEC, however with an inclusive cutoff the answer is order independent.
The size of the single particle excitation basis,  ${|u_i \rangle}$, is $m_e(N_s-m_e)$ for a noninteracting system (or a system with no hopping) when the ground state consists of exactly $m_e$ fully-occupied modes (sites) and $N_s - m_e$ completely unoccupied ones and is $N_s^2 -1$  for an interacting system with hopping when the ground state consists of modes or sites that are occupied with non-unity, non-zero probability. There are $N_s^2$ combinations of $c_i^\dagger c_j$, but the ground state itself is not part of the single-particle excitation basis.

The single-particle excitation content (SPEC) of a particular many-electron eigenstate,  $| \Psi_N \rangle $, is then
\begin{equation}
\langle \Psi_N | \hat{P}_{SPE} |\Psi_N \rangle = \left\vert \sum_i \langle \Psi_N | u_i \rangle\right\vert^2.
\end{equation}
The SPEC is thus the probability that an eigenstate can be found in the single-particle excitation subspace, or alternatively, the extent to which it can be defined entirely as a linear combination of single-particle excitations of the ground state.
To simplify the display of the information contained in the plots of single-particle excitation content, we can also plot a rolling partial sum of the SPEC over the  $N$ lowest many-body eigenstates:
\begin{equation}
 \sum_{N'=1}^N \left\vert \sum_i \langle \Psi_{N'} | u_i \rangle \right\vert^2.
\end{equation}
The single-particle excitation content is independent of which set of single-particle excitations we use, those from the site basis or the mode basis.

Because any linear combination of single-particle excitations will live entirely in this single-particle excitation subspace, the remaining subspace consists of correlated particle excitations in which two or more particles are moved coherently.  The excitation $ c_i^\dagger c_j c_k^\dagger c_l |\Psi_{GS}\rangle$ is distinct from the excitation  $ (c_i^\dagger c_j + c_k^\dagger c_l) |\Psi_{GS}\rangle$, the latter being entirely a single-particle excitation.

The single-particle excitation content is distinct from but related to the quasiparticle weight.  The relationship between the two is considered in the Supplemental Information \cite{SuppInfo} to this paper and references therin \cite{Coleman2015, Schofield1999}.

\section{SPEC: Identifying Five Regimes of Interaction Effects }\label{sect:SPEC}
\begin{figure}
\includegraphics[width=0.49\textwidth]{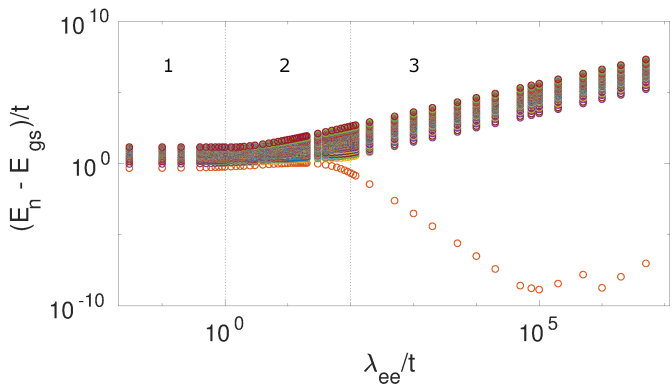}
 \caption{Excitation energies (log scale) for all many-body eigenstates of 6 electrons on a chain of 12 atoms, exchange fraction of 0.2.  The horizontal axis is the ratio of the Coulomb interaction to the hopping on a logarithmic scale.  Vertical dotted lines indicate regions of different behavior discussed in the text.}
\label{fig:Energies0.2}
\end{figure}

In the following, we will show that the SPEC can be used to characterize the effects of the electron-electron interaction. We will show that there are five regimes of behavior identified by the very different functional form of SPEC as a function of eigenstate number in each regime. As we vary the ratio of the electron-electron (and electron-nuclear) interactions to the hopping in the Hamiltonian for linear atomic chains, we use the behavior of the SPEC to identify two integrable cases ($\lambda_{ee}/t = 0$ and $t/\lambda_{ee} = 0$) and three regimes in between, as suggested by the dashed lines in the plot (Fig. \ref{fig:Energies0.2}) of the variation of the excitation energy for excited states with $\lambda_{ee}/t$.  
(Figure  \ref{fig:Energies0.2} shows the difference in energy from the ground state on a log scale for all the excited states of 6  electrons on a chain of 12 atoms when the exchange reduction is 0.2.  The horizontal axis is the ratio of the Coulomb interaction to hopping on a log scale.   Throughout the main paper we show plots for the half-filled 12 atom chain.  However results are similar for other size chains, see Supplemental Information \cite{SuppInfo}.)
In region 1, the excitation energies depend only weakly on the interaction strength. In region 2, the excitation energies exhibit significant increase with increasing interaction as well as significant crossings and mixings of levels. In region 3, the excitation energies scale linearly with the strength of the interaction as the hopping becomes much smaller than the interaction: correlation saturates because movement of the electrons is suppressed.  In addition, the ground state is becoming degenerate.  We will discuss the non-interacting regime (subesection \ref{subsect:NI}), the no-hopping regime (\ref{subsect:NH}), and then regimes 1 (\ref{subsect:Weak}), 3 (\ref{subsect:Strong}), and 2 (\ref{subsect:Int}) in turn.

\subsection{Non-interacting regime, $\lambda_{ee}/t = 0$}\label{subsect:NI}
When there are no Coulomb interactions between electrons, the ground state consists of a single Slater determinant of the $m_e$ lowest single-particle states of a finite 1-D chain.  Each excited state consists of a Slater determinant of single-particle states in which one or more of the  $m_e$ lowest single-particle states is replaced by a higher energy single-particle state.  Thus each excited state consists entirely of either a single- or a multiple-particle excitation of the ground state. This is seen in Figure \ref{fig:SPContent}a which shows the SPEC of the excited states of the non-interacting system of 6 electrons on a chain of 12 atoms.  The eigenstates that are single-particle excitations (states 1-11, 13-19, etc) have a SPEC of one and all other states have zero SPEC.  The partial sum of the SPEC of the N lowest eigenstates, shown as the black curve in figures \ref{fig:PartialSumSmallCoul}a and b rises to $m_e *(N_s - m_e)$ in the lower part of the spectrum of many-body eigenstates and then remains constant, indicating that all of the possible single-particle excitations are used up by the low-energy many-body states.  This free-fermion regime can be described by a Luttinger liquid \cite{Haldane1981}, whereas the inclusion of long-range interactions induces departures from Luttinger behavior \cite{Capponi2000}.  Likewise, the SPEC for $\lambda_{ee}/t = 0$ with  $m_e *(N_s - m_e)$ excitations is distinct from the SPEC for small $\lambda_{ee}/t$, even in the limit of very small $\lambda_{ee}/t$ because there are always  $N_s^2 -1$ single-particle excitations when interactions are included.

\subsection{No hopping regime, $t/\lambda_{ee} = 0$} \label{subsect:NH}
\begin{figure}
 \includegraphics[width=0.45\textwidth]{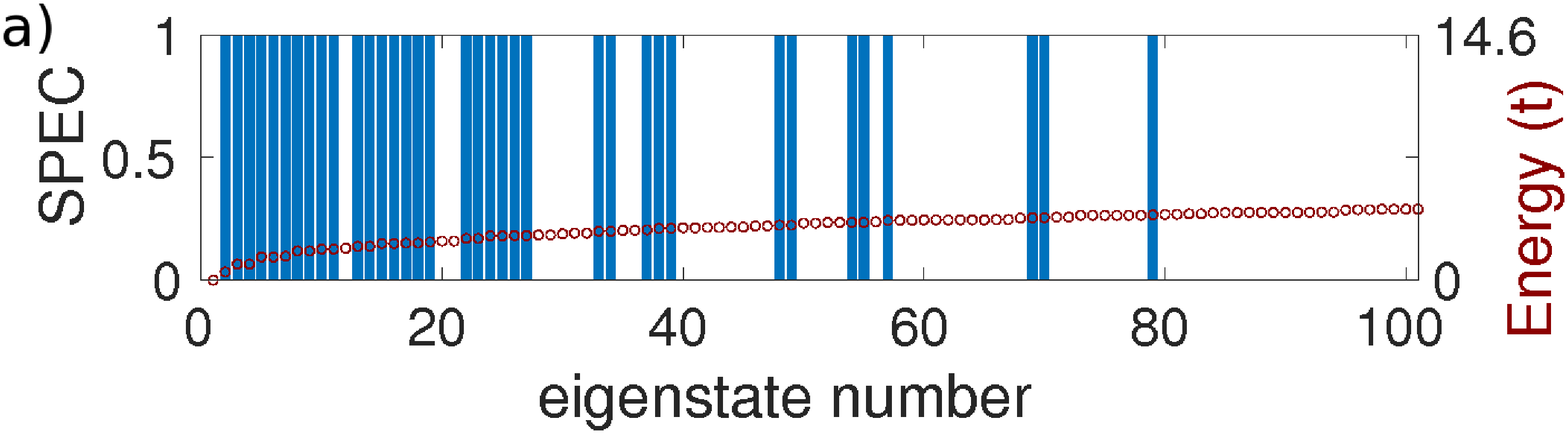}
 \includegraphics[width=0.45\textwidth]{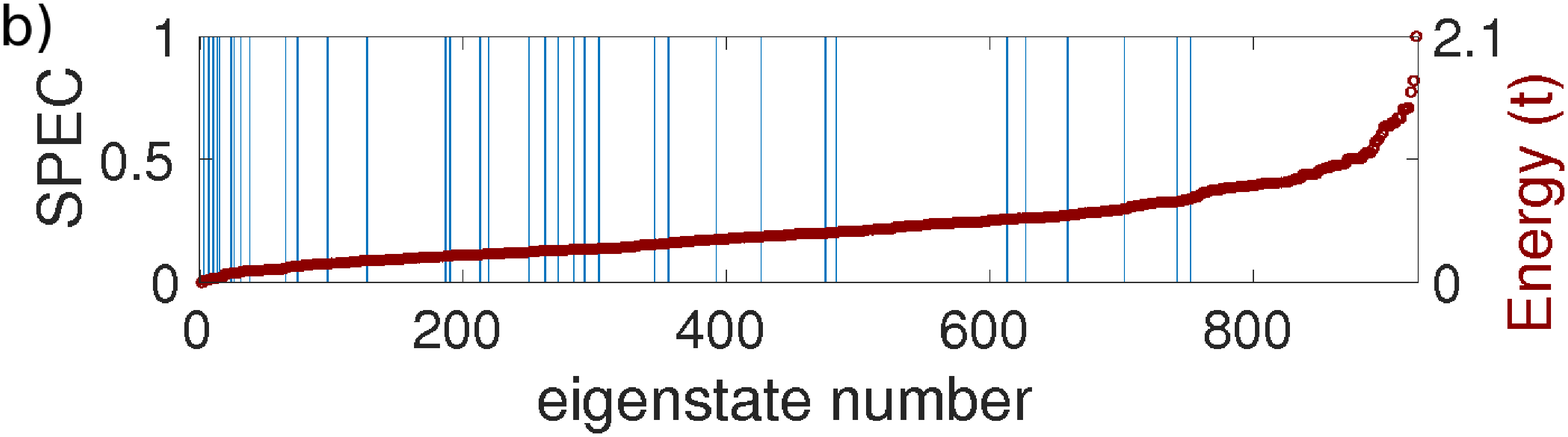}
 \includegraphics[width=0.45\textwidth]{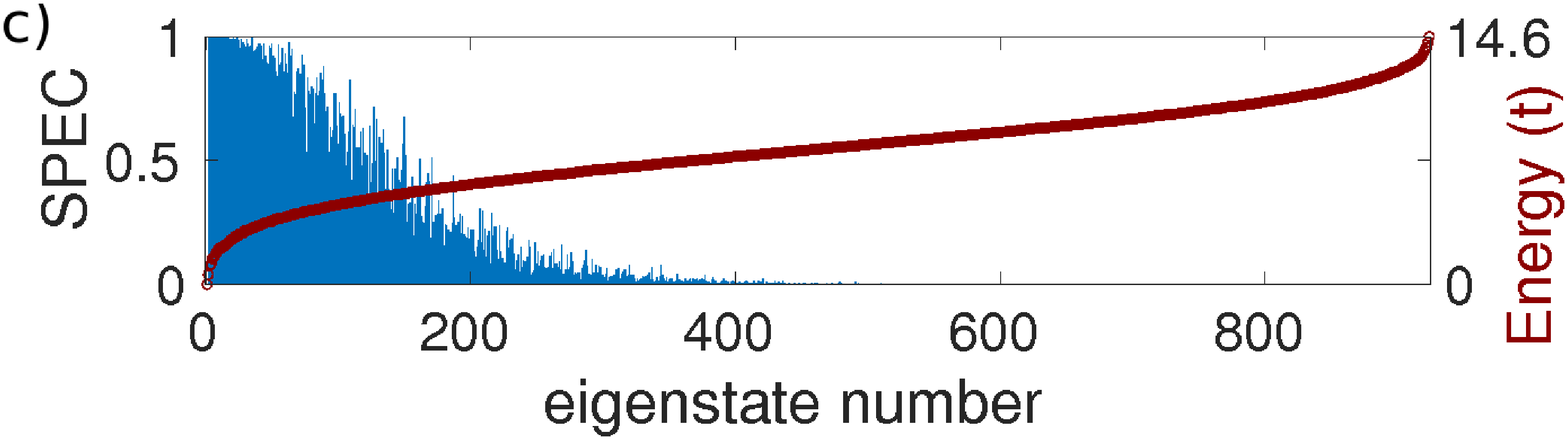}
 \caption{Single Particle Excitation Content for eigenstates of a 12 atom chain with six electrons, $|\sum_i \langle \Psi_N | u_i \rangle |^2$, for a) $\lambda_{ee} =0$, $t=1$ (no states above 100 have any single-particle content), b) $t=0$, $\lambda_{ee} =0.5$ and c) $\lambda_{ee} =t=1$. There are  ${12 \choose 6}$ $=924$ eigenstates for this system.  The red line indicates the relative excitation energy of each eigenstate.}
\label{fig:SPContent}
\end{figure}

\begin{figure}
 \includegraphics[width=0.49\textwidth]{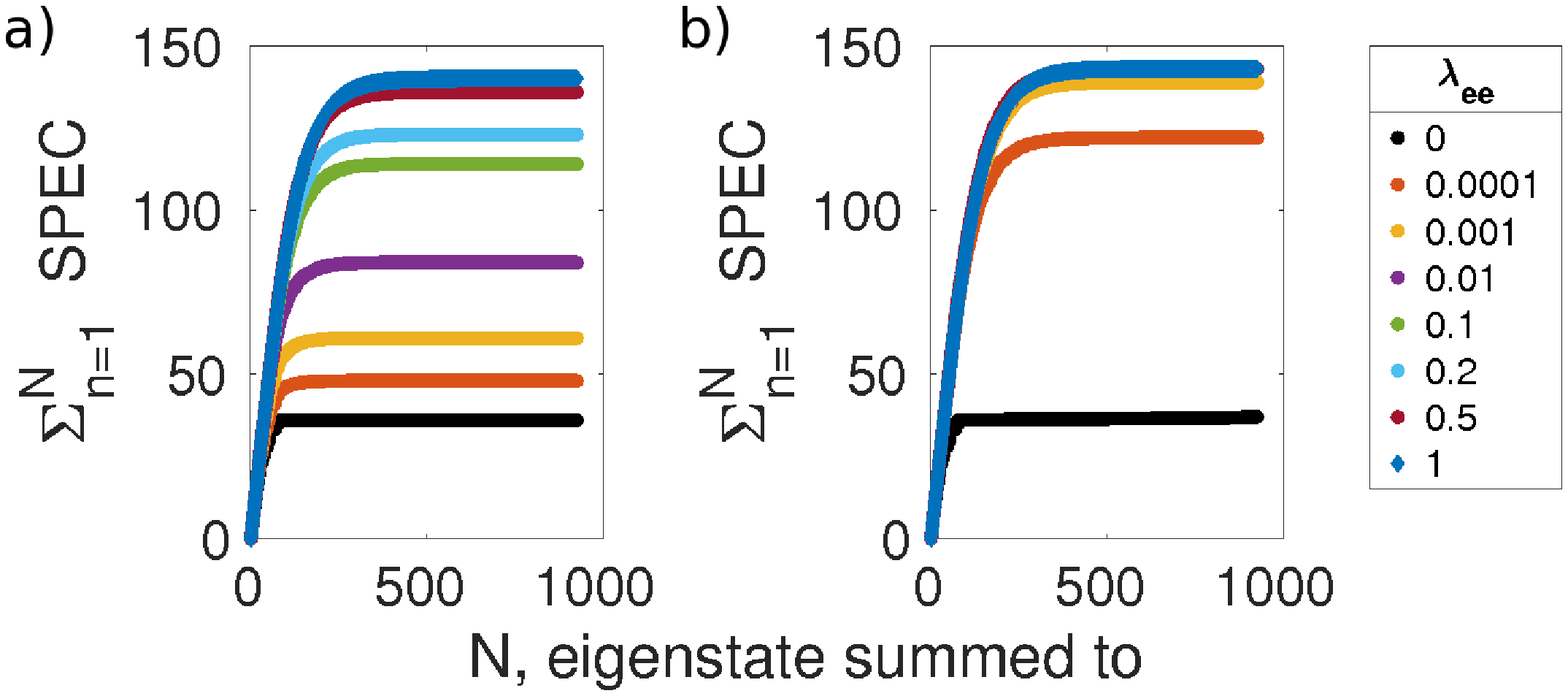}
 \includegraphics[width=0.49\textwidth]{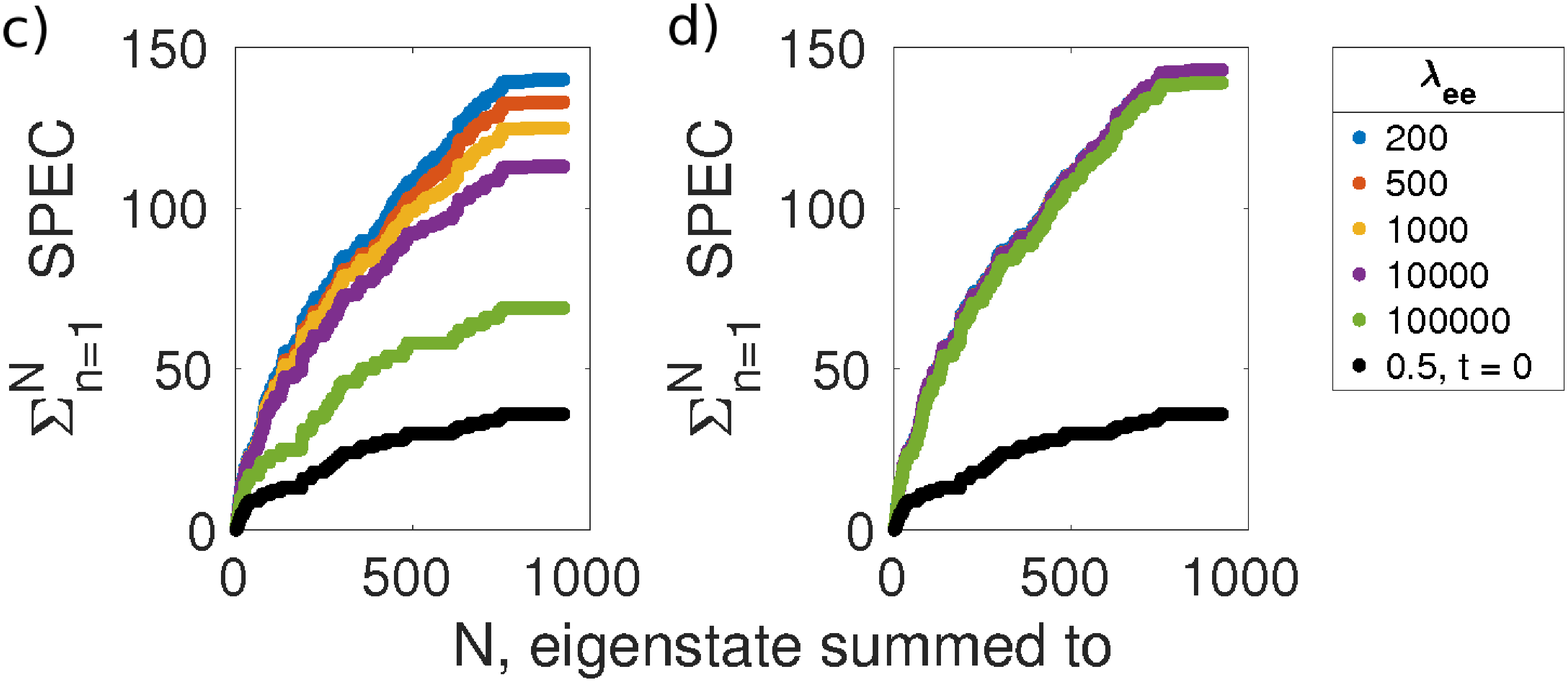}
 \caption{Rolling sum of the Single Particle Excitation Content for eigenstates of a 12 atom chain with six electrons, for the perturbative regimes near the non-interacting state (upper two panels) and near the Wigner state (lower two panels).  The right two panels have a more inclusive cutoff ($\epsilon_{\text{G-S}} = 10^{-18}$) while the left two panels have a more exclusive cutoff ($\epsilon_{\text{G-S}} = 10^{-7}$). With an inclusive cutoff, SPEC curves are universal for a wide range of interaction strengths (curves overlap for many values).}
\label{fig:PartialSumSmallCoul}
\end{figure}

When there is no hopping between sites, each eigenstate consists of a Slater determinant of single-particle states that are localized to sites.  As discussed by Hubbard \cite{Hubbard1978}, the ground state is a Wigner crystal (``generalized Wigner lattice''), the exact details of which are determined by the exchange fraction, $f_{ex}$, the number of sites and the filling.  For six electrons on 12 atoms with $f_{ex} = 0.3$ the ground state has electrons on sites 2,3,6,7,10 and 11, which we refer to as a paired Wigner crystal.  When  $f_{ex} = 0.2$ (or any value less than 0.22) the ground state is degenerate, with one state having electrons on sites 2,3,5,7,9, and 11, and the other on sites 2,4,6,8,10, and 11.  Sites 1 and 12 are not occupied in the ground state because the nuclear attraction pulls electrons toward the center of the chain.   The states which are single particle excitations (of just one of the degenerate ground states) are shown in Figure \ref{fig:SPContent}b.  The partial sum of the the SPEC (the black curve in figures \ref{fig:PartialSumSmallCoul}c and d) again shows that there are only  $m_e *(N_s - m_e)$ single-particle excitations, though they are no longer confined to the lowest part of the spectrum, since moving multiple electrons at the same time is often a lower energy excitation.  Again, SPEC clearly shows that this regime is distinct from the large $\lambda_{ee}/t$ limit where, for interacting electrons, there are  $N_s^2 -1$ single-particle excitations, no matter how strong the interaction.

\subsection{Weak interaction, $\lambda_{ee}/t \leq 1$}\label{subsect:Weak}
\begin{figure}
 \includegraphics[width=1\columnwidth]{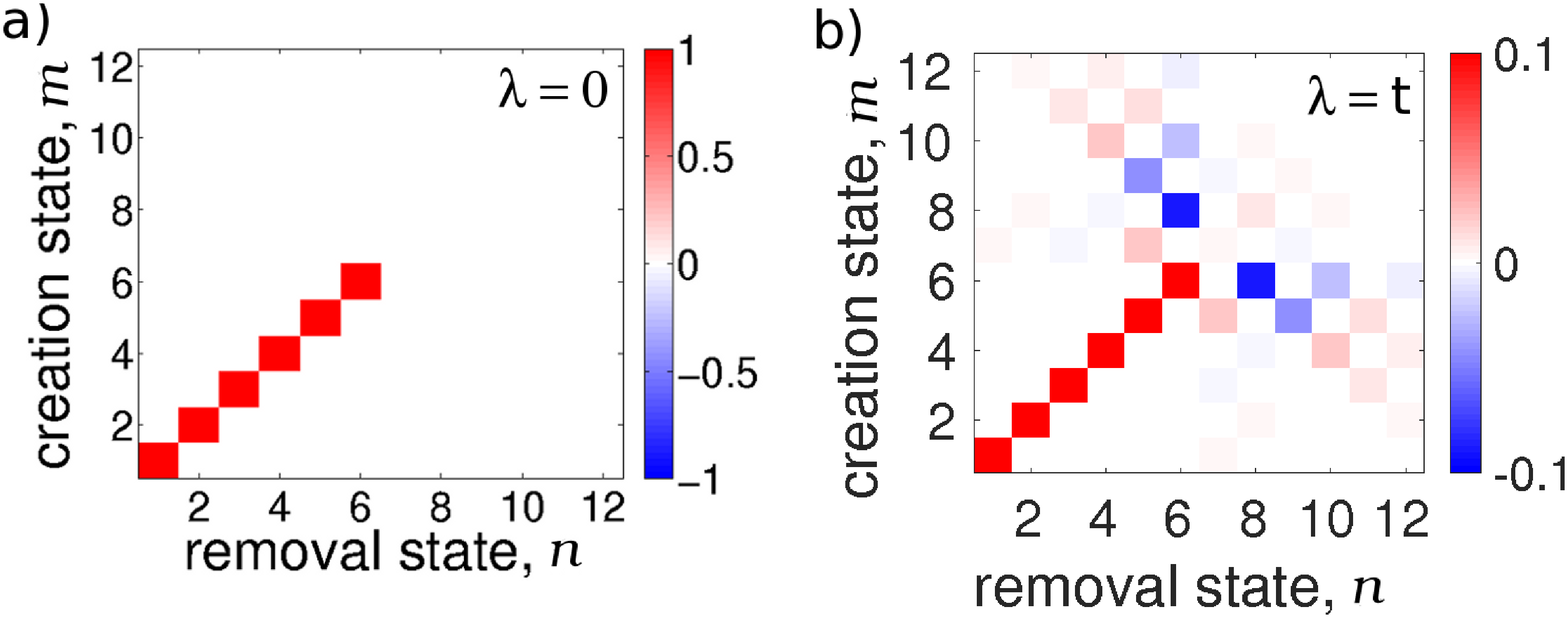}
 \caption{The expectation value of the single-particle excitation operator in mode space (projection of the ground state onto the single-particle excitations of the ground state in the wave-function basis), $\langle \Psi_{GS}|a_m^\dagger a_n | \Psi_{GS} \rangle $, for: a) no interaction and b) a Coulomb interaction equal to the hopping.  Note the color scale differs between a and b. }
\label{fig:SPExcGSCoul1}
\end{figure}

In the small interaction regime, the ground state is perturbed from the non-interacting ground state we discussed above. This can be seen in figure \ref{fig:SPExcGSCoul1}, which shows the ground state expectation value of the single particle excitation operator $a_m^\dagger a_n$ for Coulomb interaction strength of zero and $t$.  The diagonal elements show the occupancy of the non-interacting single particle modes (one or nearly one for modes one through six and zero or nearly zero for six through twelve for both cases). When the diagonal elements alone are plotted (see Supplemental Information \cite{SuppInfo}) the mode occupancy is seen to be similar to a Fermi function, with the non-interacting case a perfect step function, and the interactions smearing the Fermi sea similar to a finite temperature.  The values of off-diagonal elements are zero for the non-interacting case and increase with the Coulomb strength as the ground state becomes dressed by the interaction.

\begin{figure}
 \includegraphics[width=0.45\textwidth]{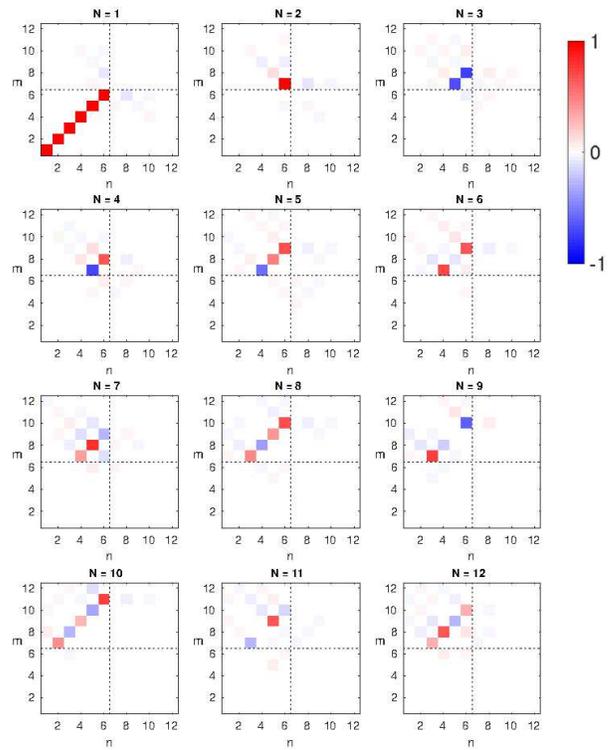}
 \caption{Coulomb strength of $\lambda_{ee} = t$.  The projection of the lowest 12 states onto each single-particle excitation of the ground state in the s.p. mode basis, $\langle \Psi_{N}|a_m^\dagger a_n | \Psi_{GS} \rangle $, with removal mode $n$ on the horizontal axes and creation mode, $m$ on the vertical axes. The off-diagonal lines are plasmonic behavior: collective excitations in which many electrons all have the same momentum shift. }
\label{fig:SPExcCoul1}
\end{figure}

The particular single-particle excitations of the ground state begin to mix together in linear combinations in this regime.  This can be seen in figure \ref{fig:SPExcCoul1} which shows the projection of the 12 lowest excited states onto each of the possible single particle excitations (in the single-particle mode basis) of the ground state for $\lambda_{ee} = t$.  The figure shows that the lowest energy eigenstates consist of linear combinations of multiple single-particle excitations. In this regime the low-energy excitations are plasmonic \cite{Bryant2020}. The first excited state is made mostly from the one way to shift one electron from the highest occupied single-particle state to the first unoccupied single particle state. The next two excitations are determined mostly by the two ways that one electron can be excited with a change in single-particle index of two. These first three excitations correspond to the fundamental plasmon mode, the plasmon mode with two nodes and the doubly excited fundamental plasmon mode  \cite{Bryant2020}. Excitations with larger changes in single-particle index correspond to higher order plasmon modes and other multiply excited plasmonic excitations. 

In the weak interaction regime the SPEC is sensitive to $\epsilon_{\text{G-S}}$, the cutoff used to decide whether a particular single-particle excitation of the ground state has enough new orthogonal components to be included in the single particle excitations basis.  
More basis states are included, capturing more of the SPEC, if a smaller cutoff is used. Genuine but small perturbations in the ground state may be excluded or not depending on that cut-off, influencing the total number of single-particle excitations of the ground state that appear (saturation value of the curve). 
 In this perturbative regime, the ground state consists of an unperturbed ground state mixed with the unperturbed excited states, making more single-particle excitations of the ground state possible.  (This is because single-particle modes are no longer completely filled or completely empty.)  However the larger the interaction, the greater the mixing and the less likely that any particular single-particle excitation of the ground state will be excluded by the cutoff.
It is the inclusive cutoff that captures the true SPEC in these cases.  Nonetheless, the behavior at other cut-off values give us insight into the meaning of the SPEC and why its sum over all states is discontinuous between the non-interacting and interacting cases.

Figures \ref{fig:PartialSumSmallCoul}a and  \ref{fig:PartialSumSmallCoul}b show the rolling sum of the SPEC for small interactions with an exclusive (large $\epsilon_{\text{G-S}}$) cutoff and an inclusive (small $\epsilon_{\text{G-S}}$) cutoff, respectively. 
For a non-zero Coulomb interaction up to and including $\lambda_{ee} =t$ 
 the behavior of the SPEC is remarkably similar for all interaction strengths, provided the inclusive cutoff is used.
SPEC is nearly a smooth function of eigenstate number (and of energy), with nearly all the non-zero SPEC occuring in the bottom quarter of the spectrum of eigenstates as shown in figure \ref{fig:SPContent}c for  $\lambda_{ee} =t$.
 When a small enough cutoff is used, the rolling sums for $0 < \lambda_{ee}/t \leq 1$ fall on the same quasi universal curve which saturates at $m_e *(N_s - m_e)$. This shows that there are distinctly different behaviors between the non-interacting case and any case with interaction, no matter how small. For any value of a perturbing Coulomb interaction (no matter how small, and up to $\lambda_{ee} =t$) the perturbation mixes excited states into the ground state, and single particle excitation content is present in many more of the excited eigenstates. 
This discrete jump in the saturation value of the sum over SPEC points to the existence of small but well-defined excitations that come into existence in the presence of even the weakest interaction.
This is consistent with the observation of Schulz \cite{Schulz1993} that the Wigner crystal (in continuous 1-D systems) occurs for any Coulomb interaction strength, and highlights the fundamental difference between weak interaction and no interactions.

As can be seen in figure \ref{fig:ProbGSinWigner} the ground state of the system in this regime still only has minute departures from the non-interacting limit: the overlap between the full ground state and the product state of the lowest six modes is nearly one.  Yet those departures already have the defining character of the interacting ground state.

Previous authors identify this regime as a ``metallic Wigner crystal'' \cite{Capponi2000, Li2019}  or a ``weakly pinned small-amplitude charge density wave'' \cite{Valenzuela2003}.  Their focus is largely on whether the system is metallic in the thermodynamic limit, and by this measure their weakly interacting regime continues all the way to $\lambda_{ee}/t = 4$ or $5$, where there is a crossover rather than a sharp transition.  As we will see below, there is also something of a crossover in the behavior of the SPEC at that value.  Characteristics ascribed to this phase include: metallic or quasi-metallic (extrapolation to long chains gives a disappearing charge gap and high charge stiffness), a charge density modulation that is small compared to the delocalized charge and disappears in the long chain limit, and having strong departures from Luttinger liquid behavior that grow with $\lambda_{ee}/t$.

\subsection{Very strong interaction, $\lambda_{ee}/t \geq 100$} \label{subsect:Strong}
The region with the highest Coulomb interaction strengths is analagous to the small interaction case.  However, now the occupation of particular sites in the chain is the applicable basis rather than the non-interacting single-particle modes being the relevant unperturbed basis. With an inclusive cut-off, the SPEC curves are all the same, up to the slight change in the order of a few of the levels. The SPEC curves for this region are shown in figures \ref{fig:PartialSumSmallCoul}c and  \ref{fig:PartialSumSmallCoul}d.  The single-particle excitations are no longer confined to the low energy part of the spectrum due to strong correlations induced by the extreme Coulomb interaction strength.  Both the universal nature of the SPEC curve and the linear scaling of the energy levels with interaction strength seen in figure \ref{fig:Energies0.2} point to the fact that the correlation is fully saturated: there is little reordering or mixing of levels as the interaction strength is varied.  The Wigner crystal found at the extreme limit of no hopping is essentially present throughout this region.

\begin{figure}
  \includegraphics[width=0.49\textwidth]{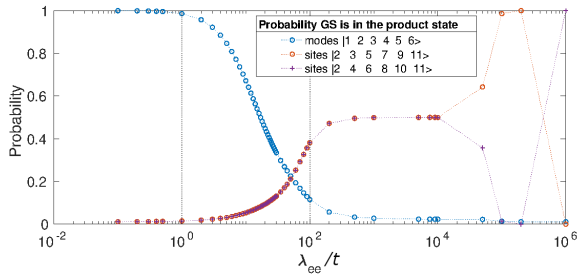}
 \caption{Probability of measuring the interacting ground state in the ground states of the two limits  $t = 0$ and $\lambda_{ee} = 0$.}
  \label{fig:ProbGSinWigner}
\end{figure}

We have defined the boundaries for the perturbative regions based on the dependence of the SPEC on the cut-off.  However figure \ref{fig:ProbGSinWigner} shows the probability of finding the ground state in the Wigner crystal of the $t = 0$ limit or the Slater determinant of the $\lambda_{ee} = 0$ limit, as a function of the strength of the Coulomb interaction.  The vertical lines indicating the perturbative regions from the SPEC analysis align with the regions of significant probability of finding the ground state in one of the limiting unperturbed ground states.

Previous authors refer to this regime as an ``insulating charge-density wave'' \cite{Capponi2000, Li2019} or a ``generalized Wigner Lattice'' \cite{Valenzuela2003}.  It is characterized by a thermodynamically significant charge density modulation and an insulating character (identified by a finite structure factor divided by chain length and a finite charge gap in long-chain extrapolation).  As mentioned in the previous section, these authors identify this regime with $\lambda_{ee}/t > 4$ or $5$.  However the charge density modulation is fully saturated for $\lambda_{ee}/t \geq 100$.  The next section will discuss the crossover regime between the regimes of weak and very strong interaction and the use of the charge-density-wave and bond-order-wave order parameters as measures for these regimes.

\subsection{Intermediate interaction: Onset and Saturation of Correlation, $1 < \lambda_{ee}/t < 100$} \label{subsect:Int}
\begin{figure}
 \includegraphics[width=0.48\textwidth]{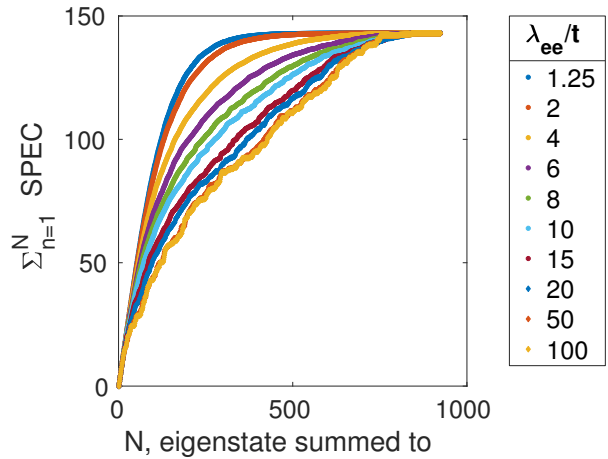}
 \caption{Rolling Sum of Single Particle Excitation Content for eigenstates of a 12 atom chain with six electrons, for non-perturbative region.}
 \label{fig:PartialSumIntCoul}
\end{figure}

In between the two perturbative regions is the region of increasing correlation.  Correlation due to Coulomb repulsion causes the many-body eigenstates to cross and mix, and the ground state of the system is fundamentally changed.  The single-particle excitation content is no longer sensitive to the cutoff used, giving the same results for a very wide range of values of $\epsilon_{\rm G-S}$.  The SPEC curves transition gradually between the quasiuniversal curves for the two perturbative regimes, as seen in figure \ref{fig:PartialSumIntCoul}.  However, as noted previously, the crossover assigned to $\lambda_{ee}/t = 5$ by  \cite{Capponi2000, Valenzuela2003, Li2019} is visible in the SPEC curves as a transition from a mostly smooth curve in which single-particle excitations primarily make up low-energy eigenstates to a curve with discontinuities in which the single-particle excitations are not confined to the low-energy part of the spectrum.

\begin{figure}
\includegraphics[width=0.45\textwidth]{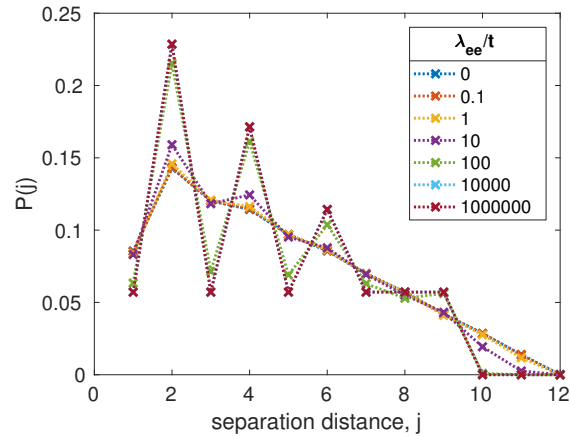}
\caption{Probability density of finding two electrons separated by a distance j for ground state of six electrons on a 12 atom chain, varying the  Coulomb interaction, $\lambda_{ee}$, measured in units of the hopping, $t$.  After reference \cite{Gambetta2014}.}
\label{fig:SepProbCorr}
\end{figure}

\begin{figure}
  \includegraphics[width=0.49\textwidth]{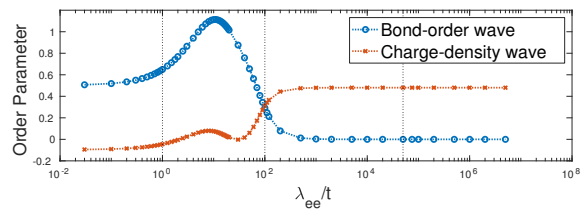}
\caption{Bond order wave (BOW) and charge-density wave (CDW) order-parameters of the ground state as a function of Coulomb interaction strength.}
\label{fig:BondOrderWave}
\end{figure}

That this is the region in which correlation sets in can be seen in figures \ref{fig:SepProbCorr} 
and \ref{fig:BondOrderWave}.  We use two measures of correlation: the first, used by Gambetta {\it et al.} \cite{Gambetta2014} and Wang {\it et al.} \cite{Wang2012}, is the probability density of finding two electrons separated by a distance j:
\begin{equation} P(j)= \sum_i \langle \Psi_{GS}| c_i^\dagger c_{i + j}^\dagger  c_{i + j} c_i + c_i^\dagger c_{i - j}^\dagger  c_{i - j} c_i |\Psi_{GS}\rangle.
\end{equation}
We see in figure \ref{fig:SepProbCorr}  that the departure from a linear decrease with distance grows in the intermediate region of Coulomb interaction ($\lambda_{ee}/t = 1 - 100$) and saturates at the upper limit of the region ($\lambda_{ee}/t =100$), with evidence of every-other site occupation. 
 
The second measure is the charge-density wave order parameter, defined by:
\begin{equation}
O_{CD} = \sum_{\Delta i=1}^{N_s-1} \frac{(-1)^{\Delta i} }{(N_s-\Delta i)} \sum_{i=1}^{N_s-\Delta i} \langle\Psi_{GS}| c_i^\dagger c_{i}  c_{i+\Delta i}^\dagger c_{i+\Delta i} |\Psi_{GS}\rangle .
\end{equation}
As can be seen in figure \ref{fig:BondOrderWave}, the charge density wave order parameter maintains a constant value in the strong correlation regime, but does not grow uniformly throughout the regime of onset and saturation of correlation.

This crossover region also sees the development and disappearance of bond-order oscillations.
A bond order wave is a state of broken symmetry in which the expectation value of the kinetic energy operator alternates between every two nearest neighbor sites.  Its order parameter is defined by 
\begin{widetext}
\begin{equation}
O_{BO} = \sum_{\Delta i=1}^{N_s-1} \frac{(-1)^{\Delta i} }{ (N_s - \Delta i)} \sum_{i=1}^{N_s -\Delta i-1} \langle\Psi_{GS}| (c_i^\dagger c_{i+1} + c_{i+1}^\dagger c_i )(c_{i+\Delta i}^\dagger c_{i+\Delta i+1} + c_{i+\Delta i+1}^\dagger c_{i+\Delta i}) |\Psi_{GS}\rangle. 
\end{equation}
\end{widetext}
 This order parameter measures the asymmetry of the bond strengths for odd and even bonds (odd bonds being those between the first and second site, the third and fourth, et cetera).  A bond order wave phase is defined as the existence of this order parameter in the infinite-chain length limit, which we do not evaluate here.  (Finite Luttinger liquid systems exhibit bond-order oscillations that are not thermodynamically significant \cite{Mishra2011}.) 
Discussions of the bond-order wave phase may be found in references \cite{Mishra2011, Liu2011, Hallberg1990}.  The Supplemental Information \cite{SuppInfo} and references therein \cite{Nakamura2000, Ejima2007, Ding2011, Zhang2004, Sengupta2002, Shao2019, Tran2017, Sengupta2002} discuss our choice of order parameter. 

One way to interpret the bond-order wave in a finite chain is as a charge-density wave which is shifted to lie between the sites.  With this interpretation, figure \ref{fig:BondOrderWave} shows the growth of charge-density order with increasing interaction strength.  Figure \ref{fig:PartialSumIntCoul} shows that for smaller interaction strength (approximately $\lambda_{ee}/t \leq 4$) all low-energy excited states are primarily single-particle excitations, and higher energy states are primarily multiple-particle correlated excitations.  This reflects the low correlation seen in the order parameter (figure \ref{fig:BondOrderWave}) and probability of separation (figure \ref{fig:SepProbCorr}).  The rolling sum of the SPEC is a smooth function of eigenstate number, implying that multiple particle excitations are being mixed into all states smoothly as a result of interactions.   As the interaction becomes stronger, lower energy states gain multiple-particle excitation content, and higher energy states gain single-particle excitation content, as correlations develop.  For the strongest interactions, neighboring excited states with nearly the same energy can have much different single-particle excitation content, and many more states in the low part of the spectrum have nearly zero SPEC (because moving two or more electrons simultaneously is both lower in energy once they are correlated and is very close to being an exact eigenstate of the system once Wigner crystallization has set in).   Thus the previously smooth curve of the rolling sum of the SPEC becomes rough as the SPEC of two consecutive eigenstates take very differing values.

\section{Conclusion}\label{sect:Conc}
We have considered small finite systems of interacting electrons with an eye toward what they can reveal about extended many-body physics.  As quantum analog simulation in small systems develops experimentally, we expect that experiment coupled with such numerical analysis will enable a clearer understanding of large-scale phenomena and greater design capability.  Understanding the nature of  excited states is crucial for designing active systems, which will necessarily leave their ground state.  

Using the single-particle excitation content of the excited states of a many-electron Hamiltonian we have identified five regimes in the parameter space of the Hamiltonian.  This identification is largely consistent with and supplements previous analyses of the ground-state phase space of this system.  This is in part because the possible single-particle excitations out of an interacting ground state reflect the nature of that ground state.  We observe unique signatures in the SPEC of the eigenstate spectrum which allow us to identify the non-interacting and no-hopping regimes, perturbative regimes around each of those, and a broad crossover region between them, in which correlation grows and saturates.  As shown in the Supplemental Information \cite{SuppInfo}, these signatures persist upon varying the range of interactions, the filling (number of electrons), the exchange fraction and the size of the chain.  This suggests that the behavior might be similar for other systems, for example in higher dimensional geometries or other forms of Hamiltonian.

When the rolling sum of the SPEC of the excited states has a universal curve as a function of eigenstate number and is sensitive to the cutoff for including minute quantities of a single-particle excitation in the vector space, the Hamiltonian is in a perturbative regime near a limit with a separable ground state (one that can be written as a Slater determinant of single-particle states).  As strongly correlated behavior reorders and redefines the eigenstates the curve of SPEC varies as well.

Small systems provide the opportunity for exact analysis of the entire spectra, finding all of the excitations of the systems. We have shown how this analysis provides new ways to define the different regimes of the many-body interactions and correlations, clearly identifying distinct differences that arise when interactions are turned on or hopping is turned off and the quasi-universal behavior that arises in the perturbative regimes of weak interaction or weak hopping. Exact analysis of small systems and determination of the full spectrum of excitations also provides an opportunity to develop a full analysis of the dynamics of these systems. 
\bibliography{MyLibrary.bib}
\bibliographystyle{unsrt}

\end{document}


\title{Supplemental Information to Many-Body Physics in Small Systems: Observing the Onset and Saturation of Correlation in Linear Atomic Chains}
\author{Emily Townsend}
\email[]{emily.townsend@nist.gov}
\affiliation{Nanoscale Device Characterization Division and Joint Quantum Institute, National Institute of Standards and Technology, Gaithersburg, Maryland 20899-8423, USA; and University of Maryland, College Park, Maryland 20742, USA}
\author{Tom\'a\v s Neuman}
\affiliation{Institut de Physique et Chimie des Matériaux de Strasbourg, Université de Strasbourg, 23, rue du Loess 67034, Strasbourg France}
\author{Alex Debrecht}
\affiliation{Nanoscale Device Characterization Division and Joint Quantum Institute, National Institute of Standards and Technology, Gaithersburg, Maryland 20899-8423, USA; and University of Maryland, College Park, Maryland 20742, USA}
\affiliation {Department of Physics and Astronomy,  University of Rochester, Rochester, New York 14627, USA}
\author{Javier Aizpurua}
\affiliation{Centro de F\'isica de Materiales CFM - MPC, Centro Mixto CSIC-UPV/EHU, 20018 San Sebasti\'an-Donostia, Basque Country, Spain}
\affiliation{Donostia International Physics Center (DIPC), 20018 San Sebasti\'an-Donostia, Basque Country, Spain}
\author{Garnett Bryant}
\affiliation{Nanoscale Device Characterization Division and Joint Quantum Institute, National Institute of Standards and Technology, Gaithersburg, Maryland 20899-8423, USA; and University of Maryland, College Park, Maryland 20742, USA}

\date{\today}

\maketitle
\section{Introduction}
In section \ref{sect:SI.MomDist} we discuss the relationship between single-particle excitation content and quasiparticle weight.  We also show the evolution of the ground state expectation value of the single-particle mode occupation with changing interaction strength.  The occupation as a function of mode number looks like a zero-temperature Fermi function for non-interacting electrons, and for small interactions takes on a character similar to a finite-temperature Fermi fuction.   
Section \ref{sect:SI.OP} discusses our choice of definition of the order parameters for charge-density and bond-order waves.  
Finally, section \ref{sect:SI.Vary} shows that the behavior of the rolling sum of the single-particle excitation content (SPEC) has broadly the same behavior for systems with varying size, filling, exchange fraction, and truncated range of Coulomb interaction.

\section{Quasiparticle weight and ground state mode occupancy} \label{sect:SI.MomDist}
\begin{figure}
    \includegraphics[width=0.5\textwidth]{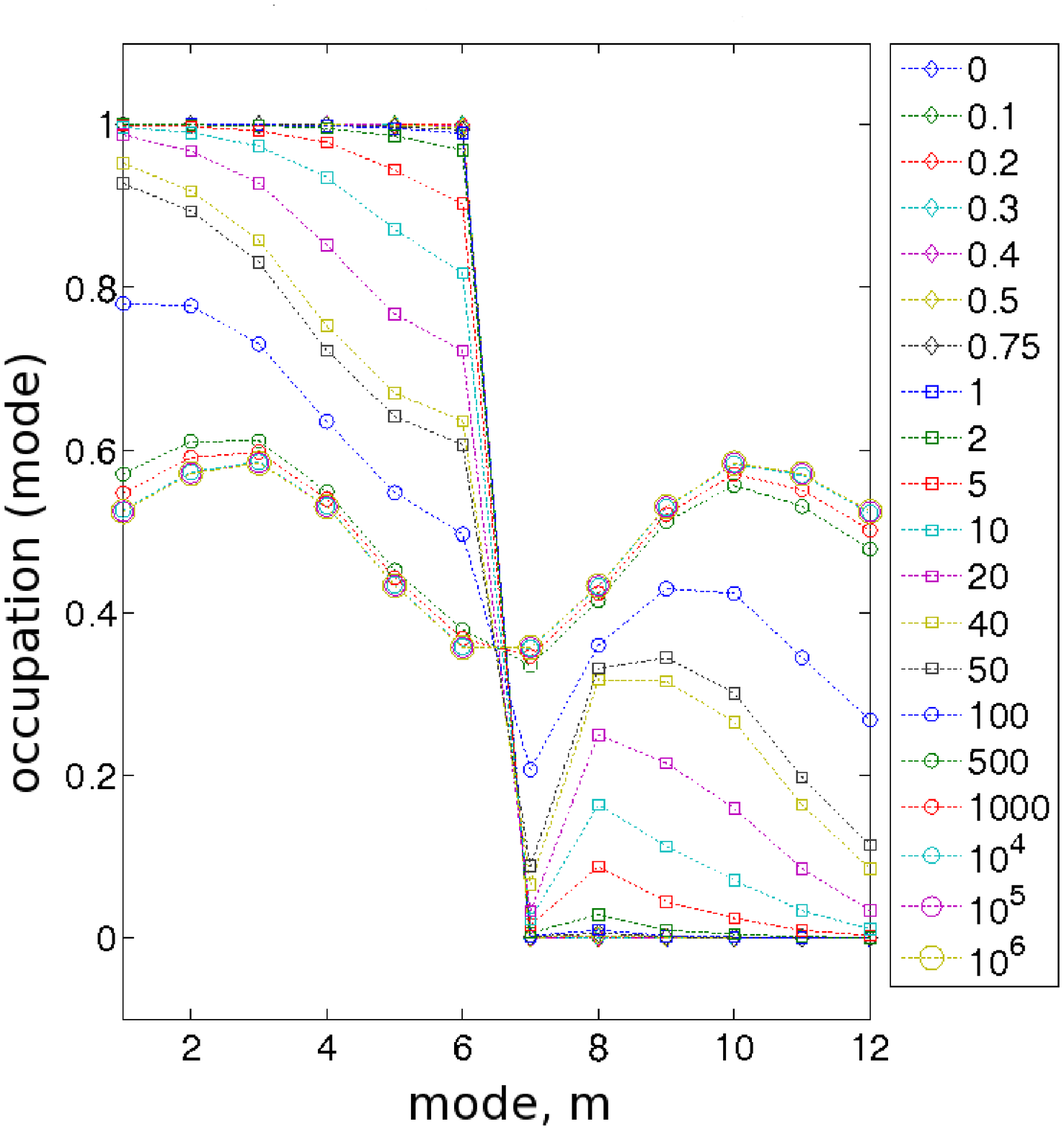}
   \caption{The occupation of single-particle modes, $m$, in the many-body ground state, $\langle \Psi_{GS}|a_m^\dagger a_m | \Psi_{GS} \rangle $, varying the interaction strength ($\lambda_{ee}/t$), for a chain of twelve atoms and six electrons. The height of the discontinuity at the Fermi-surface is $Z_m$.}
   \label{fig:MomentumDist}
   \end{figure}

Figure \ref{fig:MomentumDist} shows the expectation value of the occupation,  $\langle \Psi_{GS}|a_m^\dagger a_m | \Psi_{GS} \rangle $, of each of the single-particle modes, $m$, in the interacting ground state for varying strengths of interaction, $\lambda_{ee}/t$.  This is the same as the diagonal elements of the  ground-state expectation value of the single-particle excitations shown in Figure 4 of the main paper.  For small values of interaction ($\lambda_{ee}/t < 2$), this looks like a Fermi function, with interaction strength playing a role similar to temperature, but as correlations become important there is a significant departure from fermionic behavior.
   
The size of the discontinuity at the Fermi level (between $m=6$ and $m=7$) is sometimes referred to as the ``quasiparticle weight" or ``wavefunction renormalization", and illustrates the extent to which a quasiparticle picture is sufficient to describe the low-energy behavior of the system of interacting electrons \cite{Coleman2015, Schofield1999}.  
This quasiparticle weight, $Z_m$ is related to the probability that a bare electron added to the system with a given momentum near the Fermi momentum will overlap with a quasiparticle added to the system with the same momentum.  If this quasiparticle weight is less than one, it means that creating a single quasiparticle creates not just a bare particle but also many-particle correlations, i.e. 
\begin{equation}
    \tilde{a}_m^\dagger = \sqrt{Z_m} a_m^\dagger + \sum_{n+p-q=m} \alpha(m,n,p,q) a_n^\dagger a_p^\dagger a_q + ...
\end{equation}
where $\tilde{a}^\dagger$ creates a quasiparticle, $a^\dagger$ creates a bare particle, $m,n,p$ and $q$ are mode numbers (or momenta in an infinite/unbounded system) and $\alpha$ is a coefficient.  
Figure \ref{fig:MomentumDist} shows that these quasiparticles are a good description up to around $\lambda_{ee} = 2t$, as for these values the quasiparticle weight is nearly 1.  For larger values of interaction, the discontinuity is significantly reduced.

\begin{figure}
    \includegraphics[width = 0.65\textwidth]{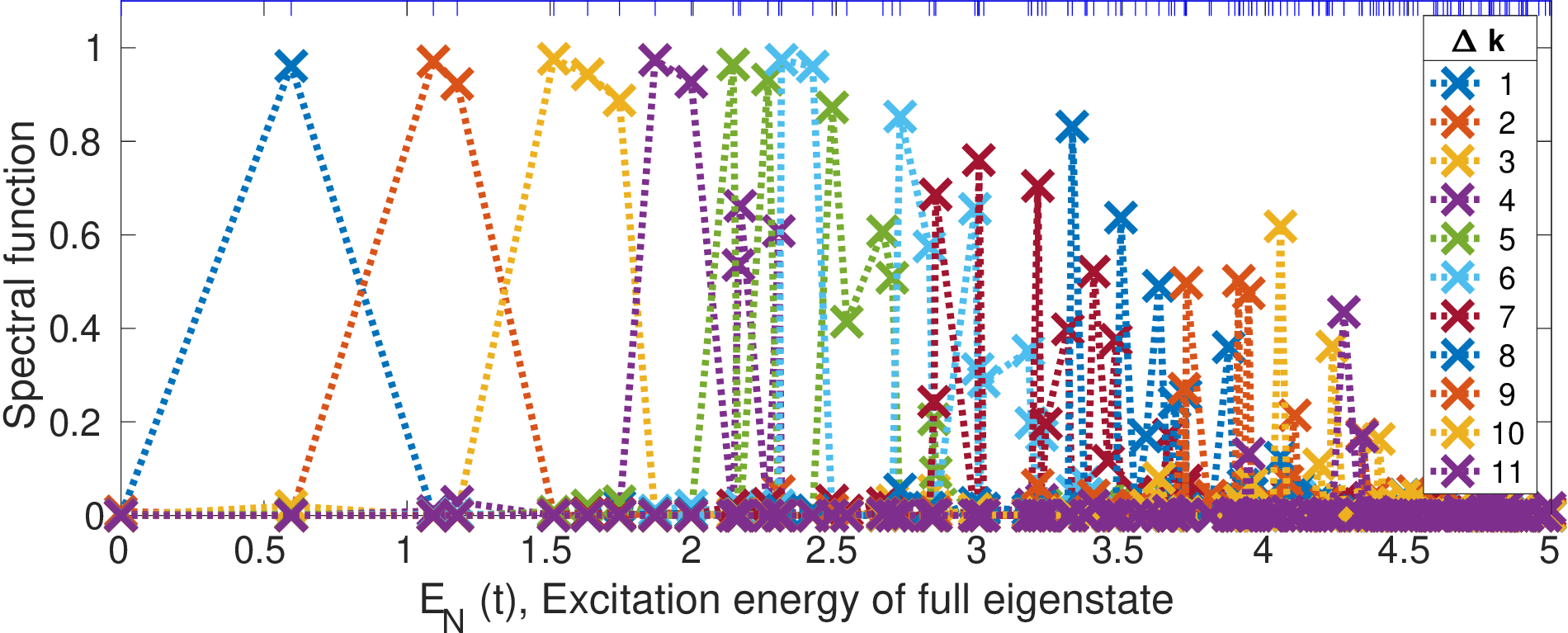}
    \caption{Spectral weight of charge-neutral excitations of momentum $\Delta k$, as defined by equation \ref{eqn:QPSpect}, for $\lambda_{ee} = t$ in a chain of twelve atoms and six electrons. The quantity as we have defined it is only defined at the (excitation) energies of the eigenstates.  These energies are marked with blue ticks along the top axis.}
    \label{fig:QPSpect}
\end{figure}

This idea is reminiscent of our distinction between single-particle excitations and many-particle excitations.  However the quasiparticle weight described above quantifies the overlap of introducing a new bare electron into the ground state with introducing a new quasiparticle into the ground state.  Since we are investigating a system with a fixed number of electrons, we would prefer to define a quasiparticle weight that describes a charge-neutral excitation in a similar way.
In our non-interacting system the ground state consists of the first $n_e$ single-particle states being filled and the remaining single-particle states being unfilled.  The low-energy excitations consist of promoting one electron from a filled state, $n$, to an empty state, $n+\Delta k$, so that $| \Psi_N \rangle^0 = a_{n+\Delta k}^\dagger a_n | \Psi_{GS} \rangle^0$ (where the 0 superscript reminds us this is the non-interacting limit and $N$ labels the eigenstate number).   One can think of this as imparting a momentum kick of $\Delta k$ to the electron.  There are multiple ways to impart any particular momentum kick to the system, i.e. the different starting values of $n$. (For example, we can move an electron from mode 6 to 8 or from 5 to 7.)  
We can define the spectral weight of a charge-neutral (constant electron number) excitation as the probability that an excited state of the interacting system overlaps with any bare ``kick" of a particular momentum given to the interacting ground state:
\begin{equation}
 A(\Delta k, E_N) = \sum_n |\langle \Psi_N|  a_{n+\Delta k}^\dagger a_n |\Psi_{GS} \rangle |^2
 \label{eqn:QPSpect}
\end{equation}
Figure \ref{fig:QPSpect} shows this spectral weight for various values of the kick, $\Delta k$ for the chain of twelve atoms and six electrons for $\lambda_{ee} = t$.  The horizontal axis, $E_N$ (the excitation energy of eigenstate $\Psi_N$), includes only the low energy part of the spectrum as higher energy states have virtually no probability of being a single-particle excitation.
Like figure 5 from the main paper, figure \ref{fig:QPSpect} illustrates that in the presence of a moderate interaction the low-energy excitations correspond well to these charge neutral momentum kicks.

\clearpage
We have no problem identifying from figure \ref{fig:QPSpect} the momentum kick, $\Delta k$ that most contributes to each energy eigenstate of the interacting system.
We could thus call the spectral function at that kick the ``quasiparticle content" of that eigenstate. 

\begin{figure}
    \includegraphics[width=0.45\textwidth]{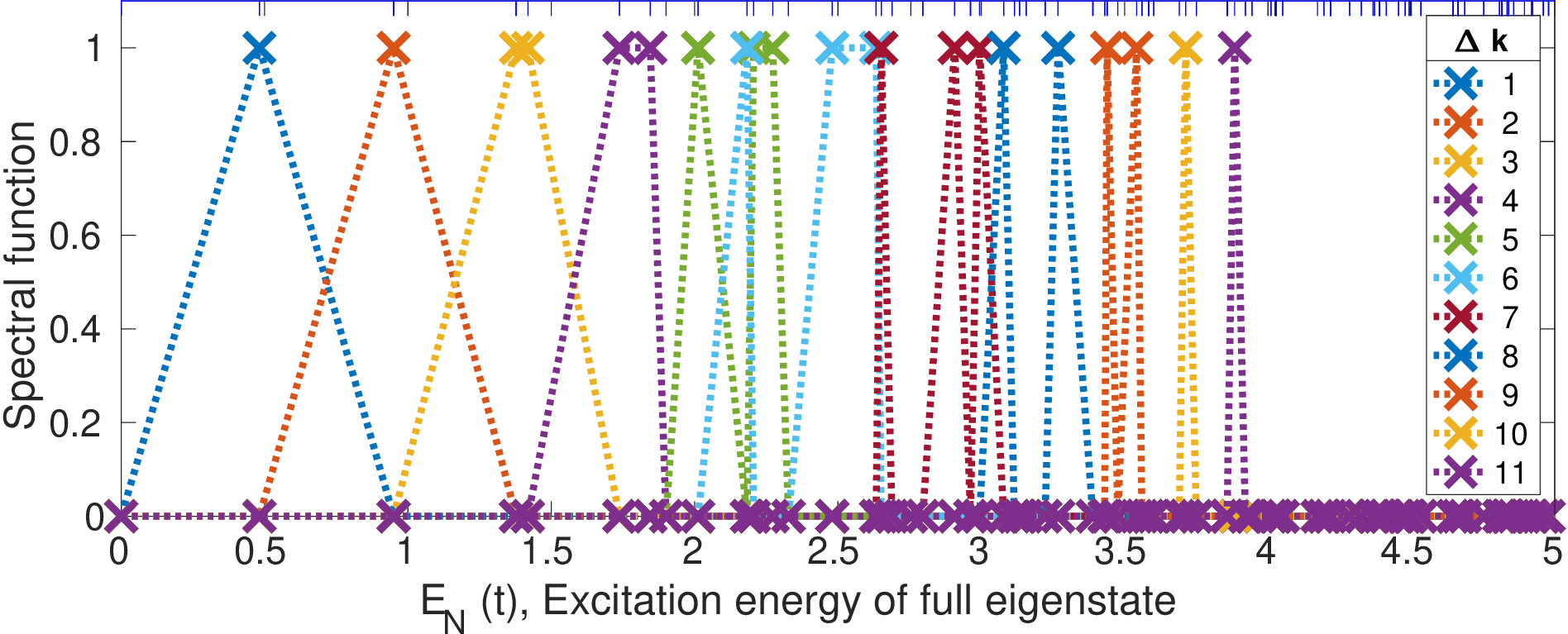}
    \includegraphics[width=0.45\textwidth]{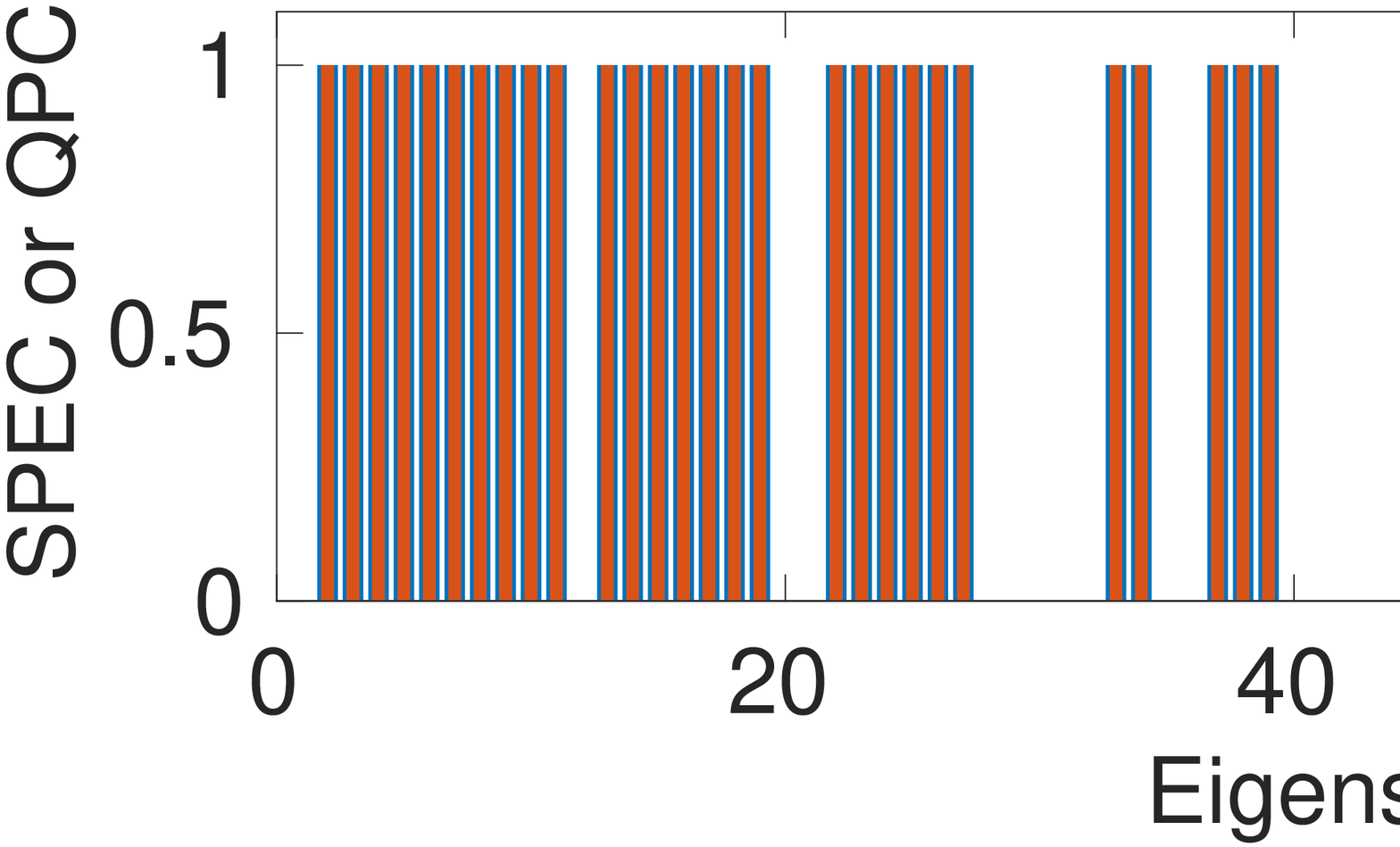}
    \includegraphics[width=0.45\textwidth]{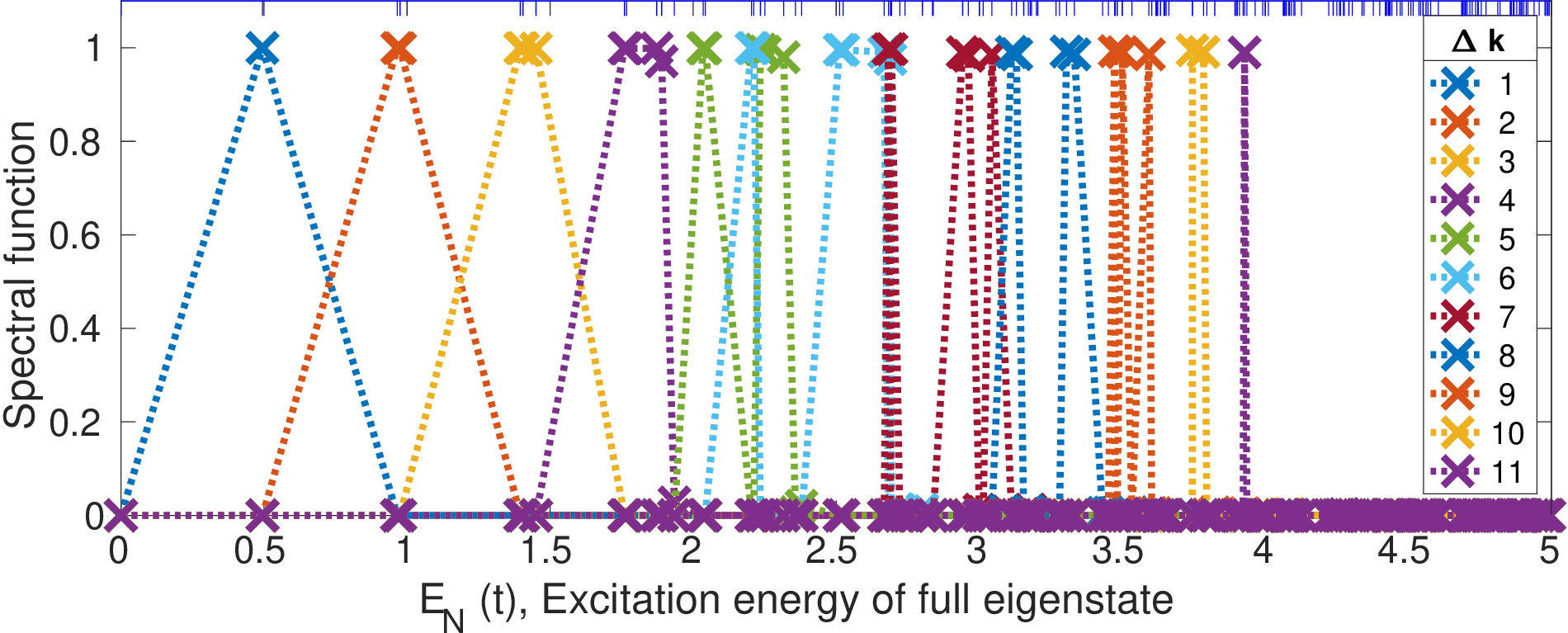}
    \includegraphics[width=0.45\textwidth]{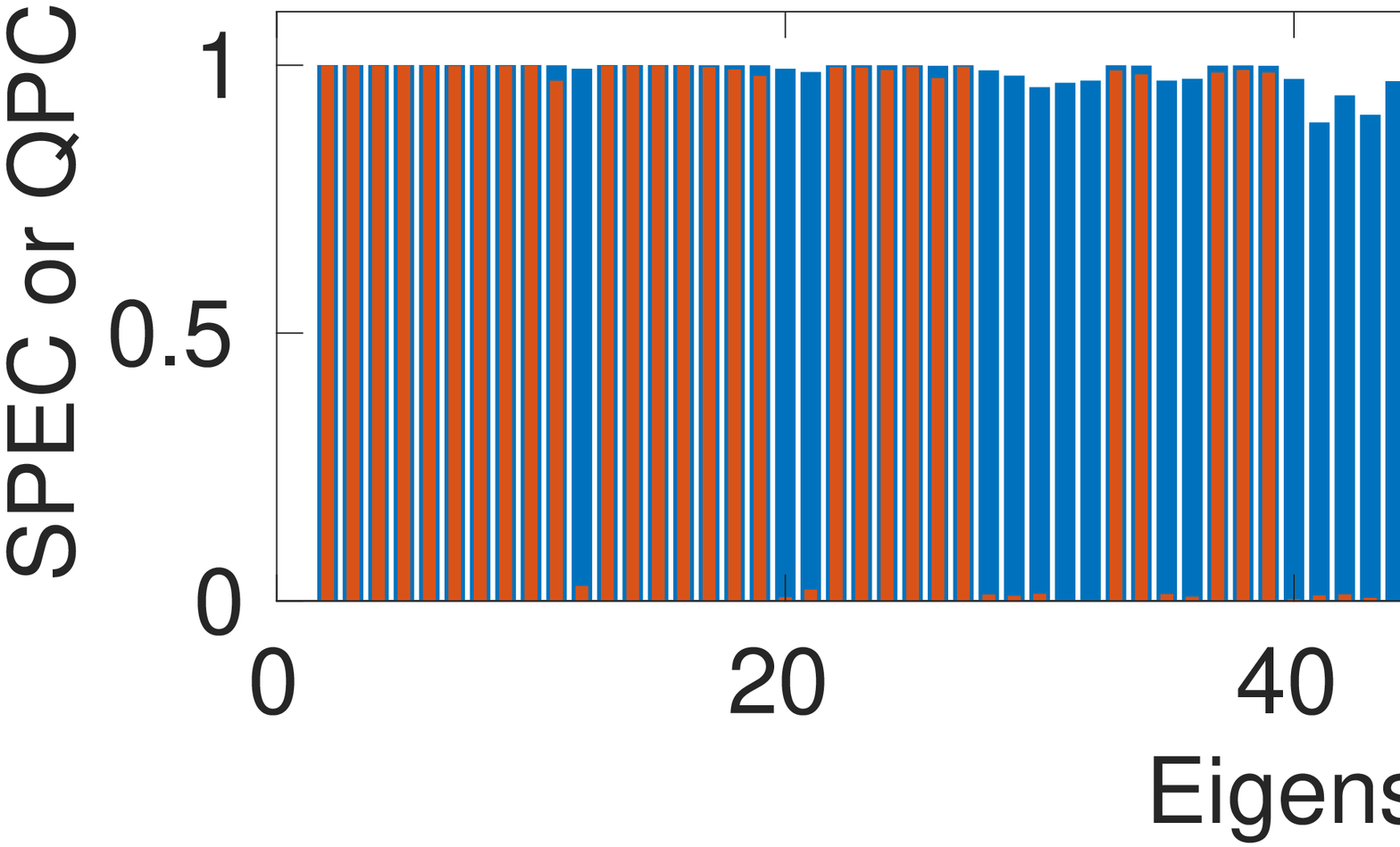}
    \includegraphics[width=0.45\textwidth]{qpSpectralFxnbyEShowEontopNoTitle1by12w6Coul1tb-1.eps}
    \includegraphics[width=0.45\textwidth]{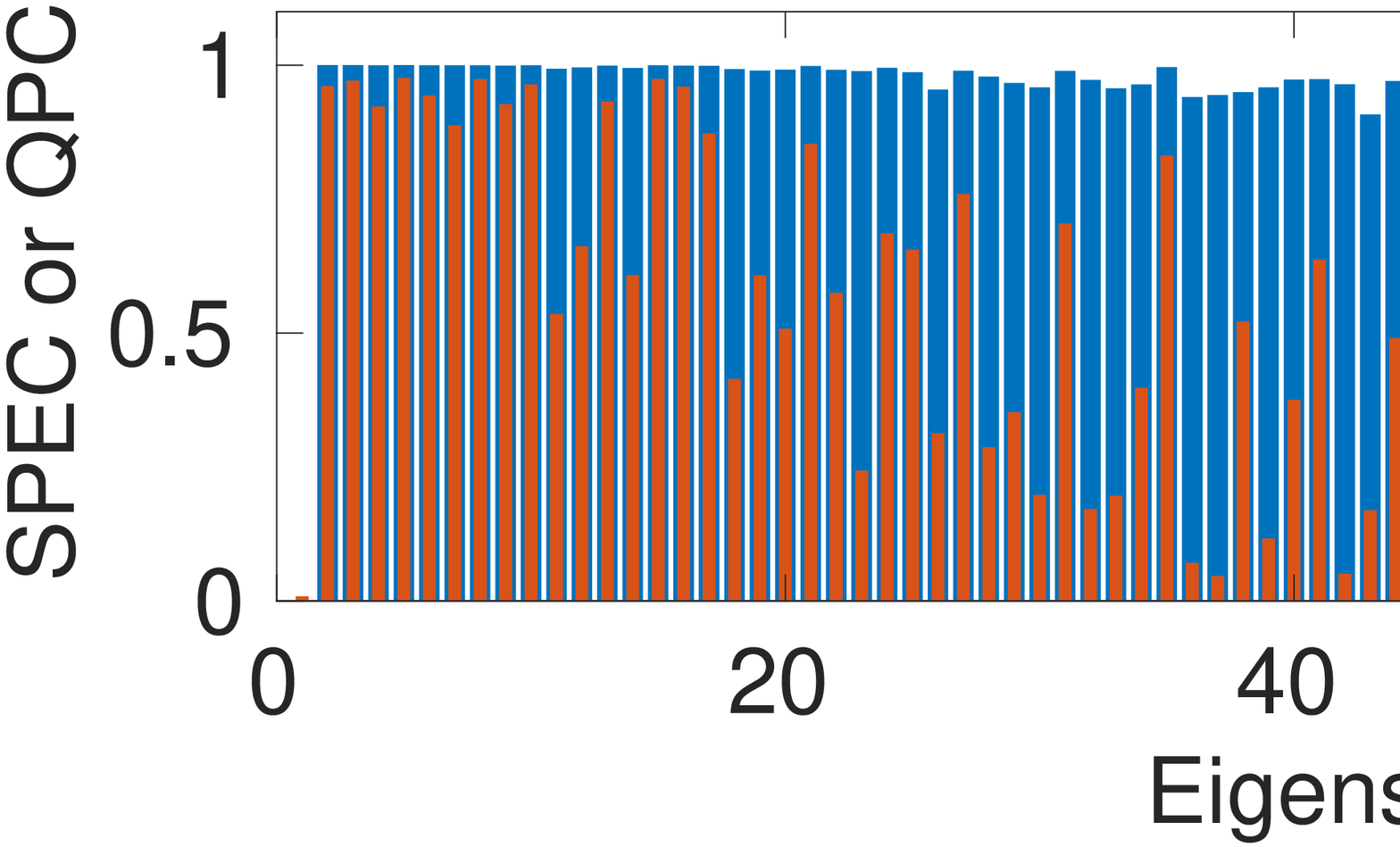}
    \includegraphics[width=0.45\textwidth]{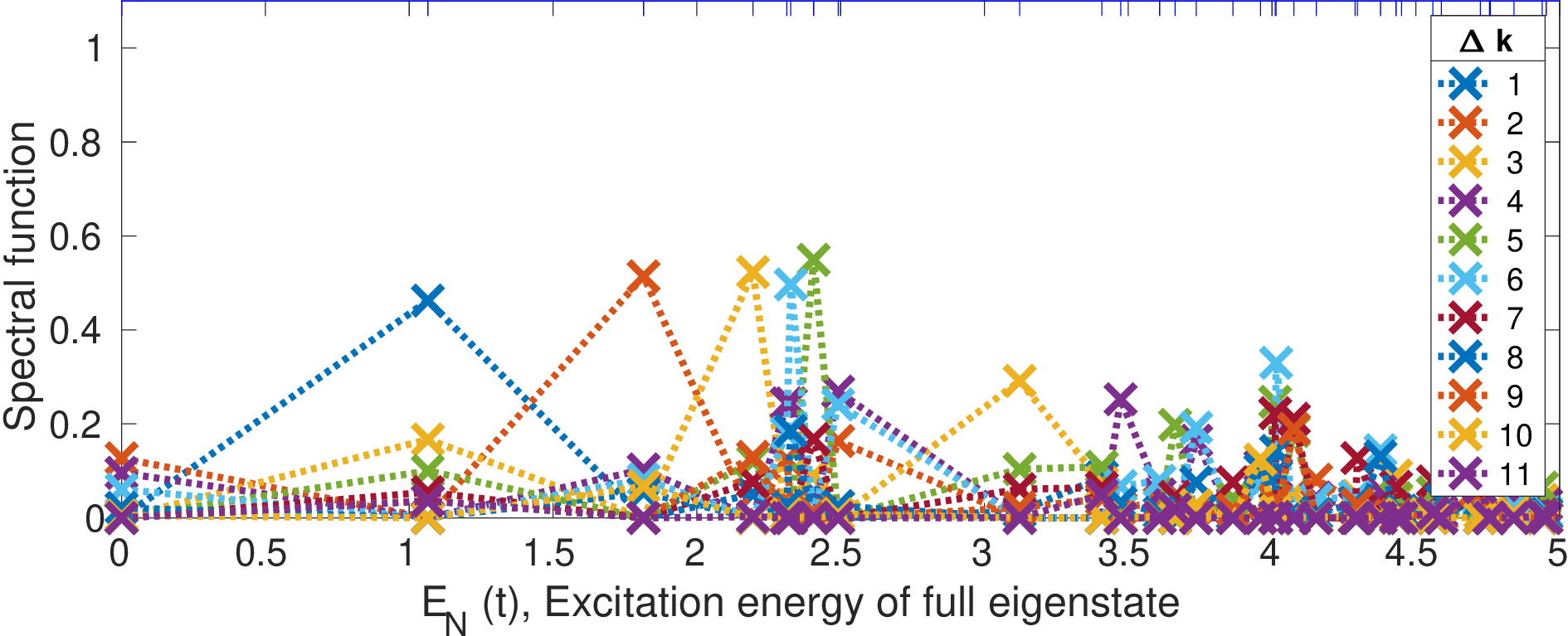}
    \includegraphics[width=0.45\textwidth]{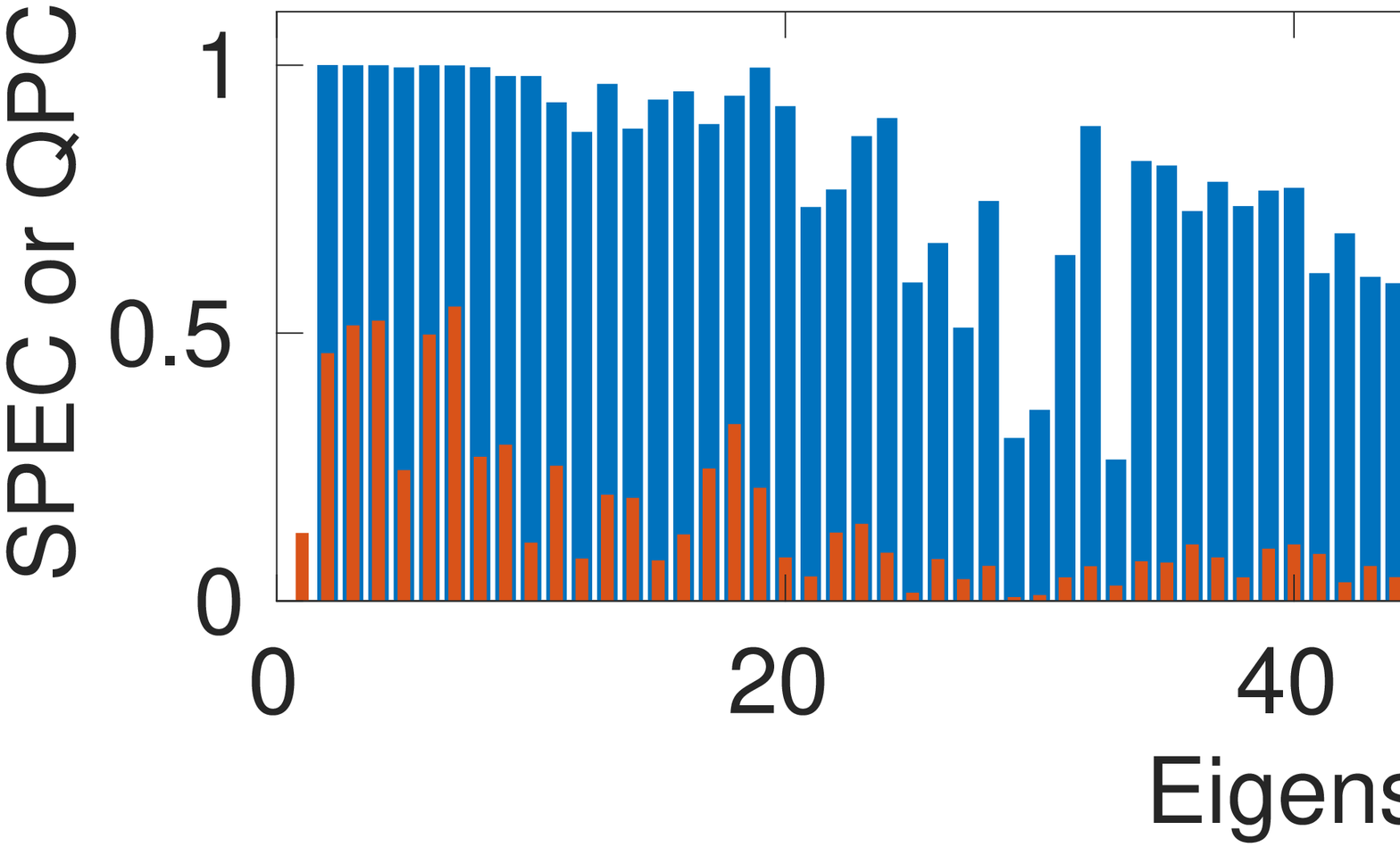}
    \caption{Evolution of charge-neutral spectral function, SPEC (blue bars) and QPC (red bars) for $\lambda_{ee}/t = $ 0, 0.1, 1, and 10 in a chain of twelve atoms with six electrons. For zero interaction (top plot) red and blue bars are identical. }
    \label{fig:SPECandQPCvaryCoul}
\end{figure}
\begin{figure}
    \includegraphics[width=0.65\textwidth]{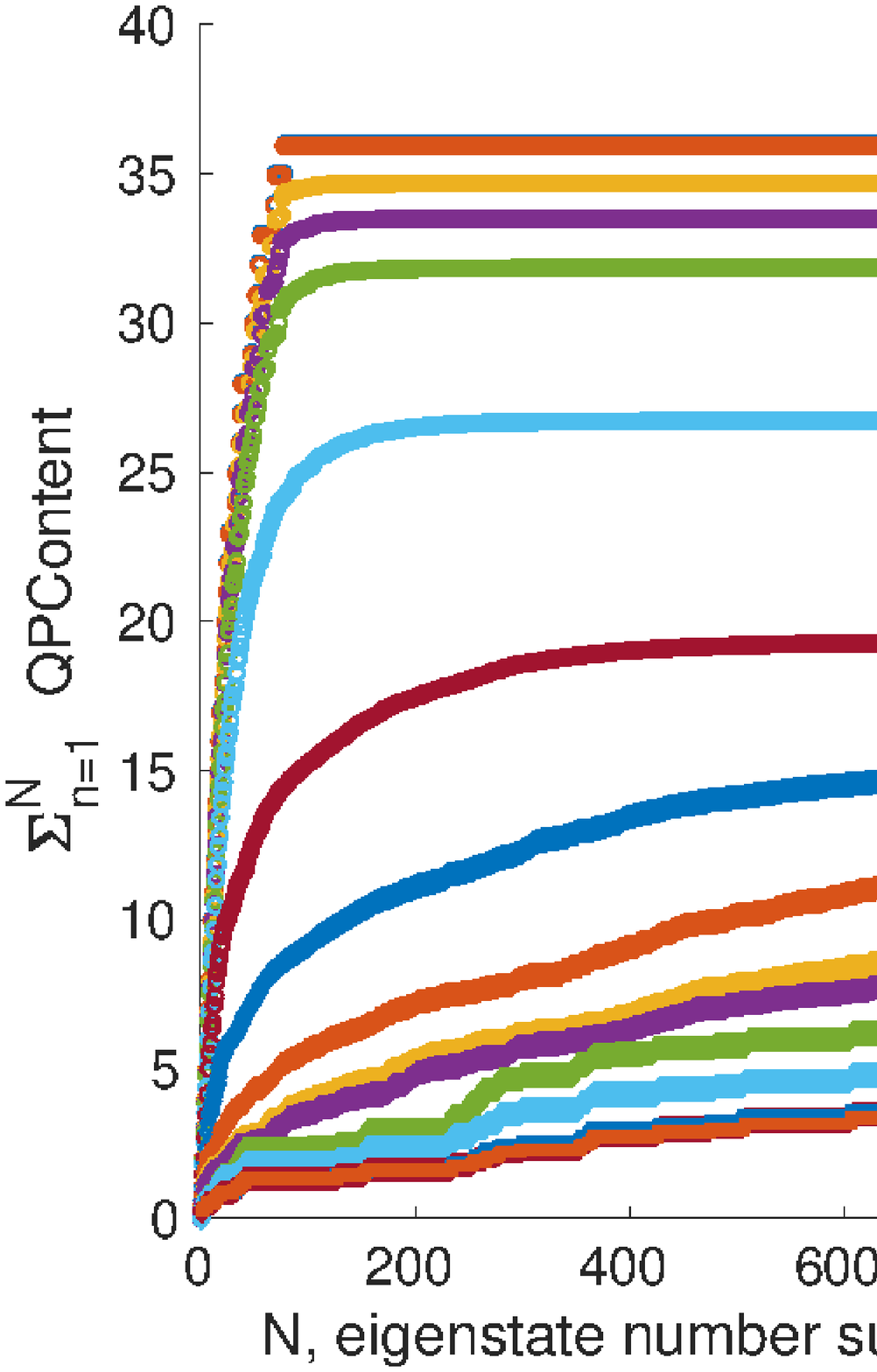}
    \caption{Rolling partial sum of QPC for varying Coulomb strengths, for a chain of twelve atoms and six electrons.  Its maximum of 36 occurs in the non-interacting limit.  The SPEC also has a maximum of 36 in the non-interacting limit.}
    \label{fig:QPCPartialSum}
\end{figure}

Figure \ref{fig:SPECandQPCvaryCoul} shows how the spectral function, SPEC and QPC evolve for different Coulomb values.  Figure \ref{fig:QPCPartialSum} shows a rolling sum for the QPC, $\sum_{n=1}^N QPC(\Psi_n) $, similar to the rolling sums for SPEC from the main paper. 
SPEC and QPC are identical for the non-interacting system.  However as the Coulomb interaction increases, the SPEC grows, while the QPC is diminished as the quasiparticles we envisioned become a less accurate description of the system.

There are three differences between the QPC and the SPEC.
First, the quasiparticle content we have just defined relies on us identifying the particular physics of the problem.  We must identify the modes and what we mean by a momentum kick in order to define what we expect the excitations to be like.  We defined it in terms of a ``momentum kick" or a change in mode quantum number that is meaningful in the non-interacting picture, but becomes increasingly meaningless as correlations become dominant.
The single particle excitation content contains every possible single-particle excitation, not merely those we expect to be relevant based on the non-interacting version of the problem.  It does not look for excitations with a particular mode shift.  As a result, in the SPEC we see parallel behavior in the two opposite regimes of the Wigner crystal perturbed by hopping and the single-particle system perturbed by Coulomb interactions.

Second, the QPC contains only excitations from a lower single-particle mode to a higher single-particle mode, whereas SPEC contains every possible transition, whether it is an excitation to a higher mode or a lower one. (And this distinction is not meaningful far away from the non-interacting picture: single particle mode number occupation is not what accounts for the difference in energy between different eigenstates.)

\begin{figure}  
    \includegraphics[width=0.4\textwidth]{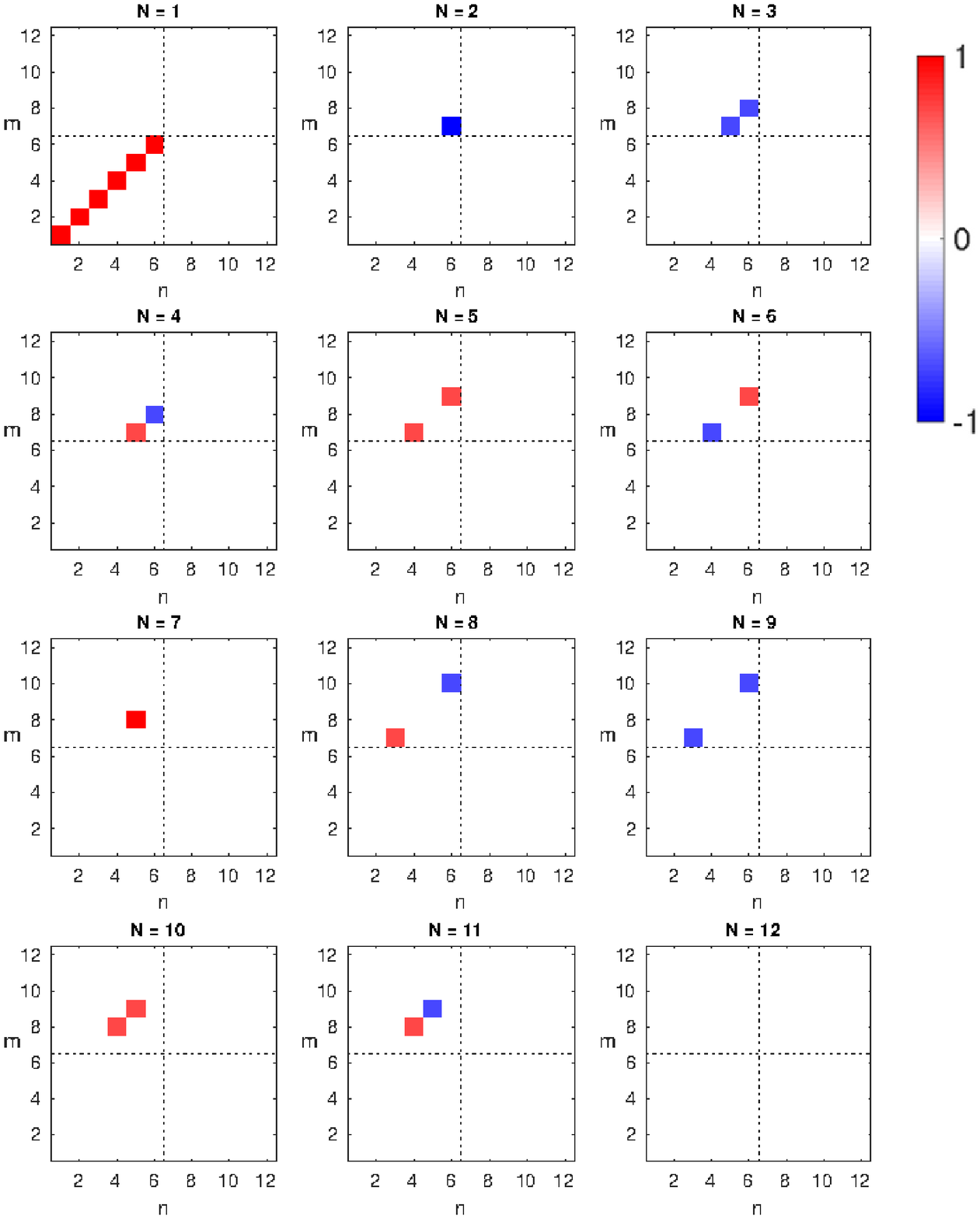}
    \hspace*{0.1\textwidth}
    \includegraphics[width=0.4\textwidth]{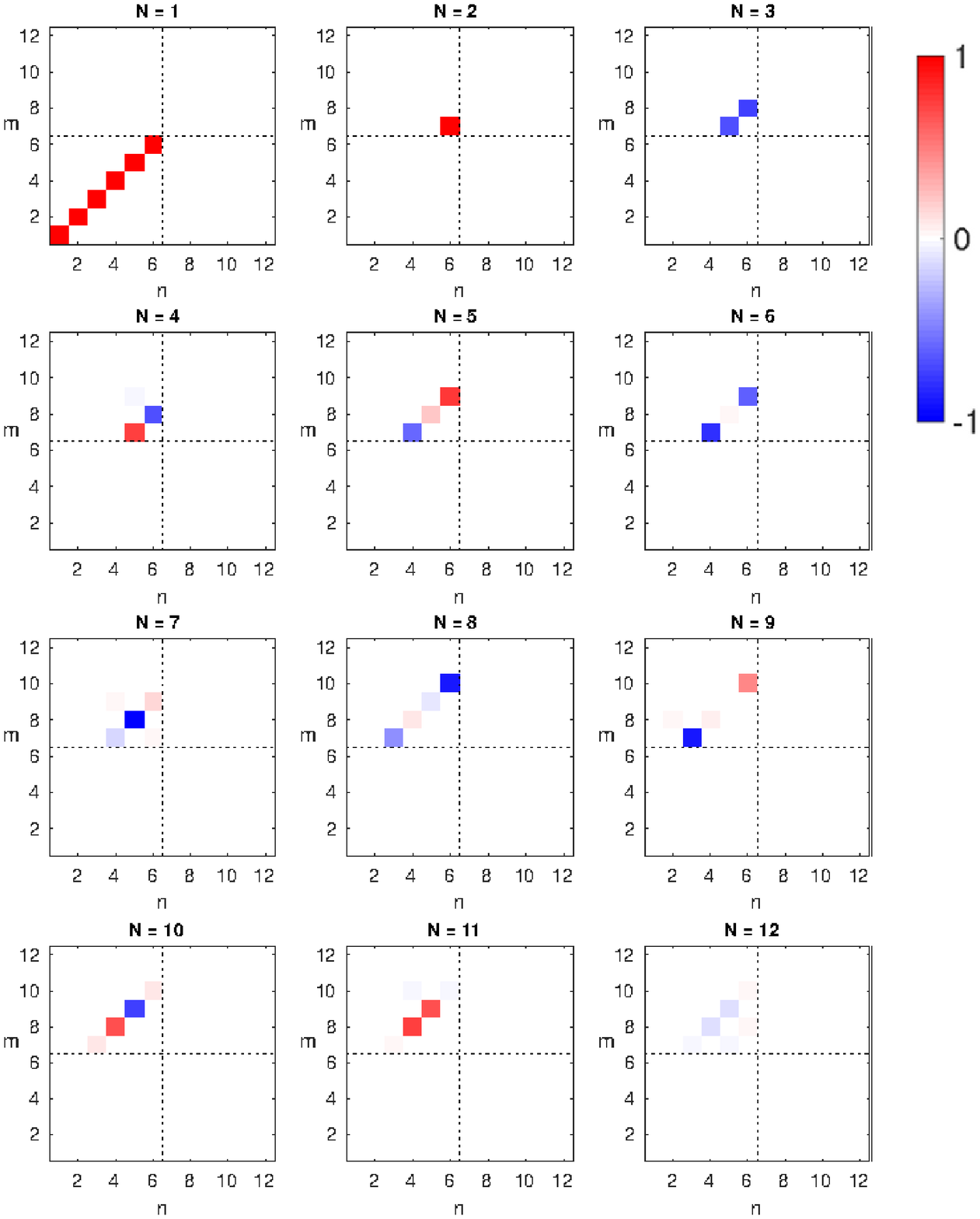} 
    \caption{Projections of excited states onto single-particle excitations: $\langle \Psi_N | a_m^\dagger a_n |\Psi_{GS}\rangle$ for $\lambda_{ee}= 0$ (left) and $0.1t$ (right), for a chain of twelve atoms and six electrons.  The SPEC and QPC of state $N=12$ are zero for $\lambda_{ee}=0$. However for $\lambda_{ee}=0.1$ the QPC is near zero, while the SPEC is close to one due to the Gram-Schmidt orthogonalization.}
    \label{fig:EigonSPwf}
\end{figure}

Finally, and most significantly, the QPC has not gone through the Gram-Schmidt orthogonalization.  The SPEC identifies the extent to which a given eigenstate fits in the space {\it spanning all} single-particle excitations from the ground state.
The QPC identifies the extent to which a given eigenstate can be identified with a particular set of single-particle excitations from the interacting ground state.  An excitation from the ground state that is only minutely possible (because the starting mode is nearly empty or the final mode is nearly full) still fully counts toward the single-particle excitation subspace.
To appreciate the impact of the Gram-Schmidt orthogonalization we can consider the difference between the SPEC and the QPC for the $N=12$ eigenstate.  
Figure \ref{fig:EigonSPwf} is similar to Figure 5 of the main paper.  It shows the projection of various eigenstates onto every possible single-particle mode change of the ground state: $\langle \Psi_N| a^{\dagger}_m a_n | \Psi_{GS} \rangle.$  The index $m$ is on the vertical axis of each subplot, and $n$ is on the horizontal axis.  The left panel is for the non-interacting system, and the right panel is for $\lambda_{ee} = 0.1t$.
As can be seen in figure \ref{fig:EigonSPwf} when there is no Coulomb interaction,  state $N=12$ is not a single-particle excitation.  And when the Coulomb interaction is $\lambda_{ee} =0.1 t$ we can see that the projection of state twelve onto every individual single-particle excitation of the ground state ($\langle \Psi_{12} | a_m^\dagger a_n | \Psi_{GS}\rangle$) is miniscule.  As a result, this state does not consist of a single-electron promotion of any $\Delta k$, and its quasiparticle content is near zero.  
However the interacting ground state has been altered by this still small Coulomb interaction.  And the space spanned by all possible single-particle excitations of the ground state now encompasses nearly all of state 12.  This says as much about the interacting ground state as it says about state 12.  There are some very small amplitude excitations of the ground state that become amplified when we ask ``what is the space {\it{spanned}} by all single-particle excitations of the ground state, and how well does state 12 fit in that space?"  

The quasiparticle content tells us whether we can renormalize the interaction and use a single-particle-like description of our system.  
The single-particle excitation content shows us the impact of the interactions on the many-body ground and excited states.
\clearpage

\section{Order parameters}\label{sect:SI.OP}

\begin{figure}
  \includegraphics[width=0.65\textwidth]{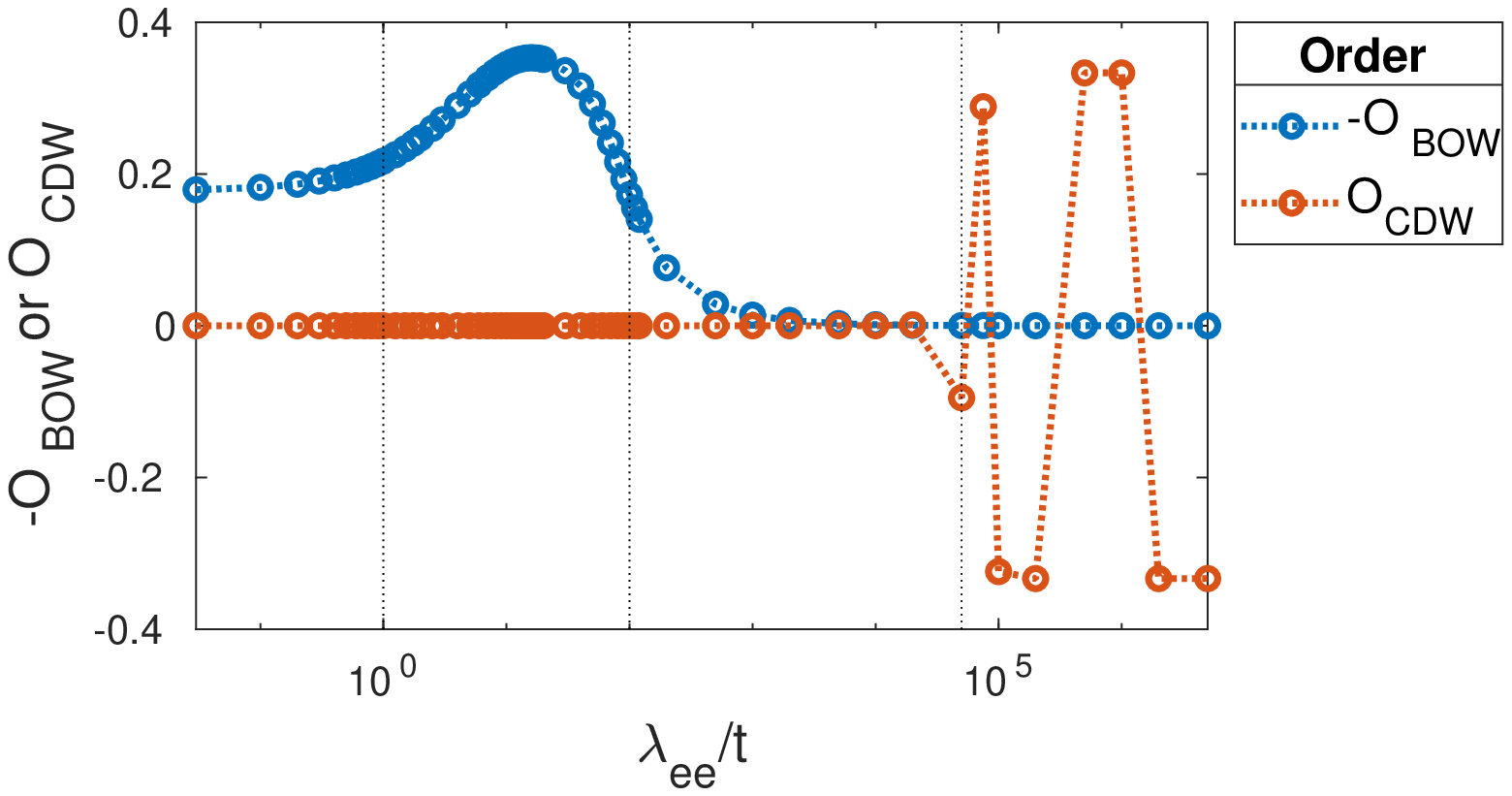}
\caption{Simplest OPs for bond order wave (multiplied by -1) and charge-density wave in the ground state as a function of Coulomb interaction strength for a chain of 12 atoms with six electrons.}
\label{fig:OPsSimple}
\end{figure}

A bond-order wave has bonds between sites in the chain which alternate in strength: e.g. a strong bond between site $i$ and $i+1$ when $i$ is even, and weak when $i$ is odd.  As such, the most straightforward way to measure its presence is to add up the strength of the bonds for even $i$, and subtract the strength of all the bonds with odd $i$.  This is a common definition of the order parameter of the bond order wave \cite{Nakamura2000, Ejima2007, Liu2011, Mishra2011, Ding2011} :
\begin{align}
O_{BOW} =  {1\over (N_s-1)} \sum_{i=1}^{N_s -1} (-1)^i   \langle\Psi_{GS}| (c_i^\dagger c_{i+1} + c_{i+1}^\dagger c_i  | \Psi_{GS} \rangle
\end{align}
and similarly for the charge density wave in which the number of particles alternates between sites: 
\begin{align}
O_{CDW} =  {1\over (N_s)} \sum_{i=1}^{N_s} (-1)^i   \langle\Psi_{GS}| c_i^\dagger c_{i} | \Psi_{GS} \rangle
\label{eq:OCDWsimple}
\end{align}
(We have adapted all order parameter definitions for short chains by dividing not by the length of the chain,  but by the actual number of terms in each sum.)
These order parameters look for order in the expectation value of the operators for the bond strength, $B_i = c_i^\dagger c_{i+1} + c_{i+1}^\dagger c_i  $ and the number operator, $n_i = c_i^\dagger c_i  $, respectively.  Figure \ref{fig:OPsSimple} shows these order parameters for our system as a function of the strength of the Coulomb interaction, $\lambda_{ee}/t$.  As discussed in the main paper, for the large interaction region where correlation is strong and Wigner crystalization is occuring the ground state of the system is doubly degenerate.  When every other site is occupied in a Wigner crystal, the odd or even sites can be occupied.    For very large values of interaction ($\lambda_{ee}/t > 5\times 10^4$) our code sometimes chooses a ground state with odd sites occupied and a first excited state with even sites occupied, and other times reverses their order. So the sign of the CDW simple order parameter (equation \ref{eq:OCDWsimple}) varies from one run to the next in this region.  However for the region $100 <  \lambda_{ee}/t < 5\times 10^4$ the two lowest states are symmetric and antisymmetric linear combinations of odd sites occupied and even sites occupied.  As a result, the expectation value of the occupancy of each site is equal in these states, and no charge density wave order is exhibited by the order parameter in equation \ref{eq:OCDWsimple} and figure \ref{fig:OPsSimple}.

\begin{figure}
  \includegraphics[width=0.65\textwidth]{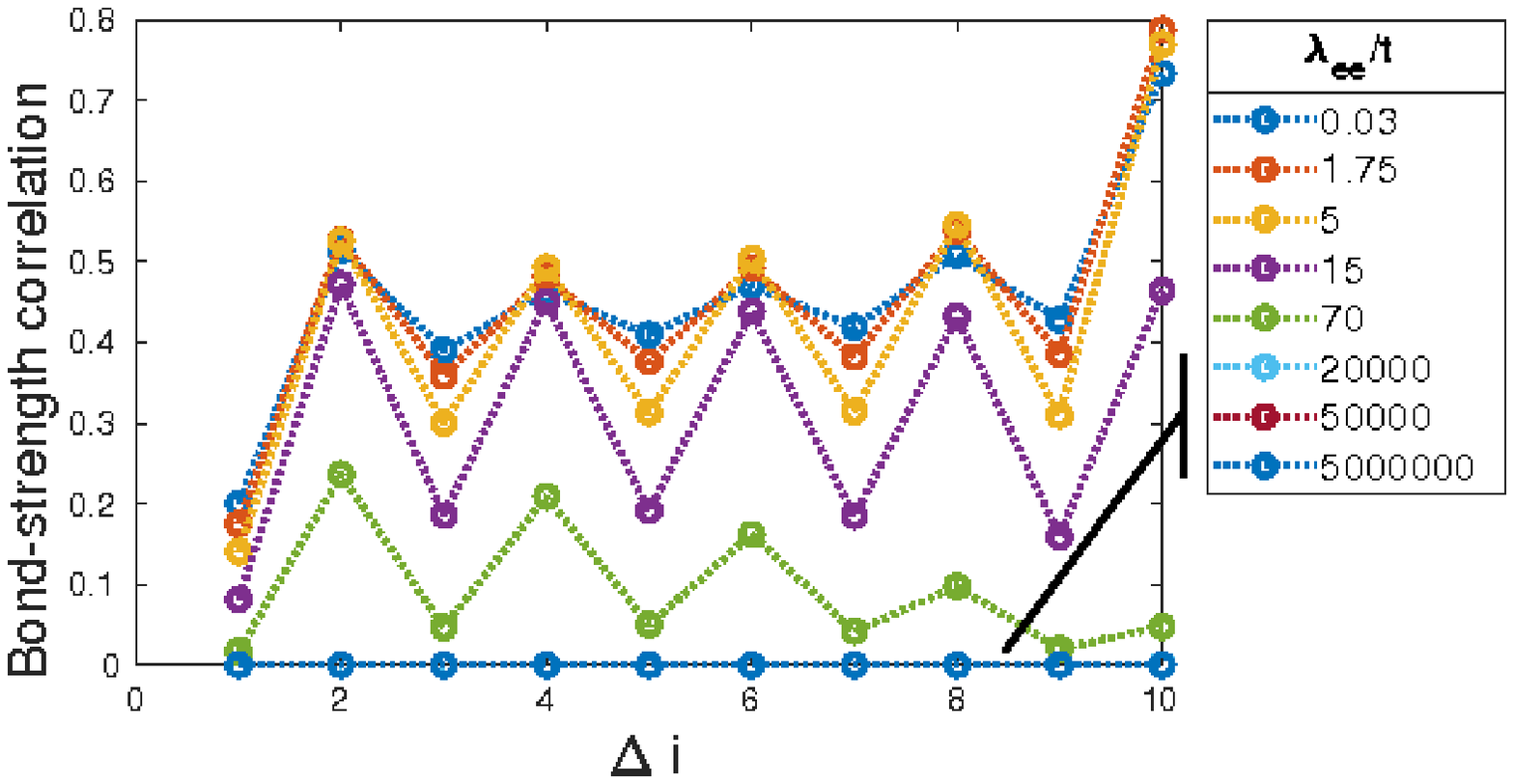}
  \includegraphics[width=0.65\textwidth]{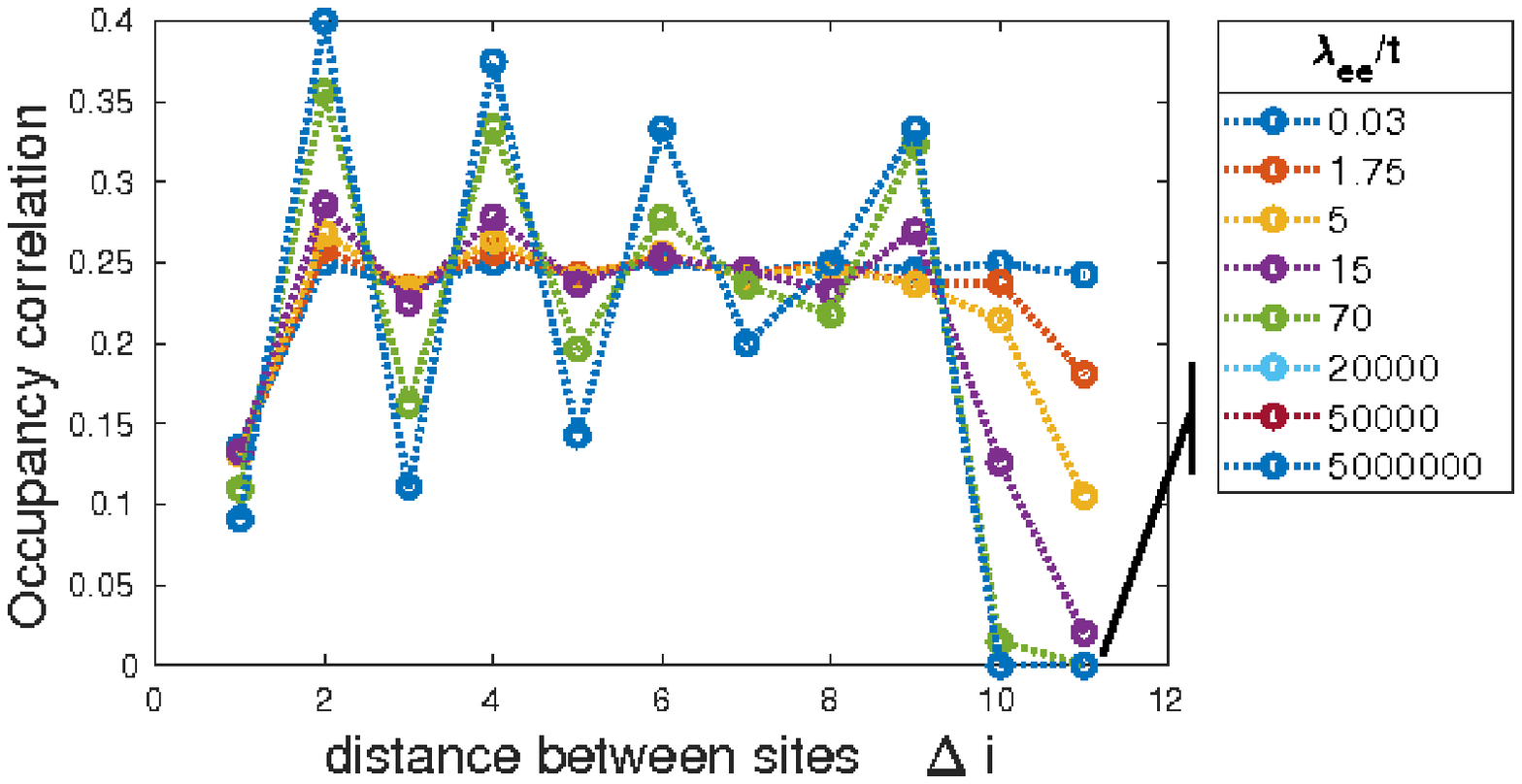}
\caption{Disconnected correlation functions for bond strength and site occupation in the ground state as a function of distance, for a chain of twelve atoms and six electrons.}
\label{fig:CorrFxns}
\end{figure}

However in quantum systems there is also the possibility that there exist correlations not captured directly in these expectation values.  As such, there are several other quantities that describe the order of a system.  
Sengupta et al. also  look for bond order by plotting the correlation function of the bond strength (kinetic energy) as a function of distance between bonds \cite{Sengupta2002},
and observing every-other site oscillations in $\Delta i$.  However in Sengupta's definition they have subtracted off the product of the individual bond strengths.  The correlation function $ \langle n_i n_{i+\Delta i}  \rangle$  is sometimes referred to as the disconnected correlation and contains both classical and quantum correlations, whereas the difference between it and the product $\langle n_i \rangle \langle n_{i+\Delta i}  \rangle$ is referred to as the connected correlation and represents only quantum correlations due to entanglement between site $i$ and $i+\Delta i$ \cite{Tran2017}. $C^{Conn} = C^{Dis} - C^{Classical}$.  
\begin{align}
    C_{BOW}^{Conn}(\Delta i) = {1 \over( N_s - \Delta i -1)} & \sum_{i=1}^{N_s - \Delta i -1}   \langle\Psi_{GS}| (c_i^\dagger c_{i+1} + c_{i+1}^\dagger c_i )(c_{i+\Delta i}^\dagger c_{i+\Delta i+1} + c_{i+\Delta i+1}^\dagger c_{i+\Delta i}) |\Psi_{GS}\rangle \nonumber
     \\
    & - \langle\Psi_{GS}| (c_i^\dagger c_{i+1} + c_{i+1}^\dagger c_i ) |\Psi_{GS}\rangle \langle  \Psi_{GS}|(c_{i+\Delta i}^\dagger c_{i+\Delta i+1} + c_{i+\Delta i+1}^\dagger c_{i+\Delta i}) |\Psi_{GS}\rangle,
\end{align}
The disconnected correlation functions for BOW and CDW are:    
\begin{align}
C_{BOW}(\Delta i) & = {1 \over( N_s - \Delta i -1)} \sum_{i=1}^{N_s - \Delta i -1}   \langle\Psi_{GS}| (c_i^\dagger c_{i+1} + c_{i+1}^\dagger c_i )(c_{i+\Delta i}^\dagger c_{i+\Delta i+1} + c_{i+\Delta i+1}^\dagger c_{i+\Delta i}) |\Psi_{GS}\rangle 
\\
C_{CDW}(\Delta i) & = {1 \over( N_s - \Delta i )} \sum_{i=1}^{N_s - \Delta i }   \langle\Psi_{GS}| c_i^\dagger c_{i} c_{i+\Delta i}^\dagger c_{i+\Delta i} |\Psi_{GS}\rangle,  
\end{align}
and are shown for a select number of interaction values in figure \ref{fig:CorrFxns}. A clear every-other site behavior is seen for smaller separations for the values of interaction strength in which there is significant order of the relevant type (BOW or CDW).  The exception is for longer range correlations of occupancy.  In our charge neutral system the electron-nuclear interaction grows proportionally to the electron-electron interaction strength.   For the larger values of Coulomb interaction strength where the CDW order occurs, the nuclear interaction draws the electrons toward the center of the chain, meaning there are effectively only ten sites available for six electrons (in low energy states such as the ground state).  This induces a defect in the preferred order of alternate site occupancy,  both on the ten sites, as well as on the two end sites.  Also apparent in figure \ref{fig:CorrFxns} is a decay in the strength of the occupancy correlation with distance, but no decay in the bond-strength correlation.

 The Fourier transform of these correlation functions gives the static structure factors \cite{Zhang2004, Sengupta2002, Shao2019}.  The disconnected or total static structure factors for a bond order wave and charge density wave for wavenumber $q$ are given by:
\begin{equation}
S_{BOW}(q) =   \frac{1}{(N_s-1)} \sum_{\Delta i=1}^{N_s-1} e^{iq\Delta i} {1\over (N_s - \Delta i - 1)} \sum_{i=1}^{N_s -\Delta i-1} \langle\Psi_{GS}| (c_i^\dagger c_{i+1} + c_{i+1}^\dagger c_i )(c_{i+\Delta i}^\dagger c_{i+\Delta i+1} + c_{i+\Delta i+1}^\dagger c_{i+\Delta i}) |\Psi_{GS}\rangle
\end{equation}
and
\begin{equation}
S_{CDW}(q) =  \frac{1}{N_s} \sum_{\Delta i=1}^{N_s} e^{iq\Delta i} {1\over (N_s-\Delta i)} \sum_{i=1}^{N_s-\Delta i} \langle\Psi_{GS}| c_i^\dagger c_{i}  c_{i+\Delta i}^\dagger c_{i+\Delta i} |\Psi_{GS}\rangle.
\end{equation}

\begin{figure}
 \includegraphics[width=0.65\textwidth]{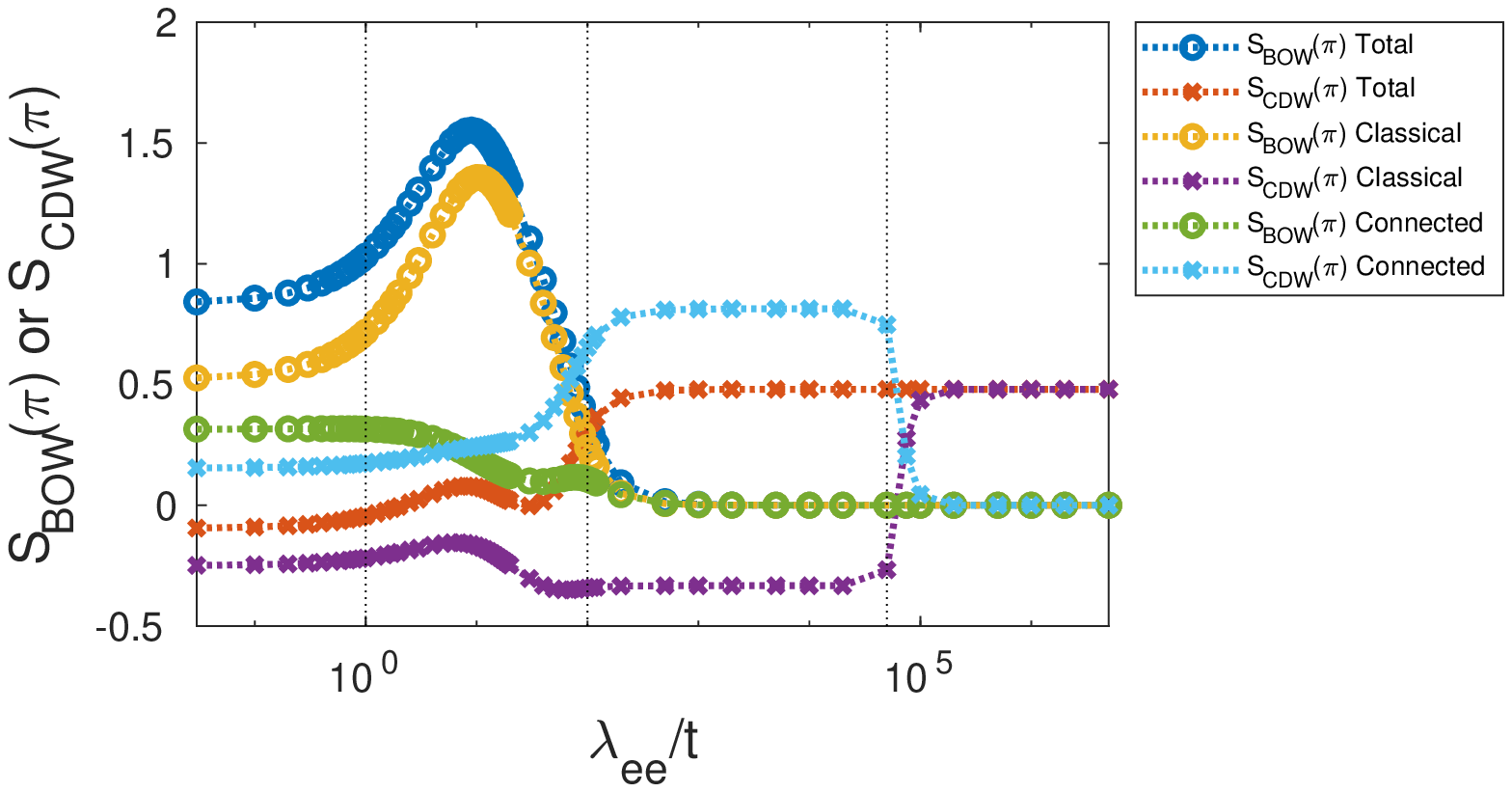}
\caption{Staggered structure factors for BOW and CDW in the ground state as a function of interaction strength. $C^{Total} = C^{Disconnected}$; $C^{Connected} = C^{Disconnected} - C^{Classical}$. The three vertical dashed lines indicate the transitions between the weakly interacting and intermediate interaction regimes, the intermediate interaction regime and the strongly interacting regime, and the point in the strongly interacting regime where the doubly degenerate ground states returned by the eigensolver transition to a different linear combination.   }
\label{fig:StagStructFacts}
\end{figure}

Since we are most interested in detecting order similar to a Wigner crystal, with modulation at every other site, the $q=\pi$ term is most relevant, giving the staggered structure factor, from which an order parameter may be defined \cite{Sengupta2002, Shao2019} as $m_{BOW}=\sqrt{S_{BOW}(\pi)}$.
(Shao \cite{Shao2019} states that this relationship between the structure factor and the order parameter applies in the $L \rightarrow \infty$ extrapolation.)
However because the staggered structure factors can be negative, we choose to plot them rather than their square root, $m$, in figure \ref{fig:StagStructFacts}, which shows both the total or disconnected staggered structure factor, as well as the staggered structure factors for the product of expectation values (classical) and the difference between these two, the connected static structure factor (due to entanglement between sites), for each type of order.
Figure \ref{fig:StagStructFacts} shows that the charge density order is present for all strong interactions, $\lambda_{ee}/t > 100$, and has some onset at the intermediate values.  
The CDW order is classically present,  ($S_{CDW}^{Classical}$, purple x's) for the largest interaction strengths (the classical no-hopping limit would have electrons living completely on every other single site), while for slightly lower values the full density-density correlation function ($S_{CDW}^{Total}$, red x's) needs to be used to identify the order.  
(The dip around $\lambda_{ee}/t = 15$  occurs because the ends of the chain start becoming depopulated at this point as the electron-nuclear interaction gives the ends of the chain a higher potential energy, as seen in figure \ref{fig:CorrFxns}.) 
The bond order, on the other hand, is apparent even in the expectation value of the bond strength ($S_{BOW}^{Classical}$, yellow circles), and the higher order correlation function ($S_{BOW}^{Connected}$, green circles) contributes only a little to the total disconnected structure factor ($S_{BOW}^{Total}$, blue circles).

Because we are interested in the total amount of order in the different regions of interaction strength, the disconnected static structure factors are those that we choose to plot as an order parameter in the main body of this paper.

\clearpage

\section{Varying Parameters} \label{sect:SI.Vary}
\begin{figure}
\includegraphics[width=0.45\textwidth]{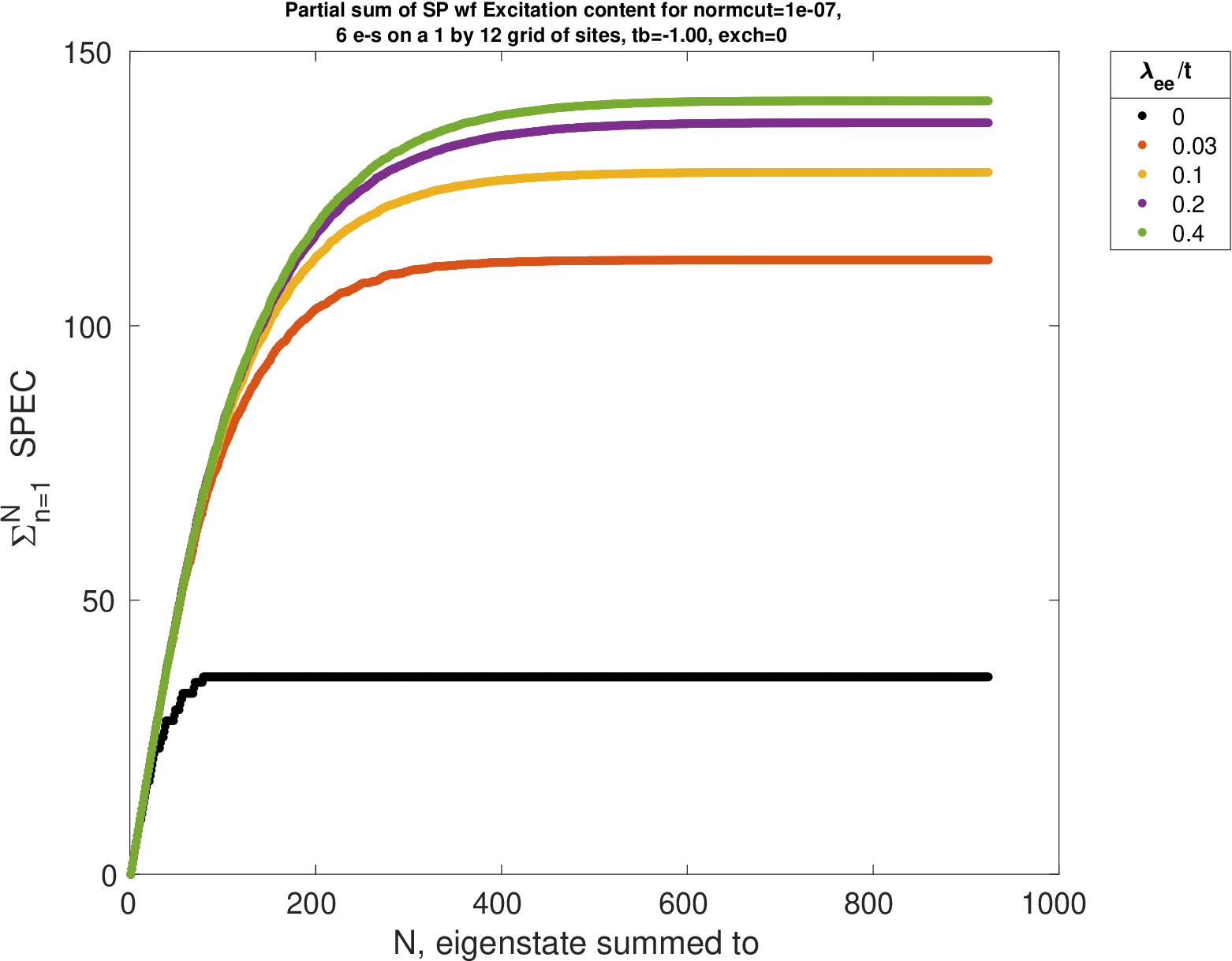}
\includegraphics[width=0.45\textwidth]{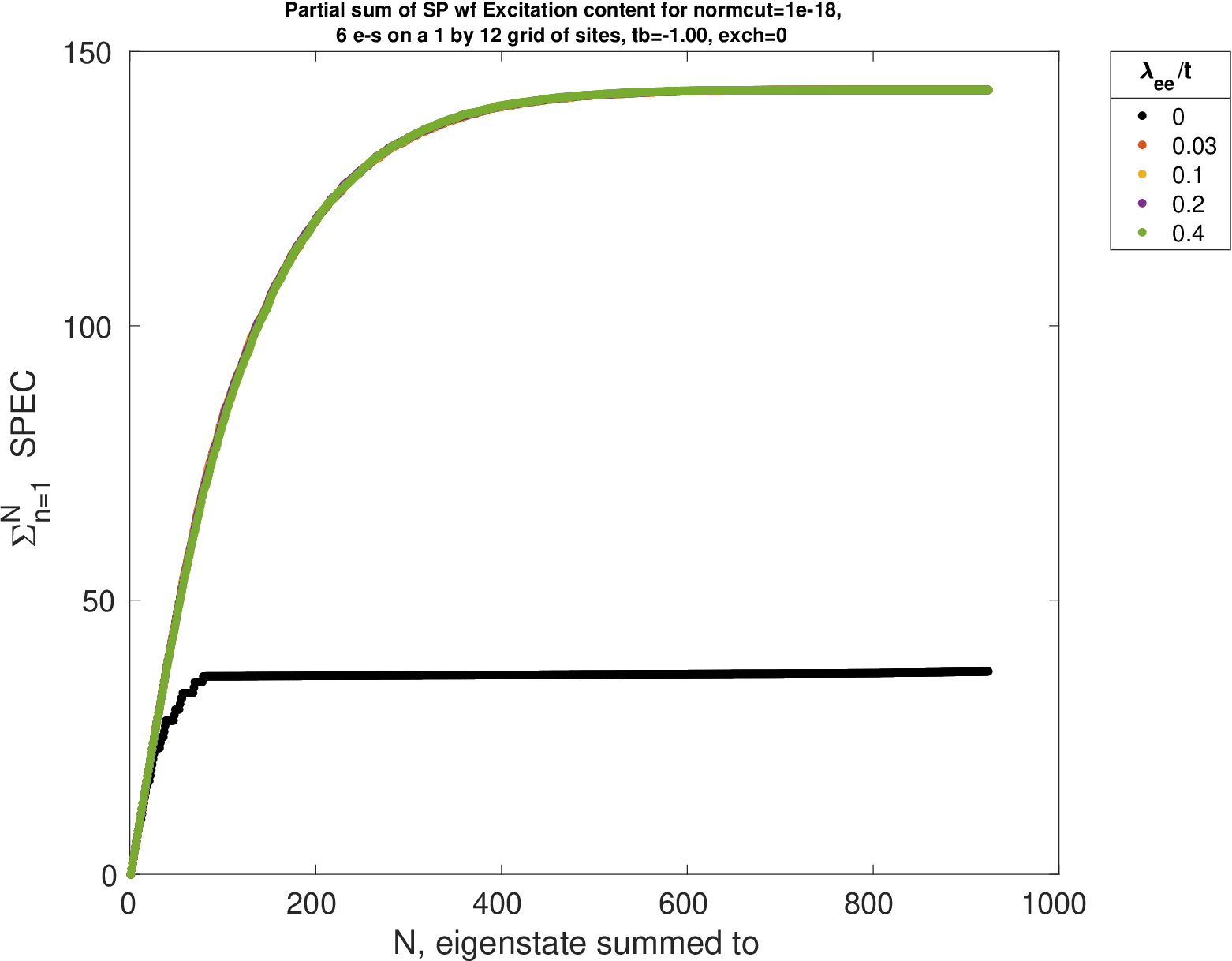}
\includegraphics[width=0.45\textwidth]{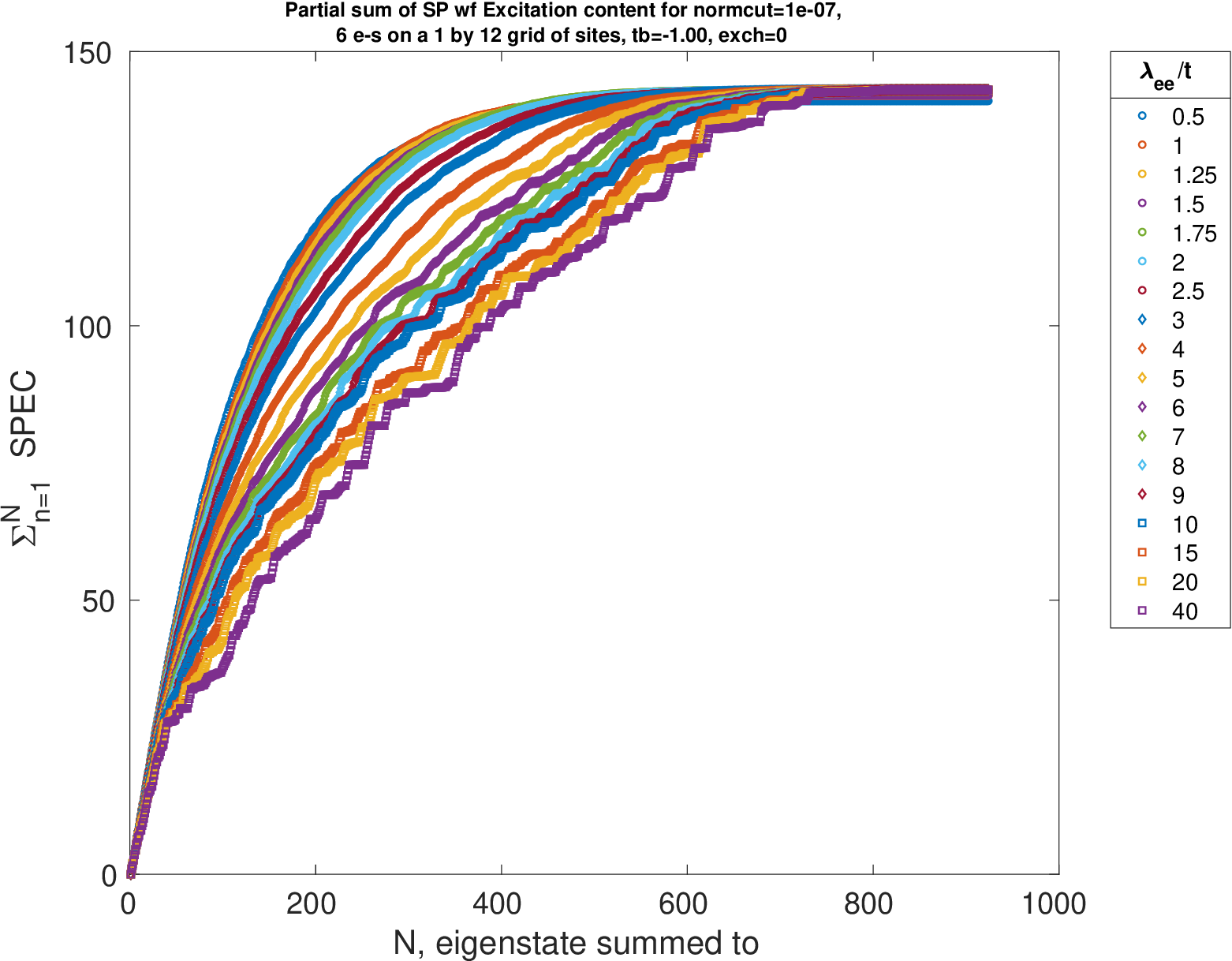}
\includegraphics[width=0.45\textwidth]{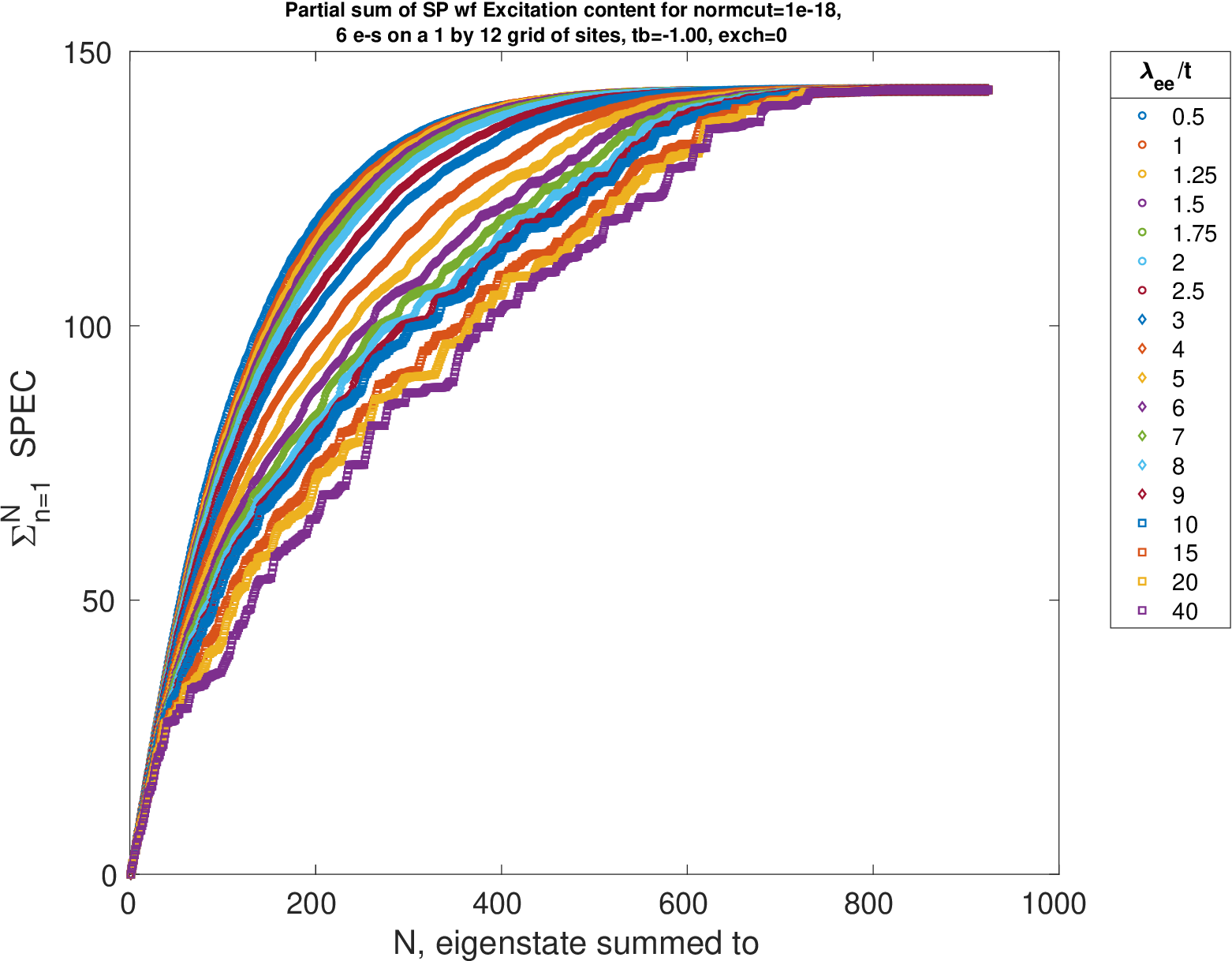}
\includegraphics[width=0.45\textwidth]{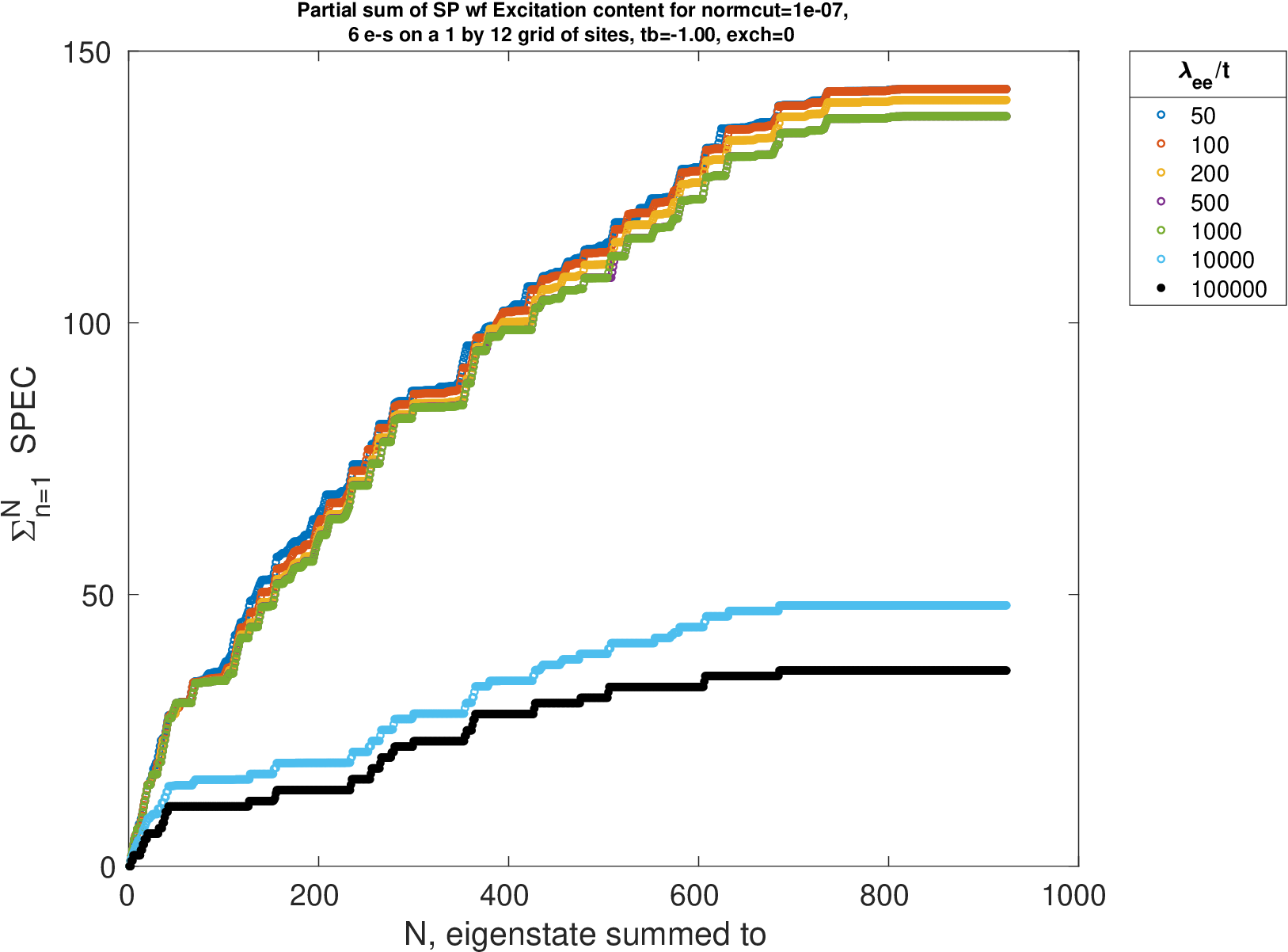}
\includegraphics[width=0.45\textwidth]{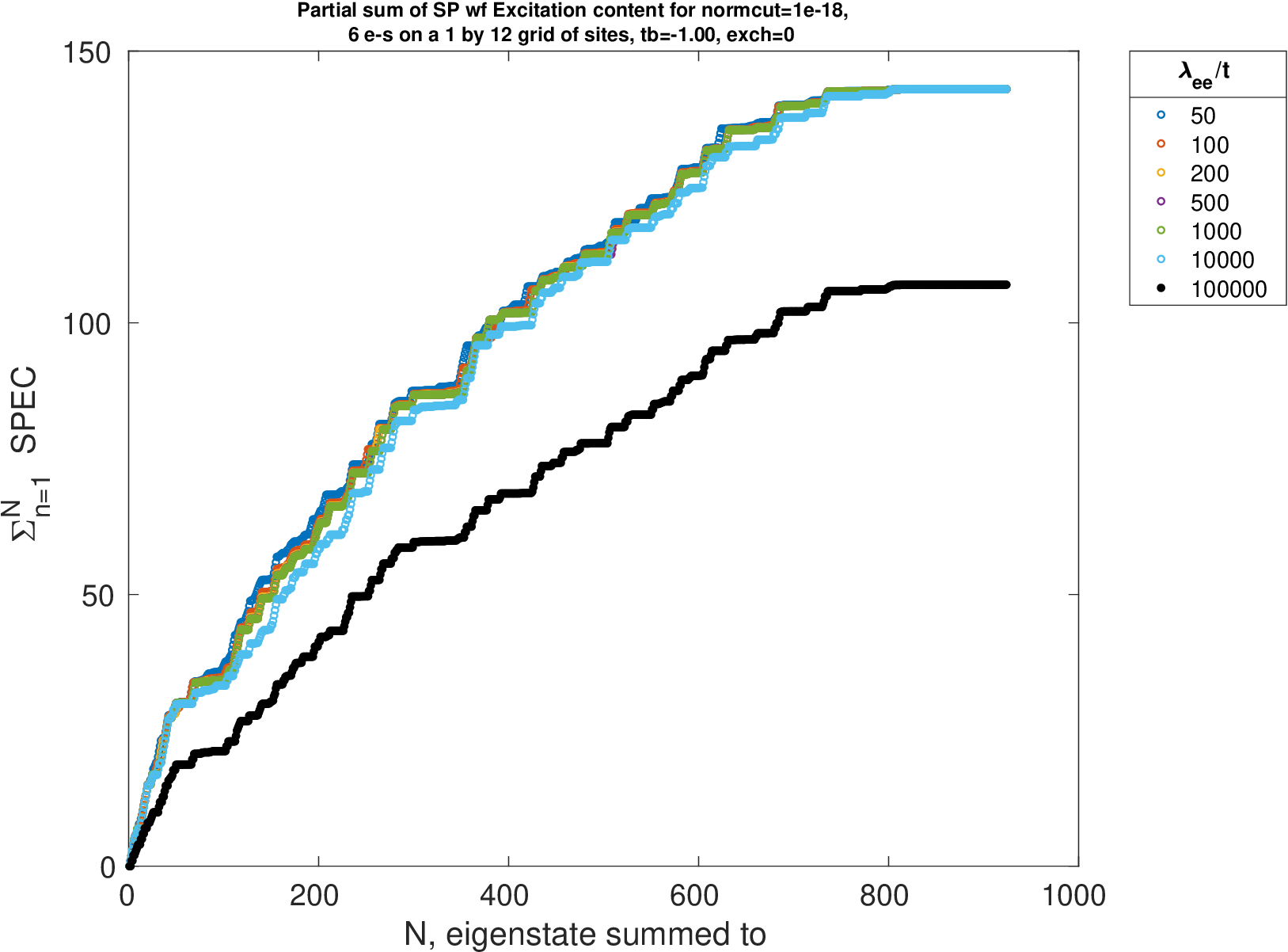}
\includegraphics[width=0.45\textwidth]{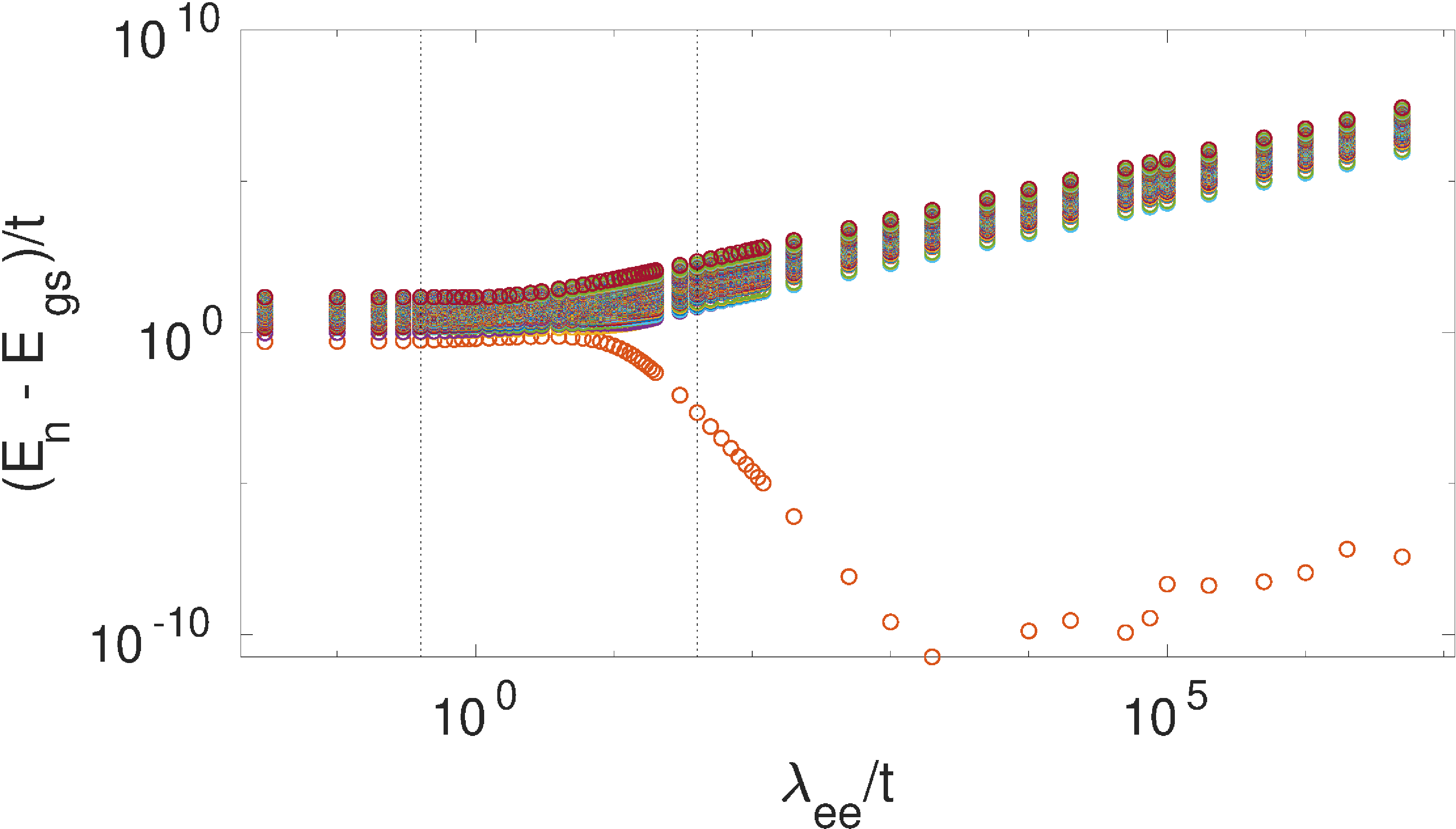}
\caption{Rolling sum of SPEC for 6 electrons in a 12 atom chain without any nearest neighbor exchange reduction ($f_{ex}=0$).   Plots show different regimes of Coulomb interaction strength, with plots in the left column having a more exclusive cutoff ($\epsilon_{G-S}= 10^{-7}$) and plots in the right column, a more inclusive one ($\epsilon_{G-S} = 10^{-18}$), as in the main text.  Bottom plot shows the excitation energy of the eigenstates as a function of interaction strength (both in units of hopping) on a log-log scale. The crossover points between strong, intermediate and weak interactions, indicated by vertical lines on the energy plot, have changed, but the behavior of the SPEC distinguishing those regions remains similar to that described in the main text.}

\label{fig:NoExch}
\end{figure}

\begin{figure}
\includegraphics[width=0.45\textwidth]{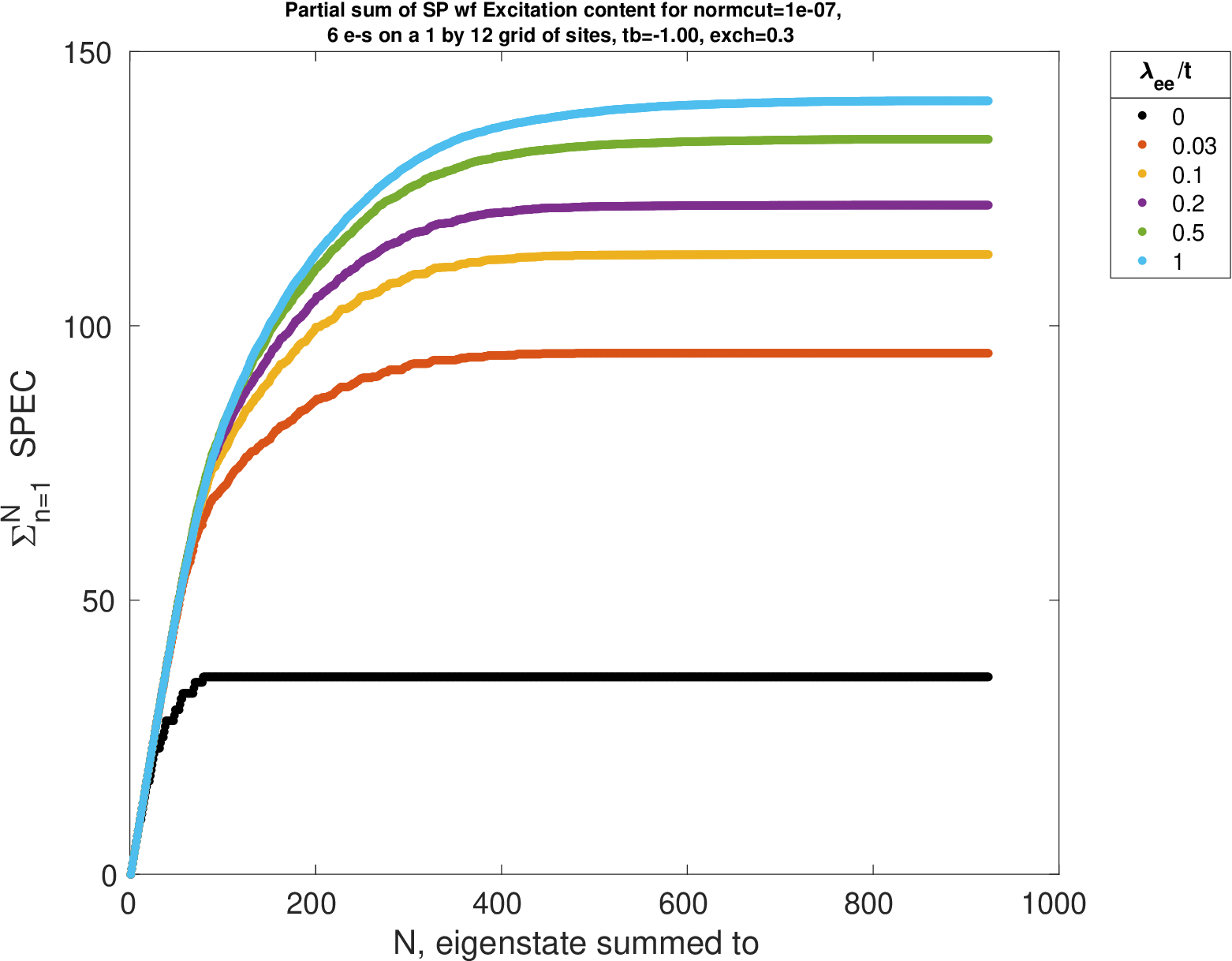}
\includegraphics[width=0.45\textwidth]{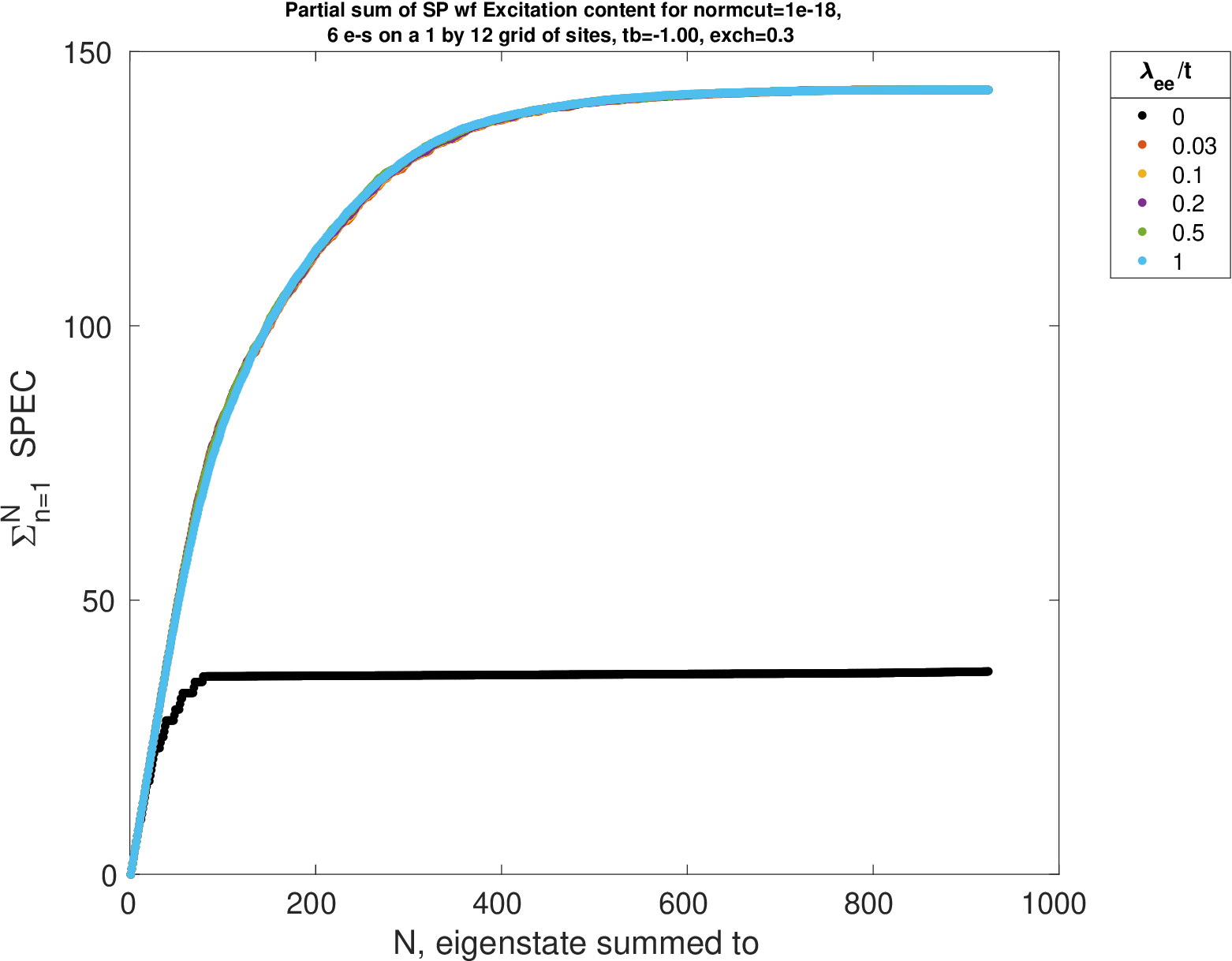}
\includegraphics[width=0.45\textwidth]{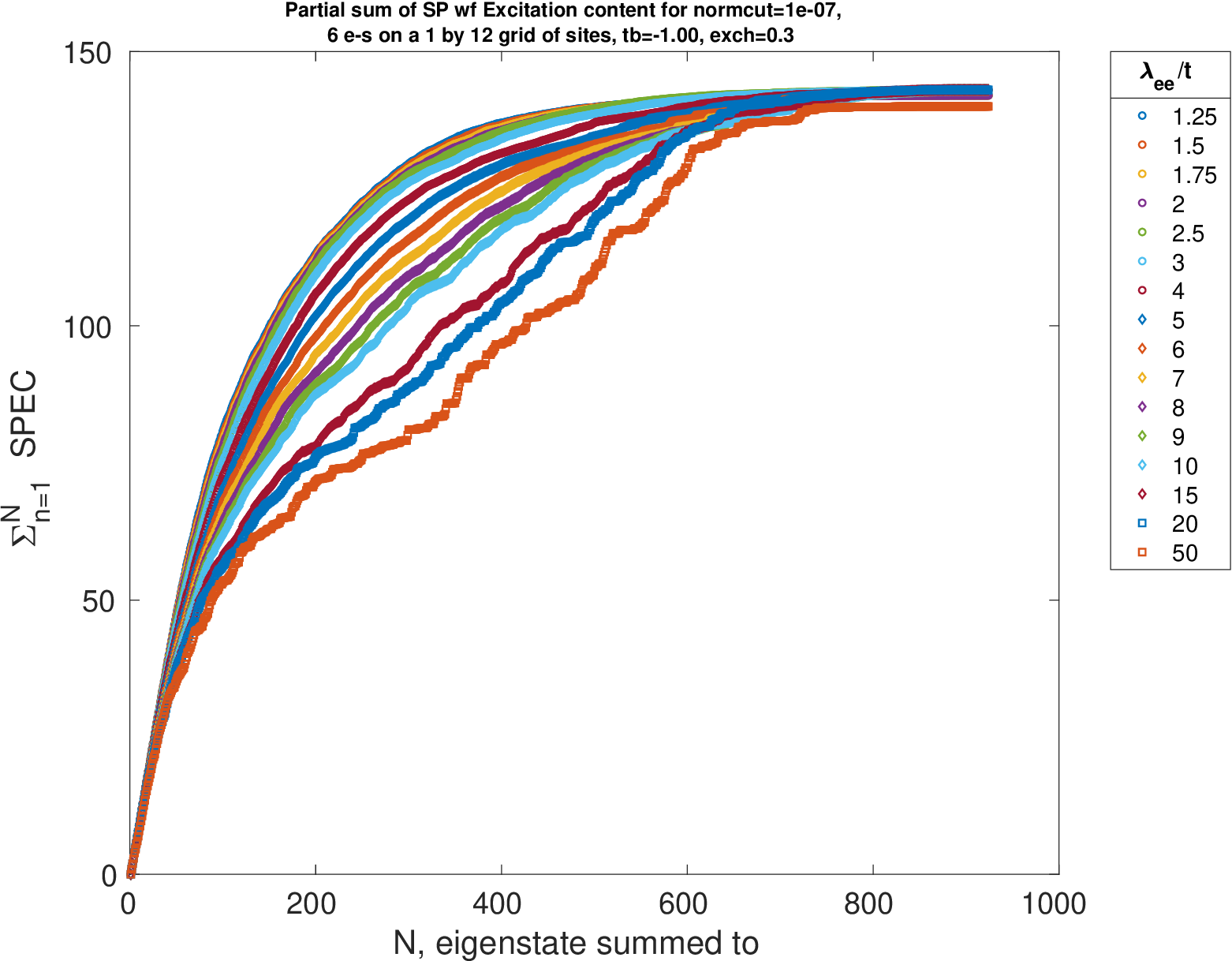}
\includegraphics[width=0.45\textwidth]{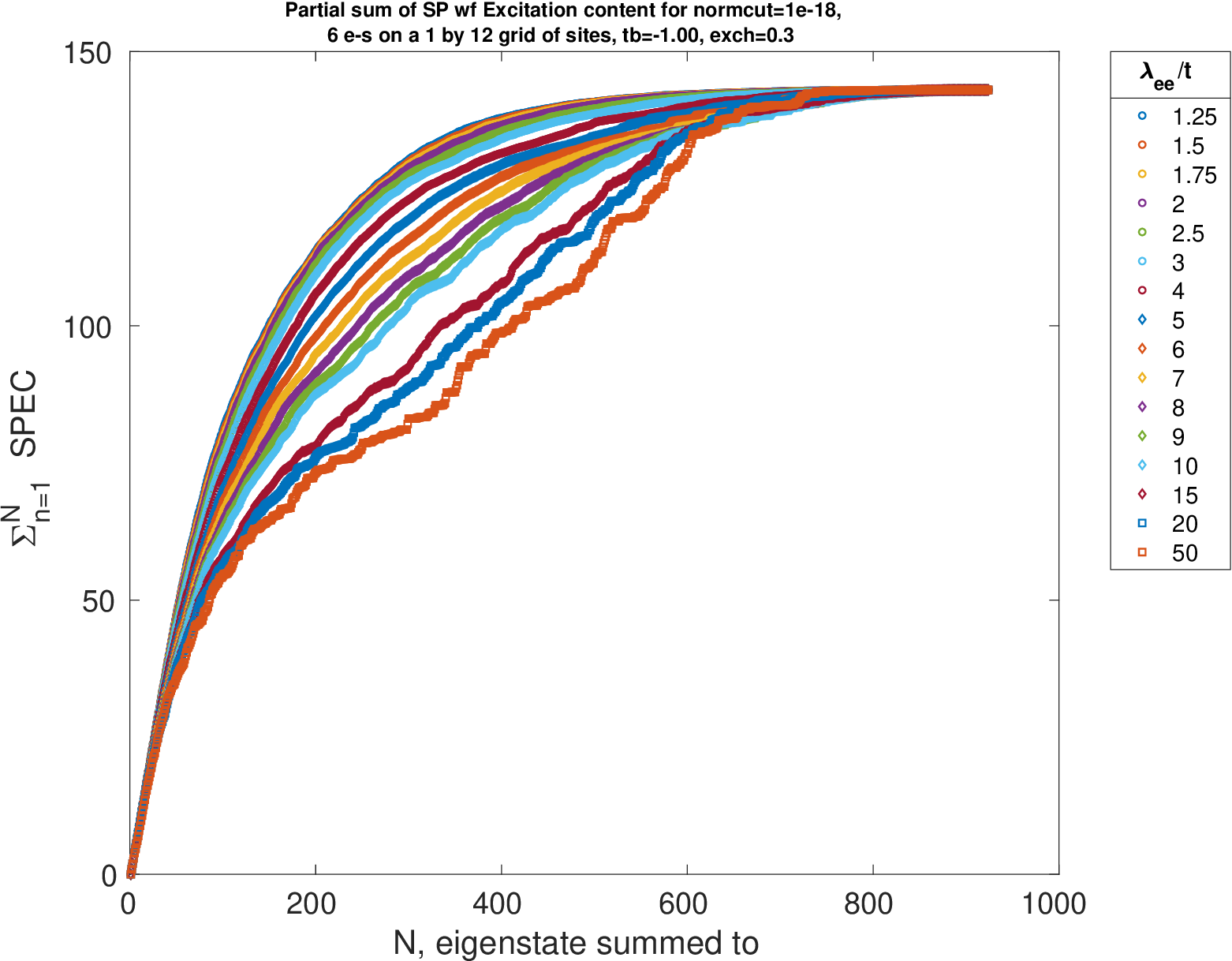}
\includegraphics[width=0.45\textwidth]{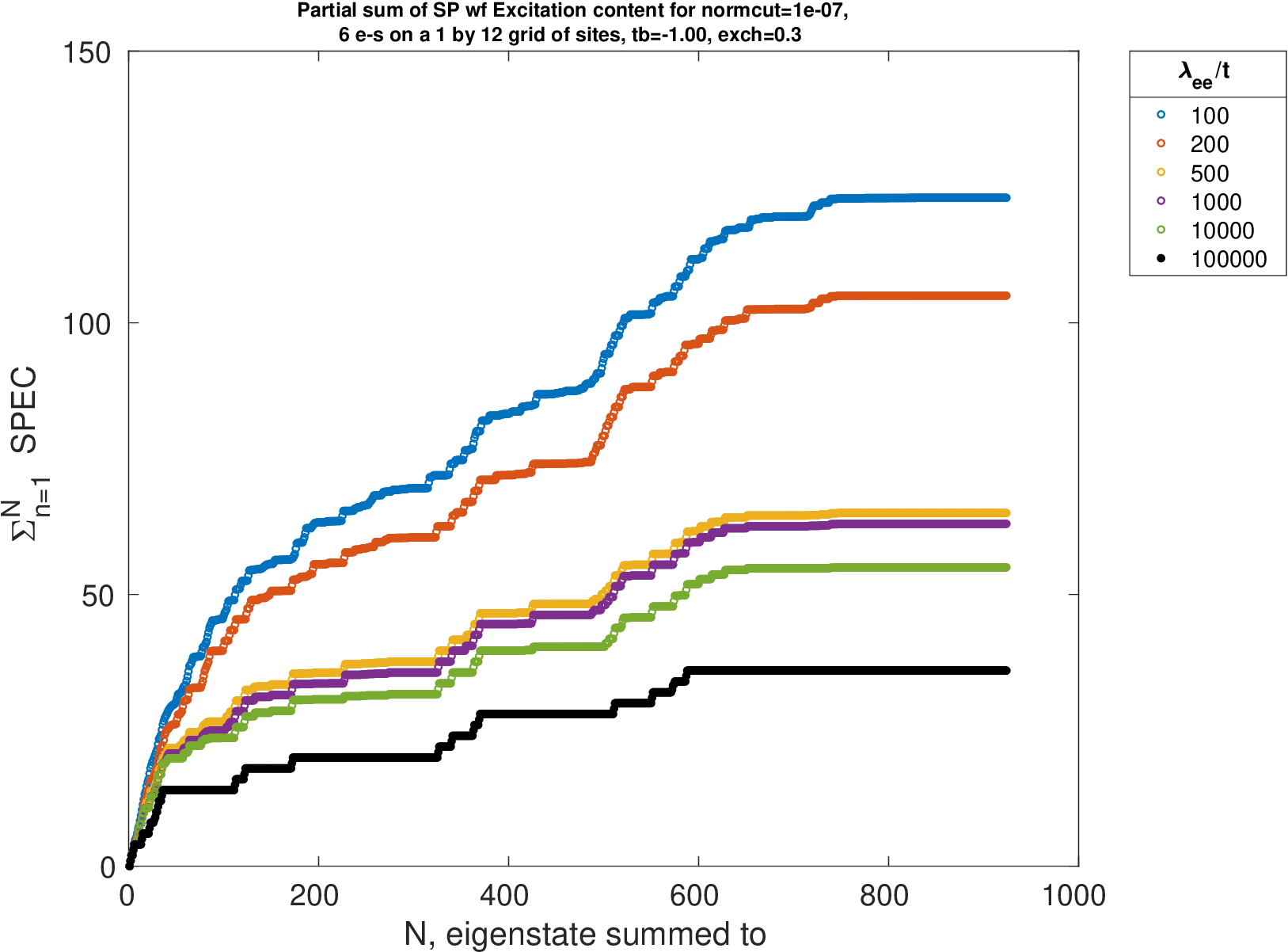}
\includegraphics[width=0.45\textwidth]{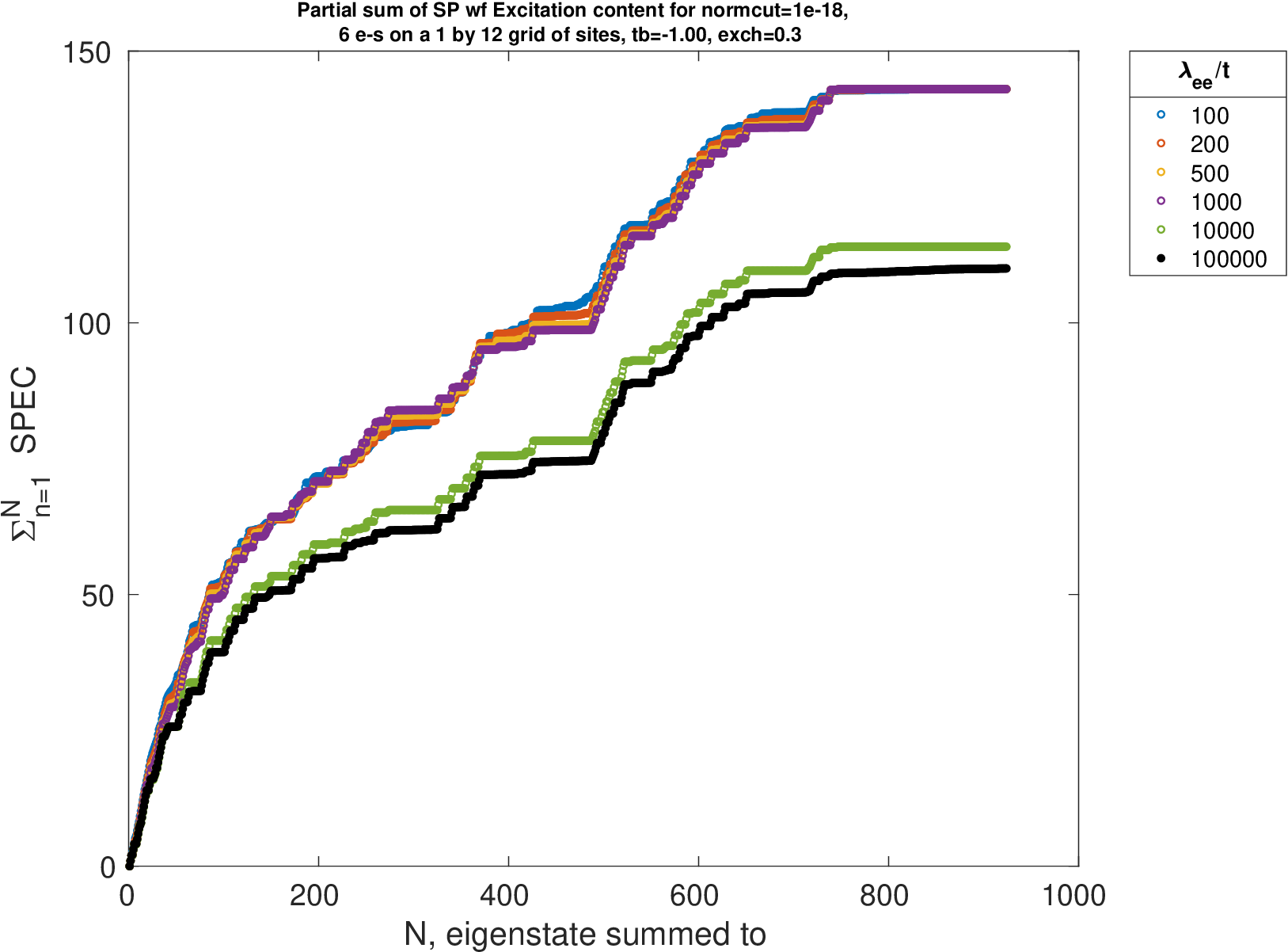}

\includegraphics[width=0.45\textwidth]{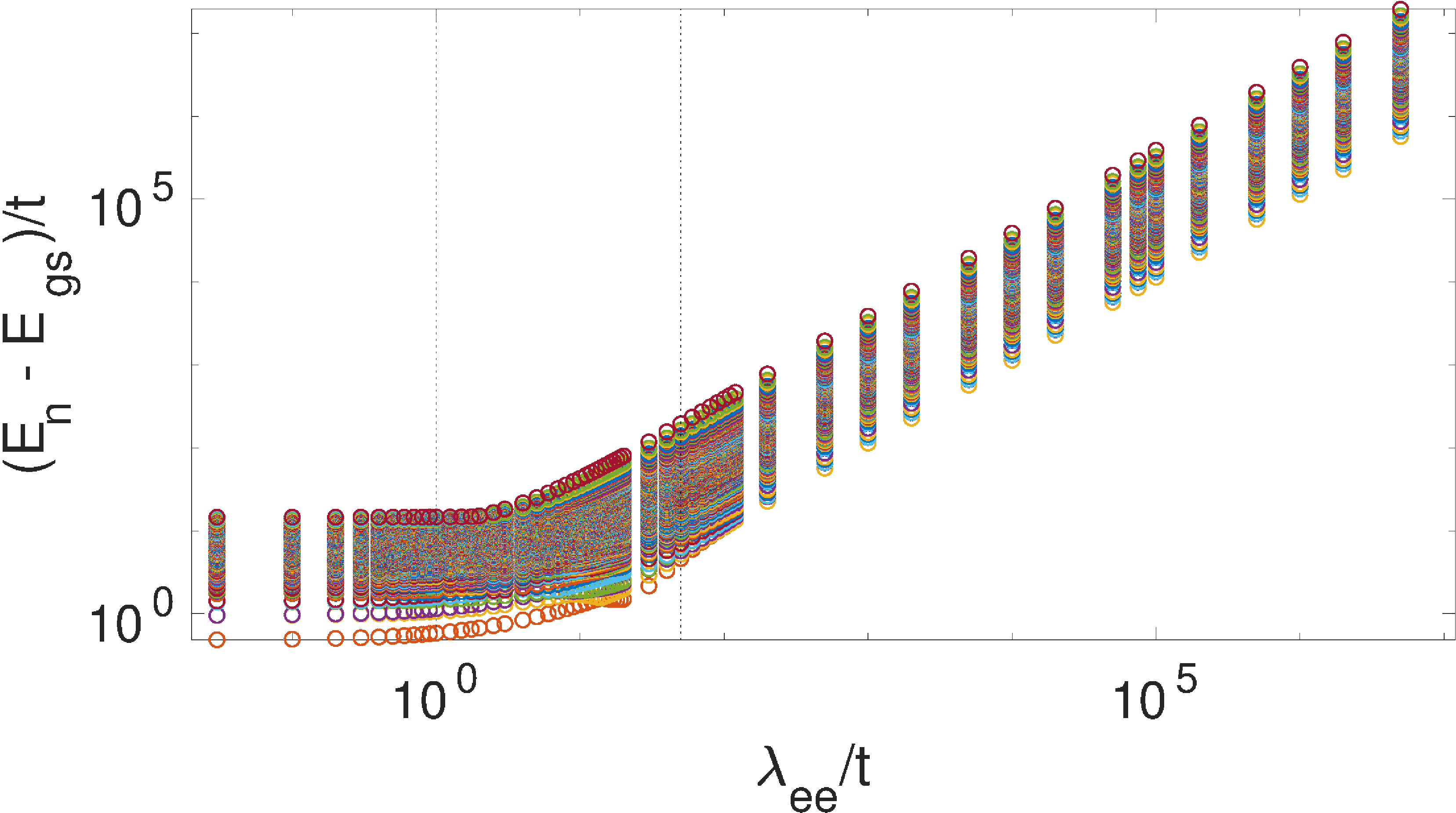}
\caption{Rolling sum of SPEC for 6 electrons in a 12 atom chain with an exchange fraction of $f_{ex}=0.3$.  Plots show different regimes of Coulomb interaction strength, with plots in the left column having a more exclusive cutoff ($\epsilon_{G-S}= 10^{-7}$) and plots in the right column, a more inclusive one ($\epsilon_{G-S} = 10^{-18}$), as in the main text.  Bottom plot shows the excitation energy of the eigenstates as a function of interaction strength (both in units of hopping) on a log-log scale.  The crossover points between strong, intermediate and weak interactions, indicated by vertical lines on the energy plot, have changed as well as the ground state degeneracy, but the behavior of the SPEC distinguishing those regions remains similar to that described in the main text.}
\label{fig:Exch0.3}
\end{figure}

We have also looked at the behavior of the rolling sum of the single-particle excitation content for chains of atoms for different chain lengths, exchange fraction, range of Coulomb interaction and filling.  Figures \ref{fig:NoExch} and \ref{fig:Exch0.3} show the rolling sum of the SPEC for 6 electrons in a twelve atom chain when the exchange fraction reducing the nearest neighbor interaction is 0 and 0.3, respectively. While the change in the exchange fraction can change the pattern of the ordered behavior in the strongly interacting limit (including the degeneracy of the ground state: doubly degenerate for exchange less than $\approx 0.224$, single above), and change the values of interaction strength that define boundaries between the different regions, the behavior of the SPEC is basically unchanged from that described in the main body of the paper.

\begin{figure}
\includegraphics[width=0.45\textwidth]{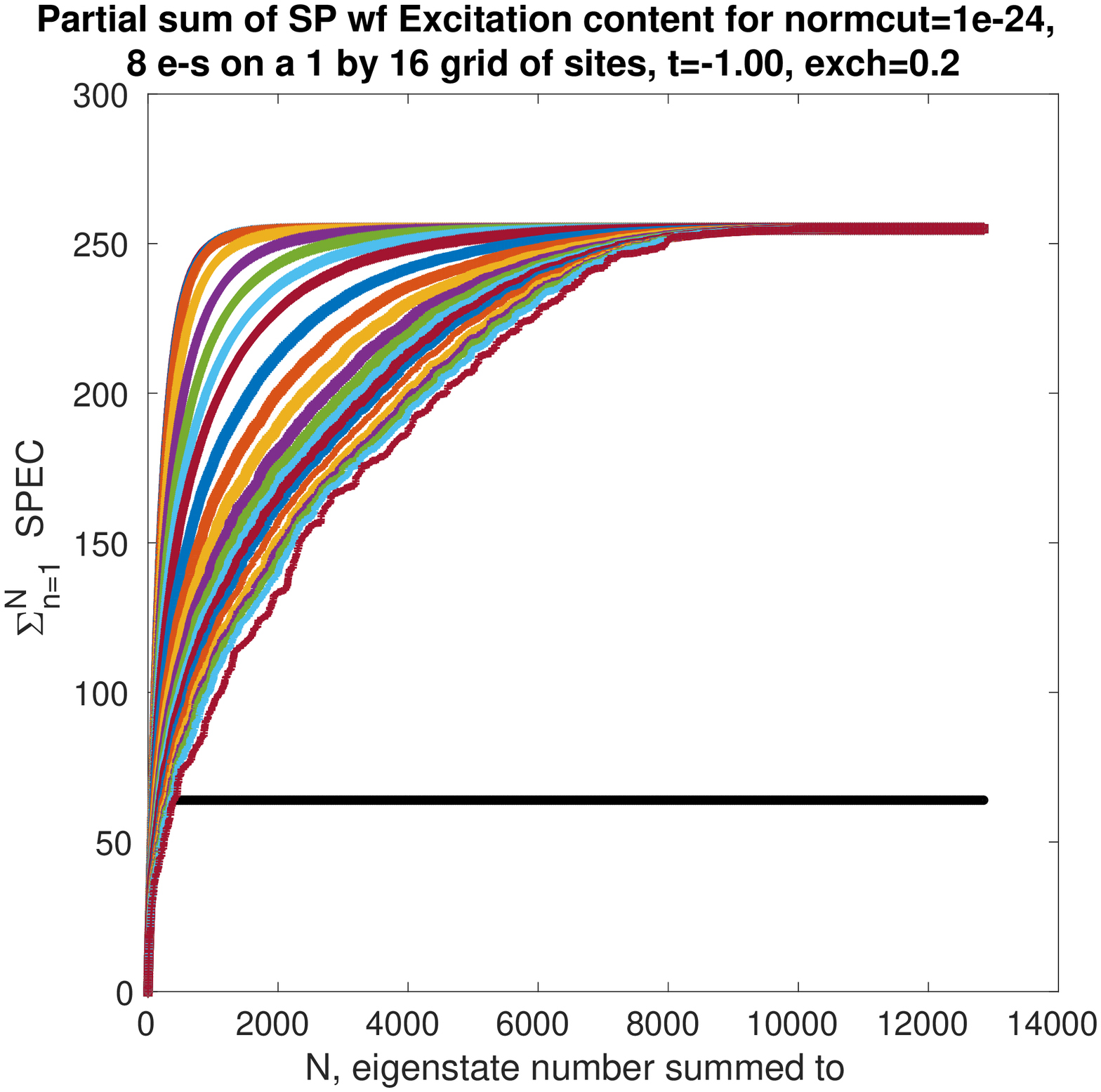}
\includegraphics[width=0.45\textwidth]{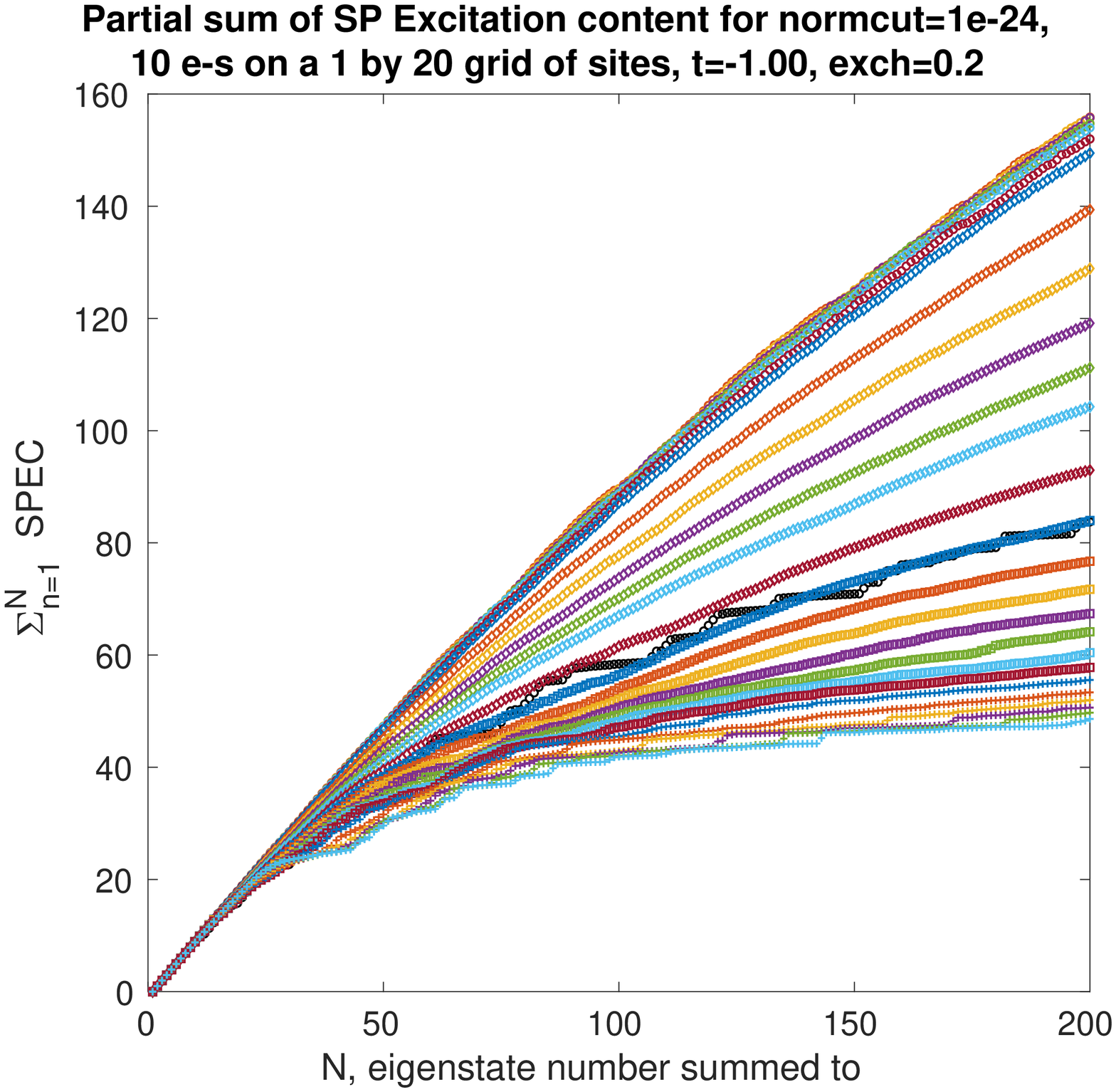}
\caption{Rolling sum of SPEC for 8 electrons in a 16 atom chain (left) and 10 electrons in a 20 atom chain (right), with inclusive cutoffs}
\label{fig:VaryLength}
\end{figure}

 Figure \ref{fig:VaryLength} shows the rolling sum of the SPEC for eight electrons in a sixteen atom chain and ten in a twenty atom chain. (The latter shows only the results for the 200 lowest eigenstates, however even with only the low energy portion of the spectrum visible, the behavior in the twenty atom chain is similar to the twelve and sixteen atom chains, with a weak interaction regime when $\lambda_{ee}/t \leq 1$, and onset and saturation of correlation for  $2<\lambda_{ee}/t \leq 100$.)

\begin{figure}
\includegraphics[width=0.45\textwidth]{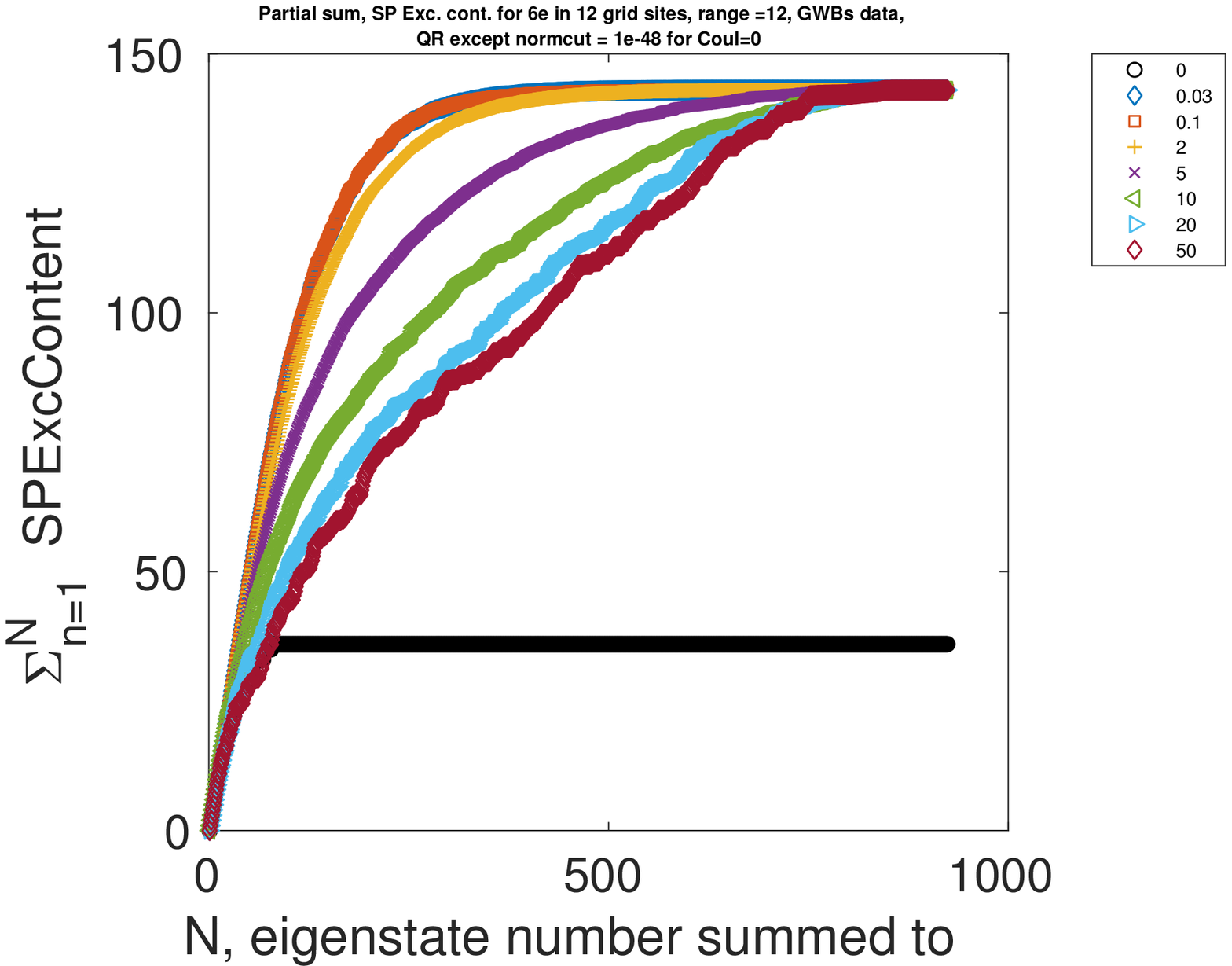}
\includegraphics[width=0.45\textwidth]{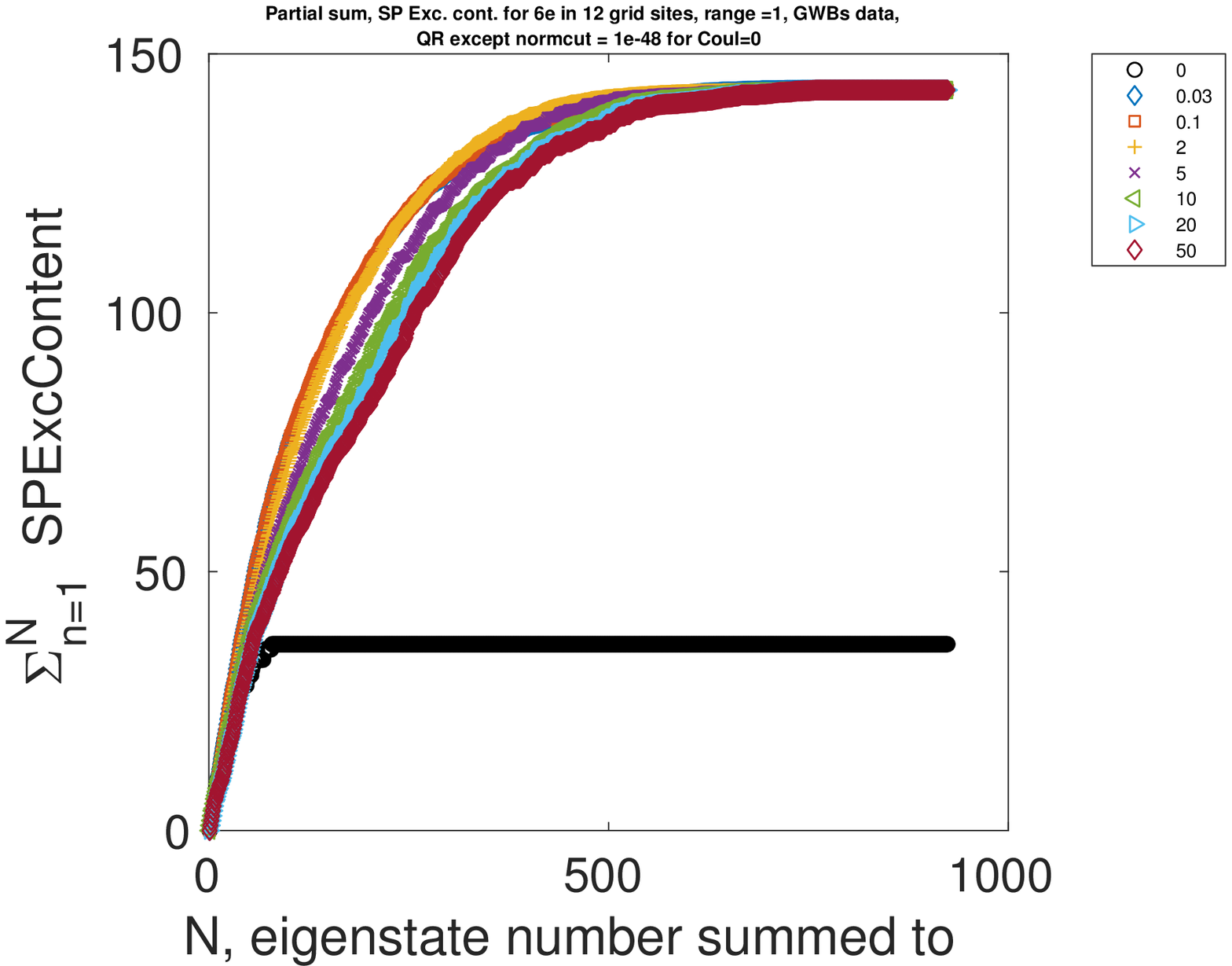}
\caption{Rolling sum of SPEC for 6 electrons in 12 atom chain when the Coulomb interaction fall of like $1/r$ for the full range of the length of the chain (range = 12, left) and when there is only a nearest-neighbor Coulomb interaction (range=1, right).}
\label{fig:VaryRange}
\end{figure}

In the extended Hubbard model studied in the main paper the Coulomb interaction between electrons fall off as $1/r$, allowing interactions that span the full length of the chain.  Figure \ref{fig:VaryRange} compares the rolling sum of the SPEC in this case (range =12) to the case where there is only nearest neighbor electron repulsion (range =1).  Without the long-range of the interaction, a Wigner crystal will not form, nonetheless the onset and saturation of correlation is still visible.  For other ranges of the interaction there are variations in the ground and excited states, some of which are discussed in \cite{Bryant2020}.

\begin{figure}
\includegraphics[width=0.45\textwidth]{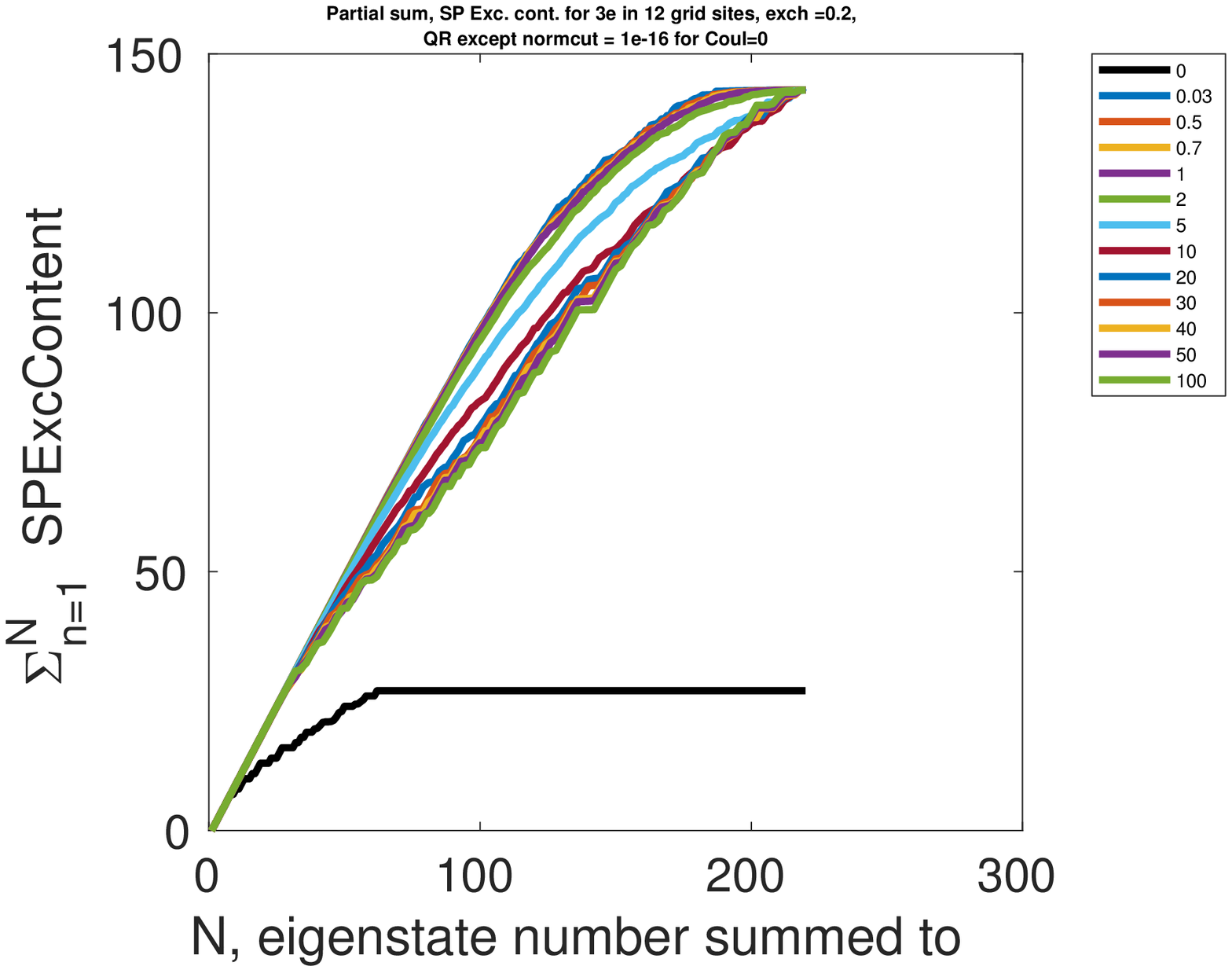}
\includegraphics[width=0.45\textwidth]{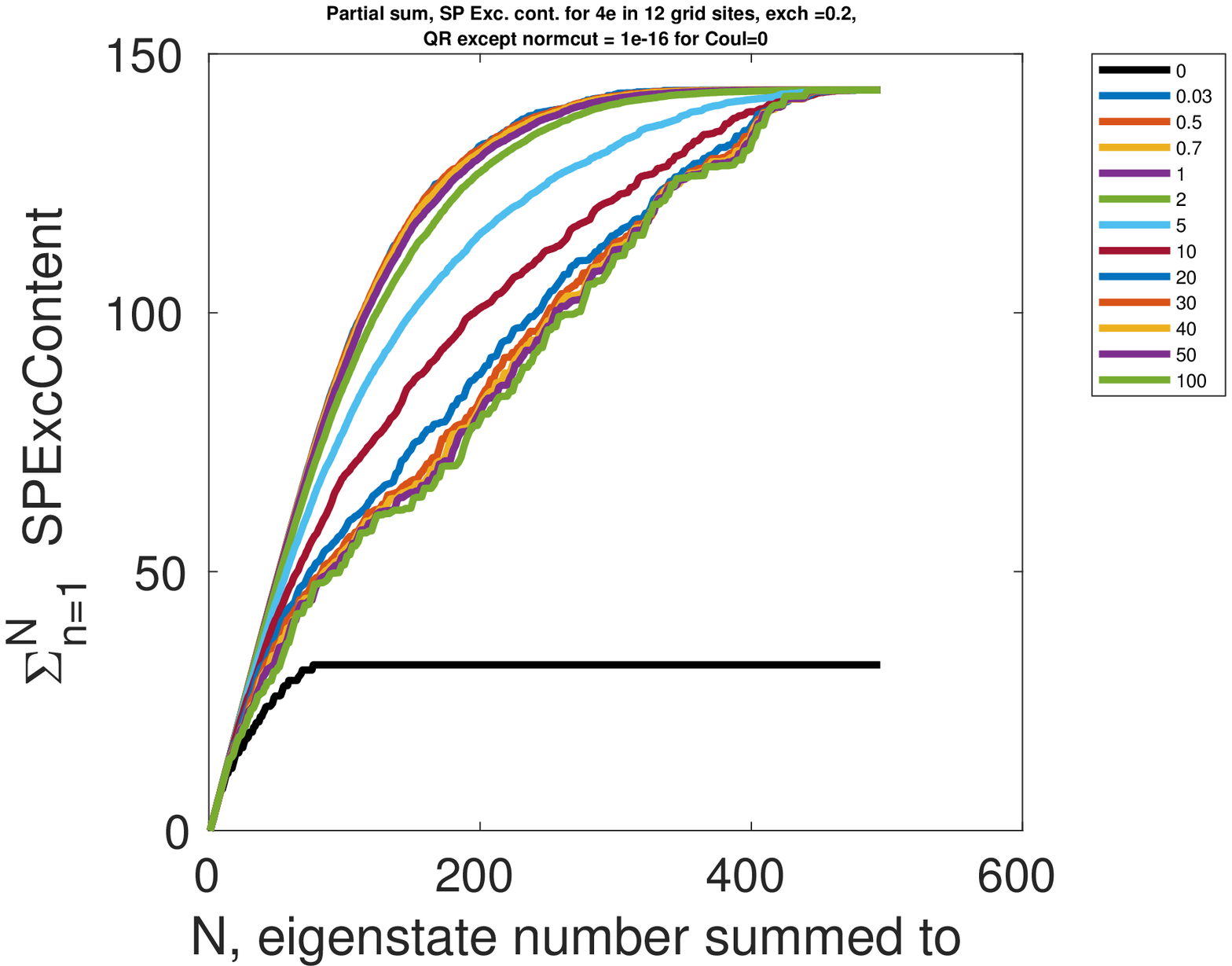}
\includegraphics[width=0.45\textwidth]{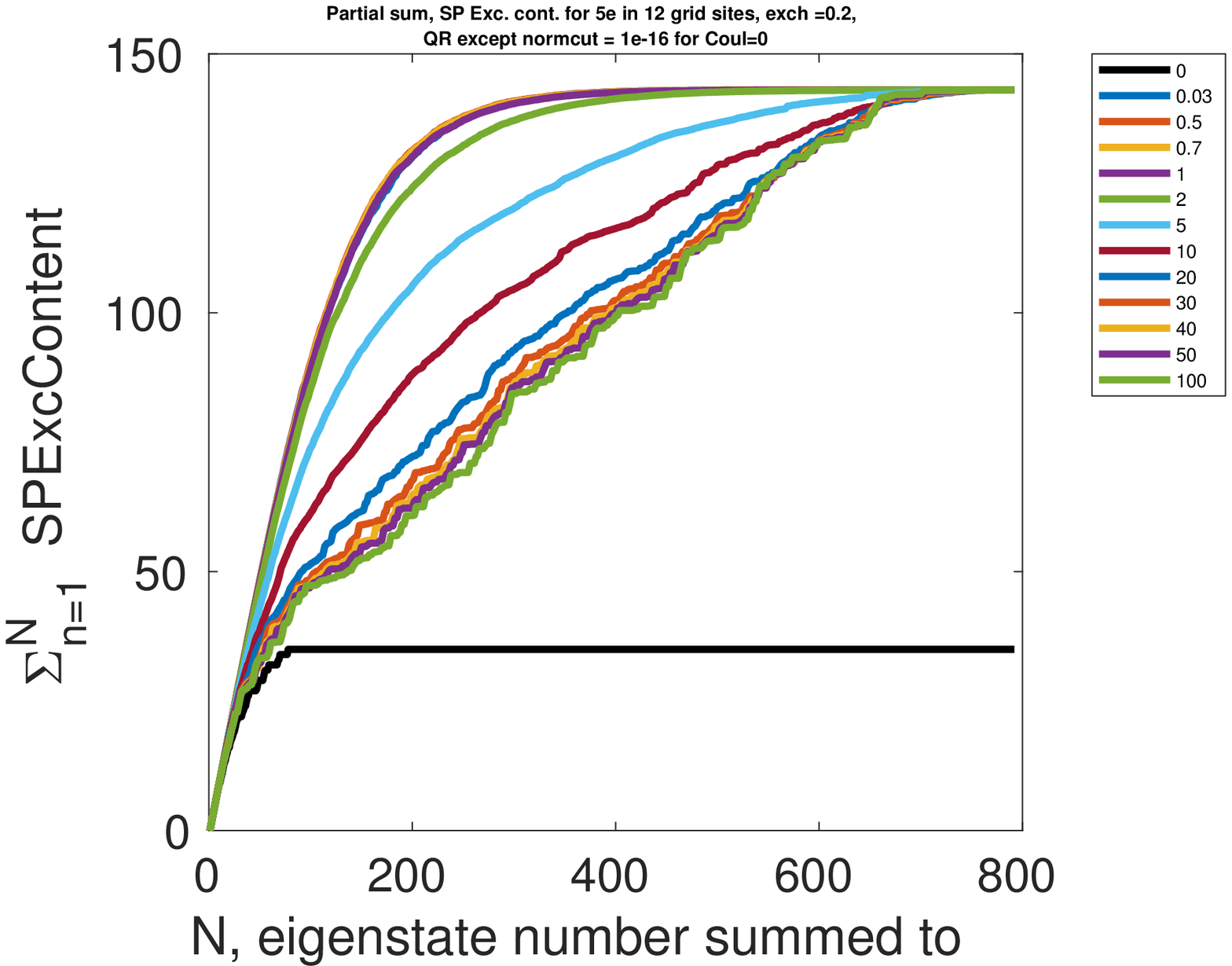}
\includegraphics[width=0.45\textwidth]{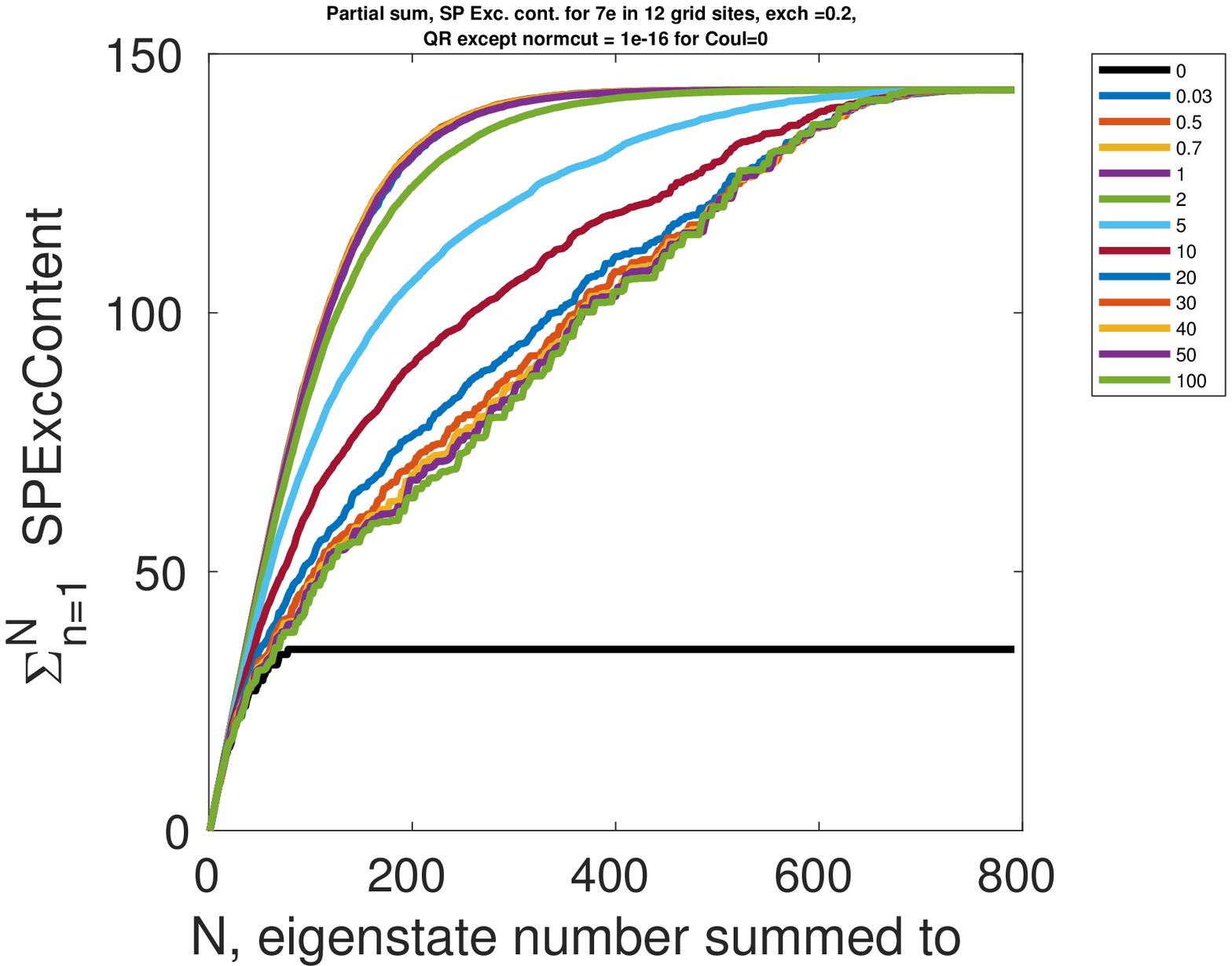}
\includegraphics[width=0.45\textwidth]{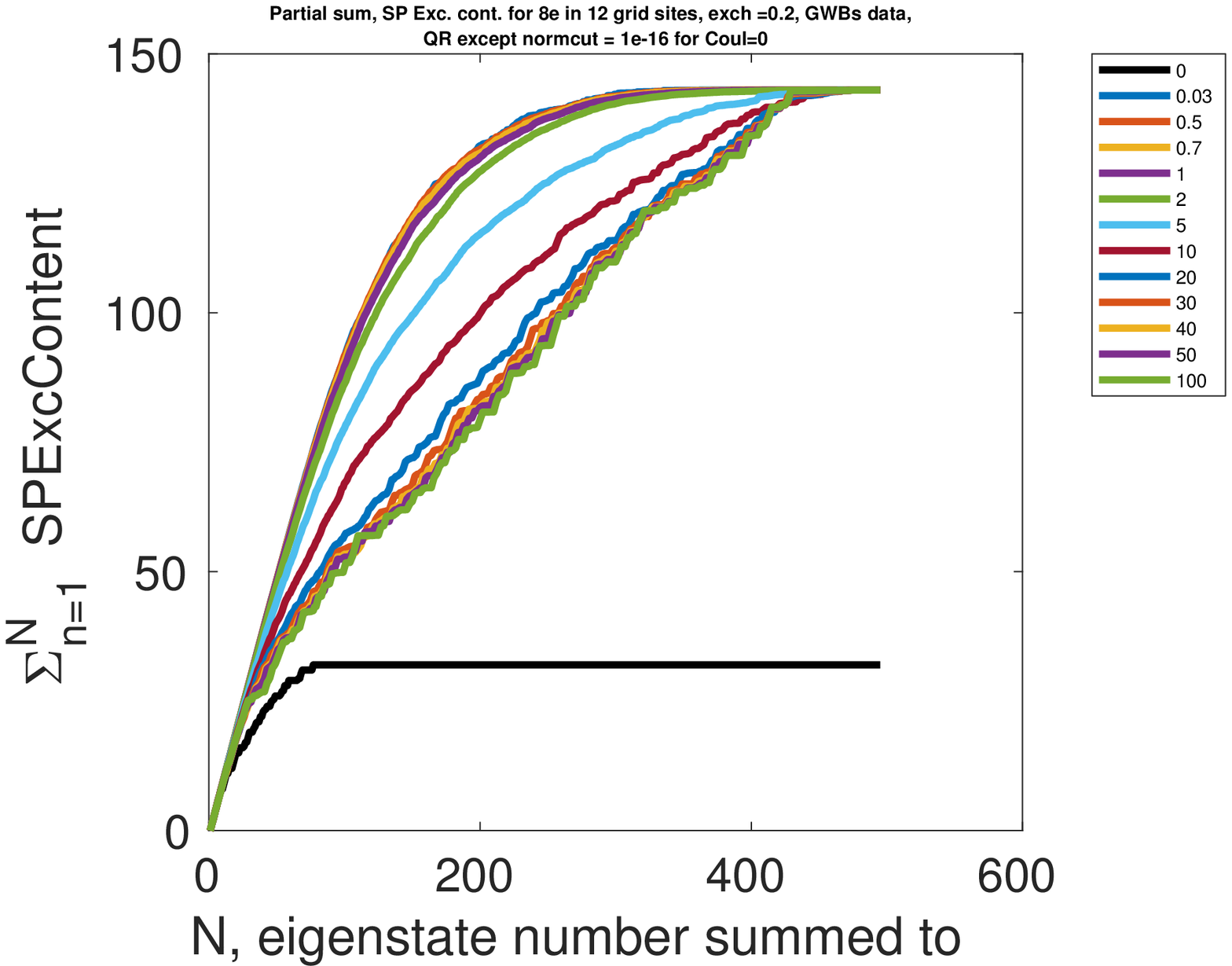}
\includegraphics[width=0.45\textwidth]{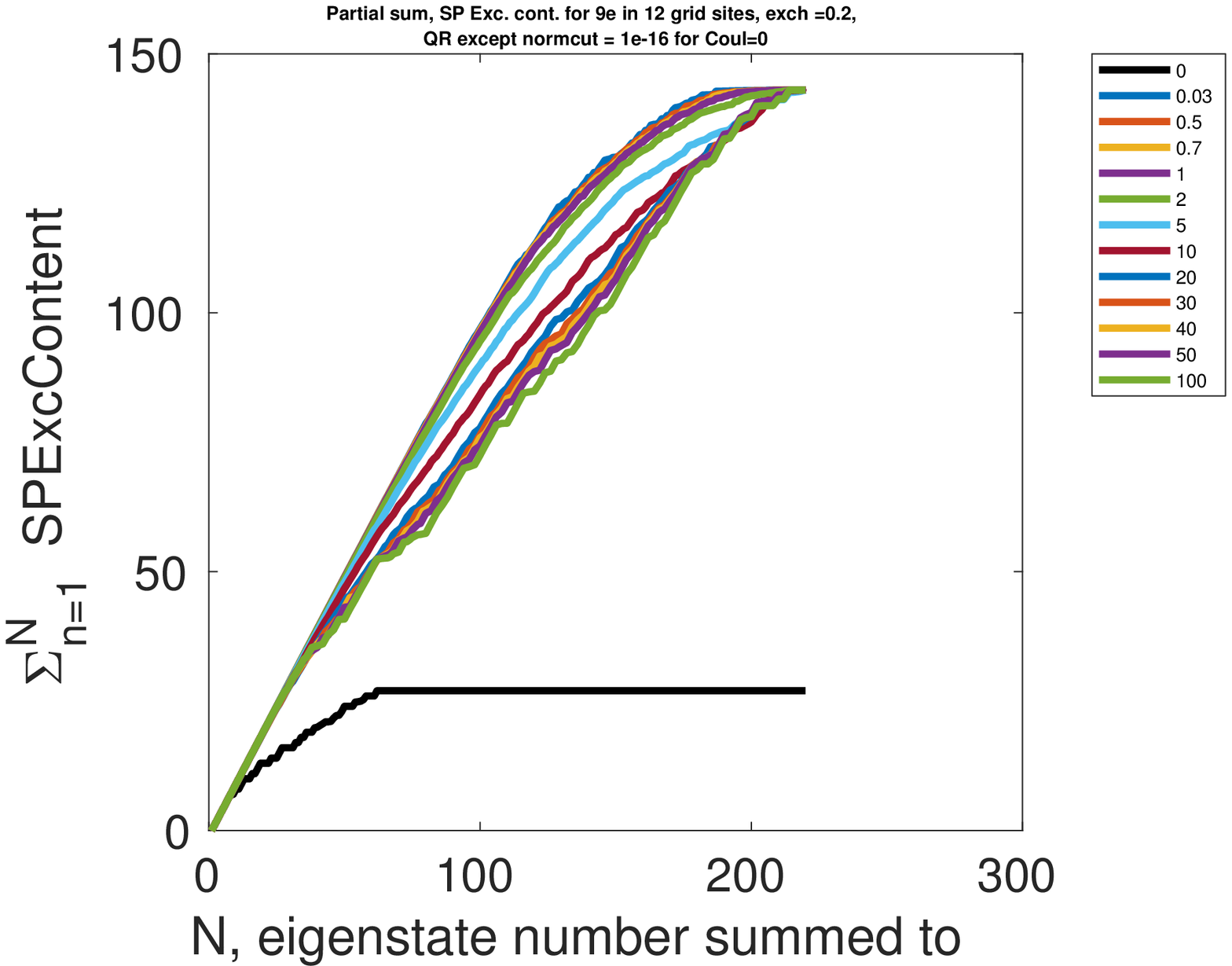}
\caption{Rolling sum of SPEC for three, four, five, seven, eight or nine electrons in 12 atom chain, varying ratio of Coulomb interaction to hopping ($\lambda_{ee}/t$) in each plot}
\label{fig:VaryFilling}
\end{figure}

The results are also not strongly dependent on the chains being half filled.  Figure \ref{fig:VaryFilling} shows the rolling sum of the SPEC for a twelve atom chain with three, four, five, seven, eight, and nine electrons.

\clearpage
\bibliography{MyLibrary}
\bibliographystyle{unsrt}